\documentclass[prd,twocolumn,superscriptaddress,nofootinbib,longbibliography,10pt]{revtex4-1}
\pdfoutput=1

\usepackage{amsmath}
\usepackage{amssymb}
\usepackage{amsfonts}
\usepackage{mathrsfs}
\usepackage[dvipsnames]{xcolor}

\DeclareMathOperator{\R}{\mathbb{R}}
\DeclareMathOperator{\C}{\mathbb{C}}

\DeclareMathOperator{\Z}{\mathbb{Z}}

\newcommand{\mz}{\mathcal{Z}}
\newcommand{\mf}{\mathcal{F}}
\newcommand{\mN}{\mathcal{N}}
\newcommand{\mG}{\mathcal{G}}
\newcommand{\mL}{\mathcal{L}}

\newcommand{\dd}{\mathrm{d}}
\newcommand{\ii}{\mathrm{i}}

\newcommand{\tr}{\mathrm{Tr}}

\newcommand{\thp}{T_{\mathrm{HP}}}

\def\XXint#1#2#3{{\setbox0=\hbox{$#1{#2#3}{\int}$}
     \vcenter{\hbox{$#2#3$}}\kern-.5\wd0}}

\usepackage[utf8]{inputenc}
\usepackage[english]{babel}

\usepackage[pdftex]{graphicx}
\usepackage{enumerate}
\usepackage{tikz}

\usepackage[colorlinks=true,citecolor=Blue,linkcolor=Red]{hyperref}

\begin{document}

	\title{Hawking--Page transition on a spin chain}

\author{David P\'erez-Garc\'ia}
\email{dperezga@ucm.es}
\affiliation{\small Departamento de An\'alisis Matem\'atico y Matem\'atica Aplicada, Universidad Complutense de Madrid, 28040 Madrid, Spain}

\author{Leonardo Santilli}
\email{santilli@tsinghua.edu.cn}
\affiliation{\small Yau Mathematical Sciences Center, Tsinghua University, Beijing, 100084, China}
	
\author{Miguel Tierz}
\email{tierz@mat.ucm.es}	
\affiliation{\small Departamento de An\'alisis Matem\'atico y Matem\'atica Aplicada, Universidad Complutense de Madrid, 28040 Madrid, Spain}

\begin{abstract}

The accessibility of the Hawking--Page transition in AdS$_5$ through a 1d Heisenberg spin chain is demonstrated. We use the random matrix formulation of the Loschmidt echo for a set of spin chains, and randomize the ferromagnetic spin interaction. It is shown that the thermal Loschmidt echo, when averaged, detects the predicted increase in entropy across the Hawking--Page transition. This suggests that a 1d spin chain exhibits characteristics of black hole physics in 4+1 dimensions. We show that this approach is equally applicable to free fermion systems with a general dispersion relation.
\end{abstract}


\maketitle
A fundamental discovery in the search for a quantum theory of gravity was the realization that quantum information is the foundation, while spacetime and the particles within it are emergent. This `it from qubit' approach, originally proposed by Wheeler \cite{Wheeler:1989ftm}, has resulted in significant progress in comprehending the quantum properties of black holes. 

The utilization of quantum information techniques in exploring the emergence of spacetime resulted in various developments, including the ER=EPR conjecture \cite{Maldacena:2013xja}, holographic quantum error-correcting codes \cite{Almheiri:2012rt,Verlinde:2012cy,Almheiri:2014lwa}, and proposals aimed at resolving the information paradox \cite{Penington:2019npb,Almheiri:2019psf}, among others.

Alongside this new method for studying quantum gravity is the advancement of quantum simulations. In brief, quantum simulators are devices designed to replicate physical phenomena impossible to study through experimentation. 

An essential query that arises from the amalgamation of these two areas is the degree to which gravity's manifestations can emerge from a set of qubits. An example in this direction was the quantum simulation of a traversable wormhole \cite{Jafferis:2022crx}.\par
\medskip
In this paper, we examine another important aspect of gravity: the formation of a black hole (BH). Unlike wormholes, which harbour innate quantum properties, BHs are prominently thermal entities. Hence, they can be analysed through conventional means such as finite-temperature lattices. A BH forms in anti-de Sitter (AdS) spacetime with the Hawking--Page transition \cite{Hawking:1982dh}. Our main point asserts that this phenomenon can be simulated on a simple spin chain, the Heisenberg XX chain.

To elaborate, we shall begin with thermal AdS$_5$. As the temperature $T$ rises, a BH solution emerges, which becomes dominant above a threshold $\thp$ \cite{Hawking:1982dh}. It can be inferred that, at $T > \thp$, the entropy would be significantly larger than thermal AdS, due to the presence of the BH horizon. This follows from the Bekenstein--Hawking formula \cite{Bekenstein:1973ur,Hawking:1975vcx} 
\begin{equation}
\label{eq:BHentropy}
	S_{\mathrm{BH}} = \frac{ A_{\mathrm{hor}}}{4 G_{\text{N}}} ,
\end{equation}
with $A_{\mathrm{hor}}$ the area of the BH horizon and $G_{\text{N}}$ the Newton constant in AdS$_5$ (in units $c=\hbar =1$).\par
The celebrated AdS/CFT correspondence \cite{Maldacena:1997re,Gubser:1998bc,Witten:1998qj} posits that gravitational asymptotically-AdS solutions are encoded into a conformal field theory (CFT) on the boundary. The most widely studied example, and the one we will focus on, is the duality between AdS$_5$ and 4d $\mathcal{N}=4$ super-Yang--Mills theory (SYM) \cite{Maldacena:1997re}.\par
\medskip
We will show that the entropy $S_{\mathrm{BH}}$ can be computed from a \emph{coupling average} of thermal, or imaginary time, Loschmidt echoes in the XX chain. Preparing the spins in the initial state 
\begin{equation}
	\lvert \psi_0 \rangle = \lvert \underbrace{ \downarrow  \downarrow \dots \downarrow }_{N}  \uparrow  \uparrow \dots \uparrow \rangle ,
\label{eq:psi0}
\end{equation}
we claim, schematically,
\begin{equation}
\boxed{\hspace{12pt}
	S_{\mathrm{BH}} \sim \ln \left( \int_0 ^{\infty} \dd J ~ e^{-\frac{J^2}{4a}} ~\langle \psi_0 \lvert e^{- J H_{\mathrm{XX}}} \rvert \psi_0 \rangle \right) ,\hspace{12pt}}
\end{equation}
with average over the coupling $J$ of the XX chain [the accurate statement is \eqref{eq:main}]. The parameter $a$, controlling how $J$ is sampled, is a monotonic function of the BH temperature $T$, cf. \eqref{eq:defaT}. In particular, we show that the averaged Loschmidt echo detects the jump in entropy predicted by the Hawking--Page transition. Our result implies that a 1d spin chain captures features of BH physics in 4+1 dimensions.\par
In the AdS$_5$/CFT$_4$ dictionary, the entry 
\begin{equation}
\label{eq:GNisN}
	G_{\text{N}} \sim 1/N^2  
\end{equation}
relates the Newton constant and the rank of SYM. The regime $G_{\text{N}} \to 0$, in which gravity becomes tractable, corresponds via \eqref{eq:GNisN} to the large $N$ limit of $SU(N)$ $\mN=4$ SYM \cite{Maldacena:1997re}. Combining the Bekenstein--Hawking formula \eqref{eq:BHentropy} with the AdS/CFT relation \eqref{eq:GNisN}, a Hawking--Page transition is characterized by 
\begin{equation}
\label{eq:lnZorder}
	S \sim \begin{cases} \hspace{12pt} s_{\text{low}} \quad & T < \thp \\ N^2 s_{\text{high}}  \quad & T > \thp \end{cases}
\end{equation}
where $S$ is the entropy and $s_{\text{low}},s_{\text{high}}$ are $\mathcal{O}(1)$ constants.
Enormous effort has been put to test and sharpen this prediction \cite{Sundborg:1999ue,Polyakov:2001af,Balasubramanian:2002sa,Aharony:2003sx,Liu:2004vy,Berenstein:2004kk,Alvarez-Gaume:2005dvb,Festuccia:2005pi,Basu:2005pj,Alvarez-Gaume:2006fwd,Biswas:2006tj,Chang:2013fba,Hosseini:2017mds,Cabo-Bizet:2018ehj,Choi:2018hmj,Choi:2018vbz,Benini:2018mlo,Benini:2018ywd,Honda:2019cio,ArabiArdehali:2019tdm,Cabo-Bizet:2019osg,GonzalezLezcano:2019nca,Larsen:2019oll,Cabo-Bizet:2019eaf,ArabiArdehali:2019orz,Benini:2020gjh,David:2020ems,Cabo-Bizet:2020nkr,Agarwal:2020zwm,GonzalezLezcano:2020yeb,Copetti:2020dil,Goldstein:2020yvj,Amariti:2020jyx,Amariti:2021ubd,Cassani:2021fyv,Aharony:2021zkr,Uhlemann:2021nhu,Hong:2021bzg,Ezroura:2021vrt,Imamura:2021ytr,Gaiotto:2021xce,Cabo-Bizet:2021jar,Murthy:2022ien,Boruch:2022tno,Honda:2022hvy,Lee:2022vig,Huang:2022bry,Imamura:2022aua,Holguin:2022drf,Eleftheriou:2022kkv,Choi:2022ovw,Chang:2022mjp,Chen:2022tfy,Liu:2022olj,Lin:2022gbu,Eniceicu:2023uvd,Beccaria:2023zjw,Chang:2023zqk,Chen:2023lzq,Ekhammar:2023glu,Bigazzi:2023hxt,Beccaria:2023hip,ArabiArdehali:2023bpq,Chang:2023ywj,Lee:2023iil,Eleftheriou:2023jxr}.\par
\begin{figure}
\centering
\includegraphics[width=0.9\linewidth]{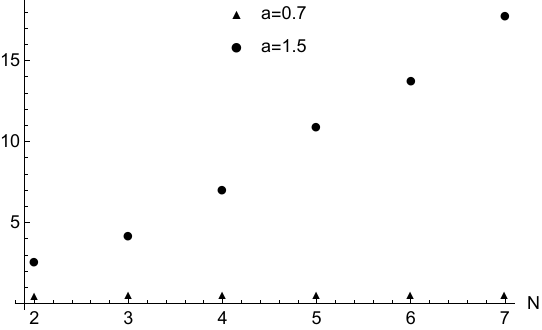}
\caption{$\ln \left\langle \sqrt{ \hat{\mL}_N} \right\rangle_{2a}$ as a function of $N$ at $T<\thp$ (${\scriptstyle \blacktriangle}$) and $T>\thp$ ($\bullet$).}
\label{fig:LvsN}
\end{figure}

\medskip
We demonstrate how the distinctive behaviour \eqref{eq:lnZorder} can be replicated on the spin chain (Fig.~\ref{fig:LvsN}), primarily focusing on weak gauge coupling. Remarkably, finite coupling corrections --- a key factor in achieving more credible results --- incur no further complexity in the setup of the spin chain (see \S\ref{app:RP1point5}), although it requires repeating the measurement multiple times.

\section{Hawking--Page transition}

In this paper we focus on the BH formation in maximally-supersymmetric AdS$_5$, whose holographic dual is $\mN=4$ SYM. On the one hand, maximal supersymmetry plus conformal symmetry tightly constrain the theory, making it the perfect playground to explore 4d physics. On the other hand, it bears universal properties akin to pure Yang--Mills and QCD \cite{Polchinski:2001tt,Bern:2004kq,Caron-Huot:2008dyw}, befitting  phenomenological applications.\par
The AdS/CFT correspondence dictates that the Hawking--Page transition \cite{Hawking:1982dh} in AdS$_5$ is dual to a first order transition in $\mN=4$ SYM on Euclidean $\mathbb{S}^3 \times \mathbb{S}^1$ \cite{Witten:1998zw} (details in \S\ref{app:SCI}), with thermal $\mathbb{S}^1$ of radius $1/T$. An effective description for this theory, in the weak coupling limit, was devised in \cite{Sundborg:1999ue,Polyakov:2001af,Aharony:2003sx}. The result, written in terms of the AdS$_5$ entropy $S$, is:
\begin{equation}
\label{eq:MMfull}
	S= \ln \left[ \oint \dd U ~ \exp \left( \sum_{n \ge 1} \frac{a_n}{n} \tr \left( U^{n} \right) \tr \left( U^{-n} \right) \right)  \right] .
\end{equation}
$U$ is the holonomy of the gauge field around $\mathbb{S}^1$, $\dd U$ is the normalized Haar measure and $a_n$ are effective couplings produced integrating out the massive modes. They depend on $T$, for instance, 
\begin{equation}
\label{eq:defaT}
	a \equiv a_1 = \frac{2(3e^{1/(2T)} -1)}{(e^{1/(2T)}-1)^3}  .
\end{equation}\par
At large $N$, \eqref{eq:MMfull} is approximated by the simpler matrix model \cite{Aharony:2003sx}
\begin{equation}
\label{eq:MM1}
	\hat{S} = \ln \left[ \oint \dd U ~ \exp \left( a \tr \left( U \right) \tr \left( U^{-1} \right)  \right) \right]
\end{equation}
with higher couplings $a_{n>1}$ being irrelevant perturbations, negligible in the analysis of the phase transition (refinements by higher couplings are dealt with in \ref{sec:improve}). The notation $\hat{S}$ stresses the difference with the exact quantity $S$.\par
The large $N$ limit of \eqref{eq:MM1} was solved in \cite{Liu:2004vy}. One rewrites
\begin{equation}
\label{eq:MM2}
	e^{\hat{S}}=  \int_{0} ^{\infty} \sigma \dd \sigma ~ \exp \left[  - N^2 \left( \frac{\sigma^2}{{4 a} } - \mf (\sigma) \right) \right] ,
\end{equation}
with $\mf (\sigma) = \frac{1}{N^2} \log \mz^{\text{GWW}} _N (N\sigma)$ the free energy of the Gross--Witten--Wadia (GWW) matrix model \cite{Gross:1980he,Wadia:1980cp,Wadia:2012fr}
\begin{equation}
\label{eq:GWW}
	 \mz^{\text{GWW}} _N (N\sigma) = \oint \dd U ~ \exp \left\{ \frac{N \sigma}{2}  \tr \left( U + U^{-1} \right)  \right\} .
\end{equation}
Then \eqref{eq:MM2} is solved in two steps: ($i$) apply the planar limit to the GWW model, and ($ii$) minimize the resulting effective action by a standard saddle point analysis. The upshot is \cite{Liu:2004vy}:
\begin{itemize}
\item If $a<1$, the saddle is $\sigma=0$ , thus $\hat{S} (a <1)  \sim \mathcal{O} (1)$;
\item If $a>1$, there is a non-trivial saddle, and $\hat{S} (a >1) = N^2 s_{\text{high}}$.
\end{itemize}
Therefore \eqref{eq:MM2} undergoes a first order transition at 
\begin{equation}
    a (\thp) =1 \quad \Longrightarrow \quad \thp \approx 0.38
\end{equation}
separating the two phases \eqref{eq:lnZorder}.\par

\section{Mapping to the spin chain}
\label{sec:chain}
Our novel claim is that the Hawking--Page transition in supersymmetric AdS$_5$ can be encoded in a 1d quantum spin chain. The simulation of such thermodynamic process requires a chain at finite temperature. The quantity detecting the sudden change in entropy is the Loschmidt echo \cite{Gorin:2006}, an observable tailored to quantify how chaotic a system is. It has already found broad application in BH physics (e.g.~\cite{delCampo:2017ftn,Chenu:2017qdv,Chenu:2018spm,Almheiri:2021jwq}). Our insight is to interpret \eqref{eq:MM2} as a \emph{coupling average} of the Loschmidt echo in the XX chain.\par
Consider a spin-$\frac{1}{2}$ chain of $L$ sites, and prepare the initial state $\lvert \psi_0 \rangle $ in \eqref{eq:psi0}. We are interested in the regime $1 \ll N \ll L$ and assume periodic boundary conditions, but the results hold more generally \cite{Perez-Garcia:2022geq}. The system is at temperature $\tilde{T}>0$ and evolves with the XX Hamiltonian \cite{Lieb:1961fr}
\begin{equation}
\label{eq:HXY}
	H_{\mathrm{XX}} = - \frac{\tilde{J}}{2} \sum_{j=0} ^{L-1}	\left(  \sigma_j ^{-} \sigma_{j+1} ^{+} + \sigma_j ^{+} \sigma_{j+1} ^{-} \right) ,
\end{equation}
where $\tilde{J}>0$ is the ferromagnetic coupling and $\sigma_j ^{\pm} = \left( \sigma_j^{x} \pm i \sigma_j^{y} \right)/2$ are the spin-flip operators on the site $j$, satisfying
\begin{align}
	&\sigma^{+} \lvert \downarrow \rangle = \lvert \uparrow \rangle , \quad \sigma^{-} \lvert \uparrow \rangle =\lvert \downarrow \rangle , \quad \sigma^{+} \lvert \uparrow \rangle =0  =\sigma^{-} \lvert \downarrow \rangle  \\
	& \left[ \sigma^+ _j, \sigma^- _k \right]  = \sigma^z_j \delta_{jk}, \qquad \left[ \sigma^z_j, \sigma^{\pm} _k \right] = \pm 2 \sigma^{\pm}_j  \delta_{jk} . \label{eq:commrel}
\end{align}\par
We consider the \emph{thermal} Loschmidt amplitude 
\begin{equation}
\label{eq:G}
	\mG_N (J) =  \langle \psi_0 \lvert e^{- H_{\text{XX}} /\tilde{T} } \lvert \psi_0 \rangle 
\end{equation}
and the corresponding echo:
\begin{equation}
\label{eq:L}
	\mL_N (J) = \left\lvert \mG_N (J) \right\rvert^2 .
\end{equation}
Note that $\tilde{T}$ is the temperature of the chain, unrelated to the BH temperature $T$. Furthermore, \eqref{eq:L} only depends on  $J\equiv \tilde{J}/\tilde{T} $. In \eqref{eq:G} we used the standard replacement $it \mapsto 1/\tilde{T}$. In this way, $\mL_N $ is not a probability, because it is not canonically normalized. It is thus convenient to work with the ratios 
\begin{equation}
\label{eq:defhatL}
	\hat{\mL}_{N} (J) = \mL_N (J) / \mL_1 (J) .
\end{equation}
One advantage of this prescription, is that $\hat{\mL}_{N}$ remains finite when $L \to \infty$.\par
Next, we map \eqref{eq:G} to the GWW model \eqref{eq:GWW} \cite{BPT,Perez-Garcia:2013lba}. We introduce $\lvert \Uparrow \rangle \equiv \lvert \uparrow , \uparrow  , \dots, \uparrow \rangle$ and
\begin{equation}
\label{eq:gjk}
	g_{j,k} (J) = \langle \Uparrow\rvert \sigma_{j}^{+} e^{- H_{\mathrm{XX}} /\tilde{T}} \sigma_k ^{-} \lvert\Uparrow \rangle .
\end{equation}
Observe that, as $L \to \infty$ \cite{BPT}
\begin{equation}
	\langle \Uparrow\rvert \left( \bigotimes_{j=0}^{N-1} \sigma_j ^{+} \right) e^{- H_{\mathrm{XX}} /\tilde{T}} \left( \bigotimes_{k=0}^{N-1}\sigma_k ^{-} \right) \lvert\Uparrow \rangle = \det_{0 \le j,k \le N-1} \left[ g_{j,k} \right] .
\end{equation}
Besides, $\lvert \psi_0 \rangle =  \bigotimes_{k=0}^{N-1}\sigma_k ^{-} \lvert\Uparrow \rangle $. Using \eqref{eq:commrel} to pass $H_{\mathrm{XX}}$ across $\sigma_{j}^{+}$, and $H_{\mathrm{XX}} \lvert \Uparrow \rangle =0 $, $g_{j,k} $ satisfies 
\begin{equation}
	\frac{\dd g_{j,k} }{\dd J}   = \frac{1}{2} \langle \Uparrow\rvert \left( \sigma_{j-1}^{+} + \sigma_{j+1}^{+}\right) e^{- H_{\mathrm{XX}}/\tilde{T} } \sigma_k ^{-} \lvert\Uparrow \rangle .
\end{equation}
Therefore $g_{j,k}$ solves 
\begin{equation}
\label{eq:dgjkdJ}
	\frac{\dd g_{j,k} }{\dd J}   = \frac{1}{2} \left( g_{j-1, k}+g_{j+1, k} \right)
\end{equation}
with initial condition $g_{j,k} (0) \propto \delta_{j,k}$. This is the recurrence relation of the Bessel function $I_{j-k} (J)$, thus 
\begin{equation}
	\frac{\mG_N (J)}{\mG_1 (J)} = \frac{1}{I_0 (J)} \det_{0 \le j,k \le N-1} \left[ I_{j-k} (J) \right] .
\end{equation}
Using the Heine--Szeg\H{o} identity 
\begin{equation}
	\det_{0 \le j,k \le N-1} \left[ I_{j-k} (J) \right]  = \mz^{\text{GWW}} _N \left( N\sigma = J \right) 
\end{equation}
we conclude that 
\begin{equation}
\label{eq:LisZ}
	\hat{\mL}_N (J) = \frac{1}{I_0 (\tilde{J}/\tilde{T})^2} \left\lvert \mz^{\text{GWW}} _N \left( N\sigma = \tilde{J}/\tilde{T} \right) \right\rvert^2  .
\end{equation}
The denominator is independent of $N$. The relation \eqref{eq:LisZ} has been previously exploited to analyze the echo of the XX chain \cite{Perez-Garcia:2013lba,Stephan:2013,Pozsgay:2013,Perez-Garcia:2014aba,Viti:2016,Krapivsky:2017sua,Stephan:2017,Santilli:2019wvq,Stephan:2021yvk,Perez-Garcia:2022geq,Parez:2022sgc,Vleeshouwers:2022eou}.\par
We revert the argument and consider the Loschmidt echo as the building block for the realization of the Hawking--Page transition \eqref{eq:lnZorder} on a quantum apparatus.\par
Comparing the exact identity \eqref{eq:LisZ} with \eqref{eq:MM2}, identifying $N \sigma = J$, the integral over $\sigma$ in $\hat{S}$ is translated into a Gaussian average over $J$. This is enforced by randomizing the coupling $\tilde{J}$ at fixed $\tilde{T}$.\par 
We thus propose the following setup. $\hat{\mL}_N (J)$ is measured in the regime $1 \ll N \ll L$, with $J$ as above. The experiment is run several times, sampling $J$ from a suitably chosen random distribution. The average echo is computed:
\begin{equation}
\label{eq:avgG}
	\left\langle \sqrt{ \hat{\mL}_N} \right\rangle_{2a} = \int_0 ^{\infty} \frac{J \dd J}{2a}  ~ e^{- \frac{J^2}{4a} } \sqrt{ \hat{\mL}_N (J) } 
\end{equation}
The dependence on $\tilde{T}$ is reabsorbed in the integration variable, thus the only free parameter is $a>0$. We refer to $2a$ as the standard deviation. The measurement can be performed at \emph{arbitrary} non-zero temperature $\tilde{T}$.\par
Altogether we arrive at the relation between the BH entropy and the averaged Loschmidt echo:
\begin{equation}
\label{eq:main}
	 \boxed{\hspace{12pt} e^{\hat{S}} \sim \left\langle \sqrt{ \hat{\mL}_N} \right\rangle_{2a}  . \hspace{12pt} }
\end{equation}
The symbol $\sim$ means equality up to details negligible when $N \gg 1$. The temperature $\tilde{T}$ is unimportant and distinct from the BH temperature $T$, which enters the left-hand side of \eqref{eq:main} via $a$. Moreover, \eqref{eq:main} is expressed using $\sqrt{\hat{\mL}_N}$, rather than $\hat{\mG}_N$, because only the former is a measurable quantity.\par
\begin{figure}
\centering
\includegraphics[width=0.9\linewidth]{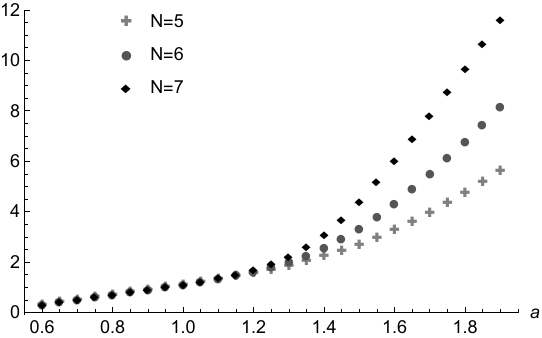}
\caption{$\ln \left\langle \sqrt{ \hat{\mL}_N} \right\rangle_{2a}$ as a function of $a(T)$ for various $N$ at $L=18$.}
\label{fig:LvsT}
\end{figure}\par
We conclude that, varying $a$ (sampling $J$ with different law), the averaged Loschmidt echo \eqref{eq:avgG} with $N \gg 1$ experiences a sharp change in behaviour, signalling the Hawking--Page transition (Fig.~\ref{fig:LvsT}).\par 
Intriguingly, to produce a jump in the echo that simulates the BH entropy, an average is introduced in the spin chain:
\begin{itemize}
\item $0 <a \ll 1$: the average over $J$ is sharply peaked, introducing ``a small amount of disorder'';
\item $a \gg 1$: the average introduces more entropy in the system because no value of $J$ is preferred, resulting in ``more disorder''.
\end{itemize}

\subsection*{Interpretation of the result}
The AdS$_5$/CFT$_4$ correspondence led us to an equality relating the large $N$ BH entropy and the coupling averaged Loschmidt echo. On the BH side of \eqref{eq:main}, the only parameter is the temperature $T$, which appears through $a$. The dependence \eqref{eq:defaT} is monotonic, thus it can be inverted and the BH temperature can be thought of as a function of $a$.\par 
On the spin chain side of \eqref{eq:main}, the free parameter $a$ determines how the coupling is sampled. It can be varied freely, and does not require tuning. The experiment is repeated many times, varying $J$ as sampled randomly according to \eqref{eq:avgG}, and computing the average of the measured Loschmidt echo. The right-hand side of \eqref{eq:main} is a controlled setup that should experience a sharp change in behaviour, when the procedure is performed sampling the coupling at $a<1$ versus $a>1$ (Fig.~\ref{fig:LvsN}).\par 
In conclusion, our formula indicates how to concretely test and validate the Hawking--Page prediction, in the spirit of quantum simulation.

\subsection*{Fermionic systems}

If the chain is fermionized using the Jordan--Wigner transformation \cite{Jordan:1928wi}, \eqref{eq:LisZ} remains valid \cite{Krapivsky:2017sua}, resulting in a completely equivalent outcome.\par
A matrix model description for the Loschmidt echo of fermionic systems corresponding to more complex spin chains is shown to hold in \S\ref{app:fermions}. The fermionic approach introduces a new set of models with implications for improving the estimation of the BH entropy, as well as for potential experimental applications.

\section{Refined probes}
\label{sec:improve}
We explain four ways to refine \eqref{eq:main} for more accurate tests of the Hawking--Page transition. We discuss:
\begin{enumerate}[1.]
    \item The analysis of the order parameters for the transition;
    \item The inclusion of corrections by higher operators;
    \item The inclusion of corrections in the 't Hooft coupling;
    \item The case of complex fugacities for the black hole charges.
\end{enumerate}
The technical details are relegated to \S\ref{app:RP}.\par

\subsection*{Order parameters}
\textbf{1.}  A fine probe of the transition, from the CFT side of the holographic duality, is the Polyakov loop. This is defined as a Wilson loop winding around the Euclidean time direction. On general grounds, its expectation value $\mathcal{P}$ vanishes if $T < \thp$ and is non-vanishing in a BH phase $T > \thp$. Therefore, the study of $\mathcal{P}$ tests the properties of the transition beyond the jump in entropy.\par
We compute $\mathcal{P}$ in the matrix model and find that indeed it is an order parameter for the phase transition (see \S\ref{app:RP2} for the derivation).\par
In the spin chain, the Polyakov loop is realized as an impurity in the preparation of the initial state. That is, the ket in \eqref{eq:G} is replaced with the state
\begin{equation}
\label{eq:psiimp}
	\lvert \psi_{\times} \rangle =  \lvert \underbrace{ \downarrow  \downarrow \dots \downarrow }_{N-1}  \uparrow , \downarrow, \uparrow \dots \uparrow \rangle .
\end{equation}
Following \cite{Perez-Garcia:2022geq}, we obtain that the resulting coupling-averaged probability, divided by the coupling-averaged Loschmidt echo, equals $\mathcal{P}$. We elaborate further on this aspect in \S\ref{app:RP2}.\par
In summary, the expectation value of a Polyakov loop can be analysed on the spin chain by testing how the results vary when a simple impurity is introduced into the initial state.\par
Even without any reference to the gauge theory, we predict the ratio of the Loschmidt echoes with and without impurities to be an \emph{order parameter} discerning between a phase with low disorder, $0 <a \ll 1 $, and a phase of large disorder, thus large entropy, $a \gg 1$. This is a novel claim for the coupling averaged Heisenberg XX chain and may be of independent interest. To quantitatively test this prediction with the current state of the art technologies seems an interesting avenue to pursue.

\subsection*{Improved estimates of the entropy}
\textbf{2.} The relation \eqref{eq:main} prevents a direct evaluation of the BH entropy from the averaged Loschmidt echo, because $\hat{S}$ is only approximately equal to the entropy $S$. Nonetheless, more accurate results are achieved extending the argument to Hamiltonians with interactions beyond nearest-neighbour (\S\ref{app:RP1}).\par\medskip
\textbf{3.} The spin chain can reproduce the perturbative corrections in the 't Hooft coupling to the BH entropy (\S\ref{app:RP1point5}). Surprisingly, this major conceptual step requires no further sophistication of the experimental setup. These corrections are incorporated without complicating the Hamiltonian nor the initial state. The implementation simply requires that the random coupling is sampled with random standard deviation --- see \S\ref{app:RP1point5} for details. The BH entropy is computed by `nested' averages of the Loschmidt echo.\par\medskip
\textbf{4.} The matrix model \eqref{eq:MMfull} corresponds to a convenient choice of fugacities for the BH charges. More general fugacities lead to complex couplings $a_n \in \C$. The model \eqref{eq:MM2} at $a \in \C$ presents more realistic features of a Hawking--Page transition compared to its counterpart with $a \in \R$ \cite{Copetti:2020dil}. In \S\ref{app:RP3} we explain how this richer behaviour is encoded in the spin chain.

\section{Discussion}
Assuming the AdS$_5$/CFT$_4$ correspondence, we showed that the Hawking--Page transition with maximal supersymmetry is reproducible on an elementary 1d spin chain. The result is based on the mathematical identity \eqref{eq:main}, relating the entropy to the Loschmidt echo of the XX chain with an average over the randomized coupling. We also studied order operators for this transition. Additionally, we enriched the chain in various ways to simulate more realistic features of the BH entropy.\par
Unlike the \emph{planar integrability} of $\mN=4$ SYM (reviewed in \S\ref{app:integrability}), the XX chain used here is an auxiliary device manifesting the phase transition, rather than being a subsector of SYM. We map one specific observable to the averaged thermal Loschmidt echo in the spin chain.\par
\medskip
The outcome highlights a \emph{universality} property: introducing entropy in the 1d chain through the average, it emulates the behaviour of (superficially unrelated) BH systems in 4+1 dimensions.\par
It is natural to inquire about the extent to which the spin chain captures structural properties of BH physics, besides the jump in entropy at the Hawking--Page transition. In \S\ref{app:outlook}, we elaborate on two concrete avenues for future research, aimed at characterizing the von Neumann algebra and the microstates of the BH via spin chain techniques.\par
This work opens the possibility for the experimental realizations of Hawking--Page transitions with simple quantum systems. A remaining challenge is to see whether the required system sizes ($N \ge 4, L \ge 11 $) and associated error estimates are feasible with the current technologies.

\vspace{12pt}
\begin{acknowledgments}
We are grateful Hamed Adami, Elliott Gesteau and especially Sofyan Iblisdir for discussions.
LS thanks the Departamento de An\'alisis Matem\'atico y Matem\'atica Aplicada, Universidad Complutense de Madrid for warm hospitality at various stages of this project.
The work has been financially supported by Comunidad de Madrid (grant QUITEMAD-CM, ref. P2018/TCS-4342), by Universidad Complutense de Madrid (grant FEI-EU-22-06), by the Ministry of Economic Affairs and Digital Transformation of the Spanish Government through the QUANTUM ENIA project call - QUANTUM SPAIN project, and by the European Union through the Recovery, Transformation and Resilience Plan - NextGenerationEU within the framework of the Digital Spain 2025 Agenda.
DPG also acknowledges financial support from the Spanish Ministry of Science and Innovation MCIN/AEI/10.13039/501100011033 (``Severo Ochoa Programme for Centres of Excellence in R\&D'' CEX2019- 000904-S and grant  PID2020-113523GB-I00) as well as from CSIC Quantum Technologies Platform PTI-001. 
The work of LS is supported by the Shuimu Scholars program of Tsinghua University and by the Beijing Natural Science Foundation project IS23008 (``Exact results in algebraic geometry from supersymmetric field theory'').
\end{acknowledgments}
\onecolumngrid
\vspace{16pt}\par\noindent\begin{center}
    \rule{0.6\linewidth}{0.4pt}\vspace{16pt}\par
    \section*{Supplemental Material}\vspace{8pt}\par
\end{center}\par
\twocolumngrid
\appendix
\renewcommand{\appendixname}{\S\hspace{-3pt}}

\section{Hawking--Page, deconfinement, and the superconformal index}
\label{app:SCI}
Our study of the AdS$_5$ entropy across the Hawking--Page transition is done in two steps: first assume the validity of the AdS/CFT correspondence, which maps the supergravity problem to a four-dimensional gauge theory problem, and then take the large $N$ limit in the gauge theory. We now present the pertinent quantities directly in $\mN=4$ SYM.\par

\subsection{Superconformal index}
Among the observables that the combination of supersymmetry and conformal symmetry allows to evaluate, a basic one is the enumeration of supersymmetric operators in the theory. The theory being conformal, the state-operator map dictates that one can equivalently count states.\par
The superconformal index (SCI) \cite{Kinney:2005ej,Romelsberger:2005eg,Romelsberger:2007ec} (for a review see \cite{Rastelli:2016tbz}) is a twisted trace over the Hilbert space of states in radial quantization. Thus, it precisely counts states in $\mN=4$ SYM, \emph{with sign} ($+$ for bosons, $-$ for fermions). It is explicitly constructed as follows (our conventions are as in \cite{Benini:2018ywd}). Put $\mN=4$ SYM on $\mathbb{S}^3$, with isometry $\mathfrak{su}(2)_+ \oplus \mathfrak{su}(2)_-$ whose Cartan generators we denote $\mathsf{J}_{\pm}$. Let $\mathsf{r}_{1},\mathsf{r}_{2},\mathsf{r}_{3}$ be the generators of the Cartan $\mathfrak{u}(1)_1 \oplus \mathfrak{u}(1)_2 \oplus \mathfrak{u}(1)_3  $ of the $\mathfrak{so}(6)_R$ R-symmetry of $\mN=4$ SYM, and define $\mathsf{r} = (\mathsf{r}_1+\mathsf{r}_2+\mathsf{r}_3)/3$. Besides, let $\mathsf{Q}$ denote the supercharge preserved by the operators of interest. The SCI is defined as 
\begin{equation}
	\mathcal{I} ^{\mN=4} = \mathrm{Tr} ~ (-1)^{\mathsf{F}} ~ e^{- \beta \left\{ \mathsf{Q}, \mathsf{Q}^{\dagger} \right\} } ~ p^{\mathsf{J}_{+} + \mathsf{J}_{-} +\frac{\mathsf{r}}{2}} q^{\mathsf{J}_{+} - \mathsf{J}_{-} +\frac{\mathsf{r}}{2}} ,
\end{equation}
where $\mathsf{F}$ is the fermion number, $\beta$ the inverse temperature and $p,q$ are fugacities. The trace is taken over the Hilbert space on $\mathbb{S}^3$ in radial quantization. Finally, the term $\left\{ \mathsf{Q}, \mathsf{Q}^{\dagger} \right\}$ is the generator of translations along the Euclidean time direction, as usual. We have worked in the preferred slice $p=q$ in the main text, so that 
\begin{equation}
	\mathcal{I} ^{\mN=4} = \mathrm{Tr} ~ (-1)^{\mathsf{F}} ~ e^{- \beta \left\{ \mathsf{Q}, \mathsf{Q}^{\dagger} \right\} } ~ q^{2\mathsf{J}_+ +\mathsf{r}} .
\end{equation}
The SCI is protected by supersymmetry and the sign $(-1)^{\mathsf{F}}$ effectively restricts the count to BPS states.

\subsection{Thermal partition function}
We are interested in a different observable than the SCI, namely the thermal partition function on $\mathbb{S}^3 \times \mathbb{S}^1$, which we denote $\mz_{\mathbb{S}^3 \times \mathbb{S}^1}$. It is defined as the path integral on the curved manifold $\mathbb{S}^3 \times \mathbb{S}^1$, but it is equivalently described as a trace over the Hilbert space in radial quantization on $\mathbb{S}^3$, 
\begin{equation}
\label{eq:thermalZtrace}
	\mz_{\mathbb{S}^3 \times \mathbb{S}^1} ^{\mN=4} = \mathrm{Tr} ~ e^{- \beta \left\{ \mathsf{Q}, \mathsf{Q}^{\dagger} \right\} } ,
\end{equation}
or refinements thereof by the fugacities $p,q$. It is similar to the SCI, but it does not involve the weight $(-1)^{\mathsf{F}} $. In practice, the boundary conditions for the fermions along $\mathbb{S}^1$ break supersymmetry --- see e.g. \cite{Witten:1998zw} for more details on placing the theory on compact Euclidean spacetime.\par
The SCI has the advantage of being independent of the 't Hooft coupling $\lambda = g_{\mathrm{YM}} ^2 N$ \cite{Witten:1982im}, therefore it can be computed in the free limit and the result is exactly equal to the strong coupling limit. By contrast, the thermal partition function does depend on $\lambda$, preventing an exact extrapolation from weak to strong coupling. However, the thermal partition function has the advantage that, due to the lack of cancellations between bosons and fermions, it is more sensitive to the deconfinement transition and hence, by AdS/CFT, to the Hawking--Page transition \cite{Witten:1998zw}.\par
The entropy in AdS$_5$ is predicted to be dual to the thermal partition function, and we have computed 
\begin{equation}
\label{eq:SCItoS}
	S \sim \ln \mz_{\mathbb{S}^3 \times \mathbb{S}^1} ^{\mN=4} ,
\end{equation}
up to normalization terms that are not relevant for our analysis. This identification follows from noting that \eqref{eq:thermalZtrace} is counting states, whence its logarithm is directly interpreted holographically in terms of microstate counting, by using the thermodynamic relation 
\begin{equation}
\label{eq:Boltzentropy}
	S = \log ( \text{\# microstates} )  .
\end{equation}
An alternative take on the relation \eqref{eq:SCItoS} uses that, in the BH phase $T > \thp$, the near-horizon geometry is well-known to be that of an AdS$_2$, which implies the dual theory to reduce to an effective quantum mechanics along a thermal circle. This effective field theory approach lies at the heart of the derivation of \eqref{eq:MMfull}, as we now review.\par

\subsection{Effective matrix model}
The insight of \cite{Sundborg:1999ue}, further extended and generalized in \cite{Aharony:2003sx}, was to consider the limit of extremely weak 't Hooft coupling $\lambda = g_{\mathrm{YM}} ^2 N$ of $\mz_{\mathbb{S}^3 \times \mathbb{S}^1} $. That is, only gauge invariance is required in \cite{Sundborg:1999ue}, and the computations are performed without accounting for $\mathcal{O}(\lambda)$ corrections. Protected by the extended supersymmetry, one is then able to interpolate to the opposite corner of the conformal manifold, in which the theory is strongly coupled and dual to a BH. No phase transition takes place in this interpolation and the process is smooth. We comment further on the inclusion of perturbative corrections in $\lambda$ in \S\ref{app:weaktHooft}.\par
Then, one considers SYM on $\mathbb{S}^3 \times \mathbb{S}^1$, for which all the fields are massive. Integrating them out, the effective description obtained is only in terms of the holonomy $U$ of the gauge field around the thermal circle $\mathbb{S}^1$. A careful account of the integration procedure leads to the effective matrix model description \eqref{eq:MMfull} for the partition function $ \mz_{\mathbb{S}^3 \times \mathbb{S}^1} $ \cite{Sundborg:1999ue,Aharony:2003sx}. Note that maximal supersymmetry requires that all the fields are in the adjoint representation of the gauge group. Therefore only effective multi-trace interactions of the form $ \tr \left( U^{n} \right) \tr \left( U^{-n} \right)$ appear in the effective action, which express the trace taken in the adjoint in terms of the standard trace $\mathrm{Tr}$ in the fundamental representation.\par
\begin{figure}[ht]
	\centering
		\includegraphics[width=0.9\columnwidth]{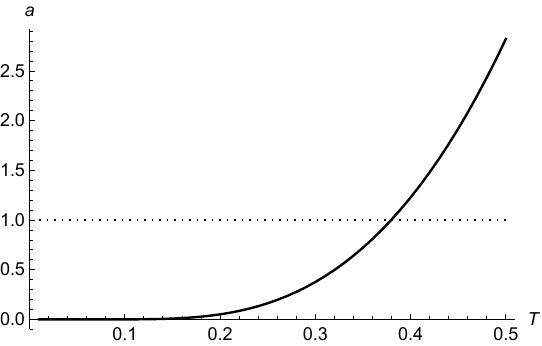}
	\caption{Coupling $a(T)$ in \eqref{eq:defaT} as a function of the temperature $T$. The dotted line is $a=1$, which defines $T_H$.}
	\label{fig:AvsT}
\end{figure}

The effective couplings $a_n$ in the matrix model \eqref{eq:MMfull} depend on the masses of the heavy modes integrated out, all of which depend on the radius $1/T$ of the thermal $\mathbb{S}^1$ as well as on the radius of the spatial $\mathbb{S}^3$. The lowest coefficient $a \equiv a_1$ was given in \eqref{eq:defaT}, and it is shown in Fig.~\ref{fig:AvsT} as a function of $T$.\par

\subsection{Weak coupling corrections and deconfinement}
\label{app:weaktHooft}

The basic ingredients on which our analysis is built are the matrix model \eqref{eq:MMfull} and the AdS/CFT correspondence. The holographic principle is used to identify the Hawking--Page transition in the bulk with a deconfinement transition on the dual $\mN=4$ SYM \cite{Witten:1998zw}. We have already explained how the matrix model is derived at zeroth order in the 't Hooft coupling $\lambda$, whereas the identification between deconfinement and Hawking--Page transition is a strong coupling effect. However, the phase transition we have studied still captures the transition from a thermal AdS to a BH phase. We refer to \cite{Aharony:2003sx} for a more exhaustive discussion on this matter.\par
Notice that, from the point of view of SYM, the jump \eqref{eq:lnZorder} has the neat interpretation of the thermal partition function being dominated by gauge-singlet bound states at low temperature, where the theory confines, and being dominated by a gas of free gluons at high temperature, after deconfinement (Fig.~\ref{fig:SYMandT}).\par
\begin{figure}[tb]
	\centering
	\begin{tikzpicture}
		\draw[->] (-3,0) -- (3,0);
		\node[anchor=south] at (3,0) {$T$};
		\node at (0,0) {$\bullet$};
		\node[anchor=south] at (0,0) {$\thp$};
		\node[align=center] at (-1.5,1) {confined\\ \tiny (gauge singlets)};
		\node[align=center] at (1.5,1) {deconfined\\\tiny (free gluons)};
		\node[align=center] at (-1.5,-1) {thermal AdS};
		\node[align=center] at (1.5,-1) {BH};
		\node[anchor=north] at (-2,0) {$S \sim O(1)$};
		\node[anchor=north] at (2,0) {$S \sim O(N^2)$};
		\node[draw,rounded corners] at (0,2) {$\mN=4$ SYM};
		\node[draw,rounded corners] at (0,-2) {AdS$_5$};
		
		\path[->] (-3,-0.8) edge[bend left] node[anchor=south,rotate=90,pos=0.5] {AdS/CFT} (-3,0.8);
		\path[->] (-3,0.8) edge[bend right] (-3,-0.8);
		
	\end{tikzpicture}
	\caption{Hawking--Page and deconfinement transitions.}
	\label{fig:SYMandT}
\end{figure}\par
Let us briefly elaborate on the physical picture, following \cite{Aharony:2003sx,Alvarez-Gaume:2005dvb}. At finite $\lambda$, starting from very low temperature and increasing it, there is a phase transition on the gravity side at a critical temperature $\thp$ at which the BH solution becomes dominant compared to the thermal AdS solution. This is the Hawking--Page transition, holographically dual to the deconfinement point in $\mN=4$ SYM \cite{Witten:1998zw}. Then, at some higher temperature $T > \thp$, there will be a second transition in which the thermal AdS solution becomes unstable, perturbation theory around that solution becomes ill-defined, and only the large BH phase remains. It was shown in \cite{Aharony:2003sx,Alvarez-Gaume:2005dvb} that the effect of taking the limit $\lambda \to 0$ while retaining gauge invariance, is that the two transitions coalesce. It follows that, for the sake of observing the signatures of the Hawking--Page transition, the picture considered in the main text suffices.\par
Including finite $\lambda$ effects, the backbone of the derivation of the effective matrix model still holds \cite{Aharony:2003sx}, but one has to improve \eqref{eq:MMfull} and include additional terms. In particular, as we are interested in the phase transition, we can still focus on the reduced matrix model \eqref{eq:MM1} and include the corrections in $\lambda$ \cite{Aharony:2003sx,Alvarez-Gaume:2005dvb}:
\begin{equation}
\begin{aligned}
	e^{\hat{S}_{\lambda}} = & \oint \dd U ~ \exp \Big[ a \tr \left( U \right) \tr \left( U^{-1} \right)  \\
	 & + \sum_{n \ge 1} \frac{b_n}{N^{2n}}  \left( \tr \left( U \right) \tr \left( U^{-1} \right) \right)^{n+1} \Big] ,
\end{aligned}
\label{eq:MMlambda}
\end{equation}
where the notation $\hat{S}_{\lambda}$ indicates that it includes the corrections in the 't Hooft coupling on the gauge theory side of the duality. The coefficients $b_{n}= b_n (T, \lambda)$ depend on $T$ as well as on the 't Hooft coupling $\lambda$ and vanish at $\lambda \to 0$. In other words, as argued in \cite{Aharony:2003sx,Alvarez-Gaume:2005dvb}, the terms $ \left( \tr \left( U \right) \tr \left( U^{-1} \right) \right)^{n}$ dominate against the corrections $\tr \left(U^n \right)\tr \left(U^{-n} \right)$ near the phase transition. Furthermore, to establish a finite-$\lambda$ phase diagram, it is sufficient to include the lowest order correction, i.e. $b_1 >0$ and $b_{n>1}=0$ \cite{Alvarez-Gaume:2005dvb}.\par
In summary, the simplified partition function \eqref{eq:MM1} captures all the distinctive properties of the Hawking--Page transition, and a more realistic approximation, which includes a coexistence region for thermal AdS and BH, is described by the improved formula \eqref{eq:MMlambda}.

\section{Refined probes of the Hawking--Page transition}
\label{app:RP}

\subsection{Polyakov loop}
\label{app:RP2}
From the point of view of the gauge theory, the Hawking--Page transition is holographically dual to a deconfinement transition at $T=T_{\rm HP}$ \cite{Witten:1998zw}. Famously, a probe of deconfinement \cite{tHooft:1977nqb} is provided by the so-called Polyakov loop \cite{Polyakov:1978vu}, placed along the circle $\mathbb{S}^1$ of radius $1/T$ in this context. Let us denote $\mathcal{P}$ the Polyakov loop expectation value, and let $A$ be the $SU(N)$ or $U(N)$ gauge field. Then, 
\begin{equation}
\label{eq:PdefN4}
	\mathcal{P} = \left\langle \mathrm{Tr} \left[ \exp \left( i \oint_{\mathbb{S}^1} A  \right) \right] \right\rangle_{\mN=4} ,
\end{equation}
where $\langle \cdot \rangle_{\mN=4}$ stands for the correlation function in the full-fledged $\mN=4$ CFT. Wilson loops in general, and $\mathcal{P}$ in particular, are charged under the centre symmetry of $\mN=4$ SYM theory. This is directly seen by the fact that \eqref{eq:PdefN4} is rotated by a phase when the gauge field $A$ is shifted by a flat 1-form in the centre of the gauge group. This makes $\mathcal{P}$ an order parameter for the deconfinement transition in $\mN=4$ SYM: it must vanish in the low temperature phase, in which the centre symmetry is preserved, and acquires a vacuum expectation value in the high temperature phase, in which deconfinement spontaneously breaks the centre symmetry.\par
While the definition \eqref{eq:PdefN4} is valid at every point of the conformal manifold of $\mN=4$ SYM, here we only consider the simplifying weak coupling limit, and consider the insertion of a Polyakov loop in the matrix model \eqref{eq:MMfull}. The computation of $\mathcal{P}$ dramatically simplifies into 
\begin{equation}
\label{PolyakovTrU}
	\mathcal{P}= \left\langle \mathrm{Tr} U \right\rangle 
\end{equation}
where now the symbol $\langle \cdot \rangle $ means the one-point function computed in the matrix model. If we are interested in signatures of the phase transition, the study of $\mathcal{P}$ simplifies further by taking the expectation value \eqref{PolyakovTrU} in the reduced matrix model \eqref{eq:MM1}, exactly as we did in passing from the entropy $S$ to $\hat{S}$.\par 
A convenient way to evaluate $\mathcal{P}$ at large $N$ is to introduce a source term 
\begin{equation}
	\exp \left( N \varepsilon  \mathrm{Tr} U \right) 
\end{equation}
in the matrix model \eqref{eq:MM2}, and compute $\mathcal{P}$ by differentiating the modified entropy with respect to $ \varepsilon $. Thanks to the charge conjugation symmetry of $\mN=4$ SYM, manifest in the $U \leftrightarrow U^{-1}$ symmetry of \eqref{eq:MMfull}, and thus in particular of \eqref{eq:MM2}, introducing the source term is equivalent to shift 
\begin{equation}
\label{eq:shiftsigmaeps}
	\frac{\sigma}{2} \tr \left(U +U^{-1} \right) \mapsto \frac{1}{2}(\sigma + \varepsilon) \tr \left(U +U^{-1}\right) ,
\end{equation}
and, denoting $\hat{S} (\varepsilon)$ the entropy modified in this way, 
\begin{equation}
\label{eq:PderS}
	\mathcal{P} = \left. \frac{1}{N} \frac{ \partial \hat{S} (\varepsilon) }{\partial \varepsilon } \right\lvert_{\varepsilon =0}.
\end{equation}
At this stage, it is straightforward to evaluate $\mathcal{P}$ at large $N$, borrowing the solution of the matrix model in \cite{Liu:2004vy} and using it to evaluate \eqref{eq:PderS} at leading order, and indeed we find $\mathcal{P}=0$ if $T < \thp$ and $\mathcal{P} \ne 0$ if $T > \thp$, in agreement with the prediction of the duality between Hawking--Page transition and deconfinement.\par
A technical caveat on the latter statement is in order. Physically, the Polyakov loop describes the insertion of a heavy quark. When considering a theory without fields in the fundamental representation, placed on a compact space without boundaries such as $\mathbb{S}^3$, the impossibility to give the loop a non-vanishing expectation value follows from the Gauss law applied to the quark flux, thus our claim may look counter-intuitive. As explained in \cite{Aharony:2003sx}, this problem is circumvented by adding an infinitesimal perturbation that breaks the centre symmetry explicitly, and setting it to zero at the end. This procedure is pretty much analogous to the study of spontaneous magnetization, in which one turns on a small external field of modulus $\varepsilon >0$ and sends $\varepsilon \to 0$ at the end.\par 
In practice, the shift \eqref{eq:shiftsigmaeps}, which for us is a convenient strategy at the level of matrix models, matches with the regularization prescribed in \cite{Aharony:2003sx}. Eventually one finds $\mathcal{P} \ne 0$ if $T > \thp$ \cite{Aharony:2003sx}.\par
\medskip
On the spin chain side, it was shown in \cite{Perez-Garcia:2022geq} that matrix model averages such as \eqref{PolyakovTrU} correspond to impurities in the preparation of the initial state $\lvert \psi_0 \rangle$. Then, $\mathcal{P}$ corresponds to the ratio of amplitudes with and without impurity. More precisely, let $\lvert \psi_{\times} \rangle $ be as in \eqref{eq:psiimp}, and denote $\mG_N^{\times} (J)$ (respectively $\hat{\mG}_N^{\times} (J)$) the analogue of $\mG_N(J)$ (resp. $\hat{\mG}_N (J)$) constructed using $\lvert \psi_{\times} \rangle $ as initial state, instead of $\lvert \psi_0\rangle $. Explicitly,
\begin{equation}
	\mG_{N} ^{\times} (J) = \langle \psi_0 \lvert e^{- H_{\rm XX}/\tilde{T}} \rvert \psi_{\times} \rangle ,
\end{equation}
and $\hat{\mG}_N^{\times} = \mG_{N} ^{\times}  /\mG_1$. Then, the spin chain definition of the observable \eqref{PolyakovTrU} is 
\begin{equation}
\label{eq:PinXY}
	\mathcal{P}= \frac{\left\langle \hat{\mG}_N^{\times} (J) \right\rangle_{2a}}{\left\langle \hat{\mG}_N (J) \right\rangle_{2a}} .
\end{equation}
The computation of $\left\langle \mathrm{Tr} U \right\rangle$ at fixed $J$ was done in \cite{Perez-Garcia:2022geq}.\footnote{In turn, the details of the computation in \cite{Perez-Garcia:2022geq} mimic the Wilson loop analysis of \cite{Hartnoll:2006,Santilli:2021rcp}.} Taking that result and performing the Gaussian average over $J$, we get the expected result for $\mathcal{P}$.\par
Notice that the subtlety mentioned in the Polyakov loop, and the necessity to introduce a symmetry-breaking deformation and the $\varepsilon \to 0$ limit at the end, have a natural solution in the spin chain. Recall that the amplitudes $\hat{\mG}_N$ are not measurable quantities, but the echoes $\hat{\mL}_N$ are. Therefore, in a hypothetical experimental setup, $\hat{\mG}_N^{\times}$ and $\hat{\mG}_N$ in \eqref{eq:PinXY} should be replaced by the square root of the averaged Loschmidt echoes, 
\begin{equation}
\label{eq:Lechoimpurity}
	\mathcal{P}= \frac{\left\langle \sqrt{\hat{\mL}_N^{\times} (J) }\right\rangle_{2a}}{\left\langle \sqrt{\hat{\mL}_N (J) } \right\rangle_{2a}} 
\end{equation}
exactly as in \eqref{eq:main}. The centre symmetry of $\mN=4$ SYM does act on the spin chain with the defect, rotating the amplitude $\hat{\mG}_N^{\times}$ by a phase. However, any physical implementation would automatically select a real saddle point for $\mathcal{P}$, since the hypothetical experiment would only be measuring the real probability (left invariant by the action of the centre symmetry) and not the amplitude. Stated differently, any concrete implementation of \eqref{eq:PinXY} would have built in the analogue of the regulating procedure described in \cite{Aharony:2003sx} to obtain $\mathcal{P} \ne 0$.\par
It is also noteworthy that the definition of $\mathcal{P}$ can be modified in both $\mN=4$ SYM and the spin chain. On the field theory side, it is possible to examine loops in higher-dimensional representations of the gauge group, such as the rank-$p$ antisymmetric representation. On the spin chain side, a feasible approach is to design an initial state $\lvert \psi_{\times} \rangle$ where the final spin $\downarrow$ is shifted $p \ge 1$ locations away from the other $N-1$ spins $\downarrow$. Notwithstanding this alteration, Identity \eqref{eq:PinXY} remains valid.\par
An open problem would be to study the spin chain avatars of other types of defects, analogous to the one considered in \cite{Chen:2023lzq}. The initial state in that case would be a (complicated) superposition of states with impurities.\par

\subsection{Improving the estimate of the BH entropy}
\label{app:RP1}
Our main result \eqref{eq:main} hinges upon the mapping of $e^{\hat{S}}$ to a coupling average of the GWW matrix model. However, as explained in \cite{Aharony:2003sx} (also \cite{Liu:2004vy,Copetti:2020dil}) and reviewed in the main text, this is far from being the end of the story for the BH entropy. $\hat{S}$ is the dominating contribution near the critical point, which therefore contains enough information to establish the phase transition, but does not accurately reproduce the entropy $S$ without accounting for the operators that have been discarded in passing from \eqref{eq:MMfull} to \eqref{eq:MM1}. Nevertheless, as long as we restrict to the regime in which the matrix model description holds, we can obtain more accurate results via mapping to more complicated spin chain Hamiltonians, with more interactions.\par
On the one hand, by repeating the Gaussian integral trick that led to \eqref{eq:MM2}, we can approximate the full BH entropy in \eqref{eq:MMfull} up to some cut-off order $K$ with $K$ Gaussian integrals over auxiliary variables $\sigma_{n}$, with standard deviation $2a_n$: 
\begin{equation}
\begin{aligned}
	e^{S} & \sim \prod_{n=1} ^{K} \int_0 ^{\infty} \sigma_n \dd \sigma_n ~ e^{- N^2 \frac{\sigma_n ^2}{4 a_n}} \\
	& \times \oint \dd U ~ \exp \left\{ \sum_{n=1}^{K} \frac{N \sigma_n}{2 n} \tr  \left( U^n + U^{-n} \right) \right\} ,
\end{aligned}
\label{eq:SgenMM}
\end{equation}
omitting uninteresting normalization factors. In fact, a (different) average point of view on \eqref{eq:MMfull} has been already advocated in \cite{Murthy:2022ien}.\par
On the other hand, the mapping to the spin chain for these more general Hamiltonians has been formalized in \cite{Perez-Garcia:2014aba,Santilli:2019wvq}. The procedure, which generalizes \cite{BPT}, is similar to the one reviewed in the main text and we skip the details, for which we refer to \cite{Perez-Garcia:2014aba}. The streamlined idea is to generalize the Hamiltonian $H_{\rm XX}$ beyond nearest-neighbour interactions: 
\begin{equation}
\label{Kmodel}
	H_{\rm gen} = - \frac{1}{2} \sum_{j=0} ^{L-1} \sum_{n=1}^{K } \frac{\tilde{J}_n}{n} \left(  \sigma_j ^{-} \sigma_{j+n} ^{+} + \sigma_j ^{+} \sigma_{j+n} ^{-} \right) ,
\end{equation}
for a collection of couplings $\tilde{J}_n$ and assuming periodic boundary conditions. For ease of exposition, we rename $\tilde{J}_n = \tilde{J}_1 \gamma_n$ and denote $J \equiv \tilde{J}_1 /\tilde{T}$ as in the main text. Then, running an argument analogous to \cite{BPT} and Sec.~\ref{sec:chain}, ones finds that the Loschmidt amplitude 
\begin{equation}
	\mG_N ^{\rm gen} (J) =  \langle \psi_0 \lvert e^{- H_{\text{\rm gen}} /\tilde{T} } \lvert \psi_0 \rangle 
\end{equation}
is computed by a Toeplitz determinant 
\begin{equation}
\label{eq:GgenToeplitz}
    \mG_N ^{\rm gen} (J) = \det_{1 \le j,k \le N-1} \left[ g^{\text{\rm gen}}_{j,k} (J) \right]
\end{equation}
where the generalization of \eqref{eq:gjk} is 
\begin{equation}
	g^{\text{\rm gen}}_{j,k} (J) = \langle \Uparrow\rvert \sigma_{j}^{+} e^{- H_{\text{\rm gen}} /\tilde{T}} \sigma_k ^{-} \lvert\Uparrow \rangle .
\end{equation}
Reasoning as in Sec.~\ref{sec:chain}, we differentiate $g^{\text{\rm gen}}_{j,k}$ and, using the commutation relations \eqref{eq:commrel}, we find
\begin{equation}
	\frac{\dd g^{\text{\rm gen}}_{j,k} }{\dd J}   = \frac{1}{2}  \sum_{n=1}^{K } \frac{\gamma_n}{n}\left( g_{j-n, k}+g_{j+n, k} \right) .
\end{equation}
This is the recurrence relation of the generalized Bessel function. Alternatively, it can be directly checked that the integral 
\begin{equation}
    \oint \frac{\dd u}{2 \pi \ii u} u^{j-k} ~ \exp \left\{ \frac{J}{2} \sum_{n=1}^{K} \frac{\gamma_n}{n} \tr  \left( u^n + u^{-n} \right) \right\} 
\end{equation}
is a solution. Plugging the result into \eqref{eq:GgenToeplitz} and applying the Heine--Szeg\H{o} identity, we have that the Loschmidt amplitude for this more general spin chain equals the matrix model 
\begin{equation}
\label{eq:spinMMgen}
	\mz^{\rm gen} _N = \oint \dd U ~ \exp \left\{ \sum_{n=1}^{K} \frac{\tilde{J}_n}{2 n  \tilde{T}} \tr  \left( U^n + U^{-n} \right) \right\} 
\end{equation}
in the limit $L \gg N$. Again, the temperature $\tilde{T}$ of the laboratory does not matter as long as it is non-zero, because the control parameters of the system are the ratios 
\begin{equation}
	J_n = \tilde{J}_n / \tilde{T} . 
\end{equation}
Thus, with the normalization as in the main text, we find the generalized thermal Loschmidt echo
\begin{equation}
\label{eq:hatLgenMM}
	\hat{\mL}_N^{\rm gen} (J_1, \dots, J_K) = \left\lvert \mz^{\rm gen} _N / \mz^{\rm gen} _1 \right\rvert^2 .
\end{equation}

Comparing \eqref{eq:SgenMM} with the generalized spin chain matrix model \eqref{eq:spinMMgen}, we observe that the entropy $S$, with irrelevant interactions\footnote{In this paper, the word ``irrelevant'' is always used in its precise RG flow sense.} cut off at some order $K$, and the generalized spin chain with up to $K$ neighbour interactions, satisfy the analogous relation as for the simplified entropy $\hat{S}$ and the XX chain. Once again, the spin chain coupling constants $\tilde{J}_n$ are related to the Hubbard--Stratonovich fields $\sigma_n$ through $N \sigma_n = \tilde{J}_n/\tilde{T} \equiv J_n$, and in the generalized setting one is instructed to take a Gaussian average over all the spin-spin couplings $J_n$.\par
In principle, one is able to approximate the entropy $S$ (and hence the black hole entropy $S_{\mathrm{BH}}$ in the phase $T > \thp$) to arbitrarily high order $K$ of irrelevant couplings, by measuring the Loschmidt echo in the generalized chain and then taking a Gaussian average over all the couplings.\par
We underline that the average perspective on the entropy of a BH in the holographic dual to $\mN=4$ SYM was already undertaken in \cite{Murthy:2022ien}, but here we are pointing out a very concrete and explicit way to implement the average in a simple 1d auxiliary device. 

\subsection{Fermionic point of view}
\label{app:fermions}
\begin{figure}[ht]
    \centering
    \begin{equation*}
        \begin{matrix} \hspace{12pt} \\ {\scriptstyle n_j=} \ \end{matrix} \begin{matrix} \ \cdots & \ \circ & \ \circ & \ \bullet & \ \bullet &  \ \cdots  & \ \bullet  & \ \circ & \ \circ & \ \cdots \\ \ \cdots & \ \scriptstyle{0} &  \ \scriptstyle{0} & \  \scriptstyle{1} & \ \scriptstyle{1} & \ \cdots & \ \scriptstyle{1}  & \ \scriptstyle{0} & \ \scriptstyle{0} & \ \cdots \end{matrix}
    \end{equation*}
    \caption{The initial state \eqref{eq:psi0} is equivalent, via a Jordan--Wigner transformation, to place $N$ adjacent fermions on a lattice of $L$ sites. $\circ$ indicates an empty site, $\bullet$ indicates a site occupied by a fermion.}
    \label{fig:freeferm}
\end{figure}

The XX model, also known as isotropic XY model, is well-known to be equivalent to a system of free fermions, via the Jordan--Wigner (JW) map \cite{Jordan:1928wi}. We can identify the spin at site $j$ of the chain with a fermionic occupation number, according to
\begin{equation}
\label{eq:mapspin01}
    \lvert \uparrow \rangle \mapsto \lvert 0 \rangle , \qquad  \lvert \downarrow \rangle \mapsto \lvert 1 \rangle 
\end{equation}
where $\lvert n_j \rangle $ indicates $n_j \in \left\{ 0,1 \right\}$ fermions on the site $j$. The initial state \eqref{eq:psi0} thus corresponds to place $N$ adjacent fermions on a periodic lattice of $L$ sites (Fig.~\ref{fig:freeferm}),
\begin{equation}
\label{eq:psiferm}
	\lvert \psi_0 \rangle = \lvert \underbrace{ 1,1 \dots ,1 }_{N} ,0,0 \dots ,0 \rangle .
\end{equation}
This fermionic model leads to the same Loschmidt echo used in the main text \cite{Krapivsky:2017sua}.\par
\medskip
It is possible to extend this result and to show that for a 1d fermionic model%
\begin{equation}
    H_{\text{\rm ferm}}= \frac{1}{L}\sum_k\epsilon _{k} \tilde{c}_{k}^{\dag } \tilde{c}_{k}  \label{ferm}
\end{equation}%
with a general dispersion relation $\epsilon _{k}$, the transition amplitude is still a unitary matrix model (for fermions with periodic boundary conditions) with the dispersion relation as a confining potential.\par
The Hamiltonian \eqref{ferm}, is expressed in momentum space, with $k$ running over the dual lattice,
\begin{equation}
    k=\frac{2\pi}{L} q, \qquad q = 0, \dots, L-1 ,
\end{equation}
and $\tilde{c}^{\dagger}_k, \tilde{c}_k$ the fermionic creation and annihilation operators in momentum space, which satisfy the canonical anti-commutation relations
\begin{equation}
    \left\{ \tilde{c}^{\dagger}_k, \tilde{c}_{k^{\prime}} \right\} = 2 \pi \delta_{q,q^{\prime}} .
\end{equation}
They are related to the creation and annihilation operators $c^{\dagger}_j, c_j$ for fermions on the lattice site $j$ via the discrete Fourier transform 
\begin{equation}
    \tilde{c}_k = \sum_{j=0}^{L-1} e^{\ii j k } c_j = \sum_{j=0}^{L-1} e^{-\ii 2\pi q j /L} c_j .
\end{equation}\par
Fermionic models of the form \eqref{ferm} have been analyzed in \cite{Suzuki1971}, to which we refer for a careful treatment of the periodic boundary conditions. As a side remark, we note that the fermionic setup allows non-isotropic XY interactions \cite{Suzuki1971}, but for concreteness we do not consider such scenario.\par
The result equating the Loschmidt echo of the fermionic model to the Loschmidt echo of the spin chain holds true even for cases where $\epsilon _{k}$ is not equal to $\cos(k)$, which corresponds to the image of the XX model by the JW transformation and results in the GWW matrix model for the Loschmidt amplitude.\par
To derive the matrix model description, we extend the result of \cite{Krapivsky:2017sua}. With the initial state $\lvert \psi_0 \rangle$ as in \eqref{eq:psiferm}, we consider the amplitude 
\begin{equation}
    \mG_N^{\text{\rm ferm}} =  \langle \psi_0 \lvert e^{- H_{\text{\rm ferm}} /\tilde{T} } \lvert \psi_0 \rangle .
\end{equation}
Inserting $N$ times the resolution of the identity on the left and on the right and approximating for $L \to \infty$ (for the finite-$L$ expression, simply replace the integrals by Riemann sums), we get
\begin{equation}
\begin{aligned}
     \mG_N^{\text{\rm ferm}} = & \int_{0}^{2\pi} \frac{\dd k_1}{2 \pi} e^{- \epsilon_{k_1}/\tilde{T}} \cdots \int_{0}^{2\pi} \frac{\dd k_N}{2 \pi}  e^{- \epsilon_{k_N}/\tilde{T}} \\
     & \times ~\lvert \langle k_1, \dots, k_N \vert \psi_0\rangle \rvert^2 .
\end{aligned}
\end{equation}
This expression heavily relies on the fact that \eqref{ferm} describes non-interacting fermions. The amplitude $\lvert \langle k_1, \dots, k_N \vert \psi_0\rangle \rvert$ is computed by a Slater determinant. We also make a change of variables $k_j = \theta_j + \pi$, which is not necessary in principle but is required for consistency with our conventions in the spin chain, as we have (implicitly) worked with variables $\theta_j \in [-\pi, \pi]$ throughout. Then, one finally arrives at 
\begin{equation}
	\mG^{\text{\rm ferm}} _N = \oint \dd U ~ \exp \left\{\frac{1}{\tilde{T}}\tr ~ V_{\epsilon}  \left( U \right) \right\} ,
\end{equation}
where $U$ is a unitary matrix with eigenvalues $e^{\ii \theta_1},\dots, e^{\ii \theta_N}$, and $V_{\epsilon}$ is the matrix model potential defined by the dispersion relation, 
\begin{equation}
    \tr ~V_{\epsilon} \left( U \right)=\sum_{j=1}^{N} \epsilon_{k_j=\theta_j + \pi} .
\end{equation}
This equals \eqref{eq:spinMMgen} upon identification of the dispersion relation 
\begin{equation}
\label{eq:epsilondr}
\epsilon_k =  \sum_{n=1}^{K} (-1)^n \frac{\tilde{J}_n}{2 n} \left( e^{\ii k} + e^{-\ii k} \right).
\end{equation}
Therefore, with this choice, $\mG^{\rm ferm} _N =\mG^{\rm gen} _N $.\par
We emphasize that, despite the equality of this specific observable, the free fermion system with generalized dispersion relation and the spin chain with interaction beyond nearest neighbour \emph{are not} equivalent. It is a known fact that the JW transformation cannot map these two models for any $K>1$.\par
\medskip
We have obtained that the spin chains in \S\ref{app:RP1} and the fermionic models \eqref{ferm} have equal return amplitude in the suitably chosen initial state \eqref{eq:psi0}.\par 
We now proceed to show this fact explicitly, thus yielding a consistency check of the previous derivation and demonstrating as a proof of concept that two models that are not equivalent can nevertheless have equal Loschmidt echo.\par
It is well-known that, for an open chain, the inverse JW map [in the conventions \eqref{eq:mapspin01}] transforms the fermionic bilinear terms according to \cite{Jordan:1928wi,Suzuki1971}
\begin{equation}
 \label{eq:inverseJW}
     \frac{(-1)^{n}}{4} c_j^{\dagger} c_{j+n} \mapsto - \sigma^{-}_j \left( \prod_{\nu=1}^{n-1}\sigma_{j+\nu}^z \right) \sigma^{+}_{j+n} ,
\end{equation}
where the sign on the left-hand side comes from the $(-1)^n$ in \eqref{eq:epsilondr}, and the overall minus sign on the right-hand side agrees with the spin chain conventions in \eqref{Kmodel}. The spin chain Hamiltonian resulting from the inverse JW transformation differs from the Hamiltonian \eqref{Kmodel} by the presence of the intermediate $\sigma_{j+\nu}^z$, enforcing fermionic statistic. We refer to \cite{Suzuki1971,derzhko2008jordan} for more details as well as for the subtleties arising from the periodic boundary conditions.\par
Since \eqref{ferm} only contains fermion bilinears, the fermion parity is a conserved quantity. For the observable of interest, where $H_{\text{\rm ferm}}$ acts only on \eqref{eq:psiferm}, this equals $(-1)^N$.\par
Consider first an Hamiltonian with only $n^{\text{th}}$-nearest neighbour interaction, and map it to spin variables via \eqref{eq:inverseJW}. To compute the return amplitude for this term, we expand the exponential and compute, for every $m\ge 0$,
\begin{equation}
    \langle \psi_0 \lvert \left( \sum_{j=0}^{L-1} \sigma^{-}_j \left( \prod_{\nu=1}^{n-1}\sigma_{j+\nu}^z \right) \sigma^{+}_{j+n} + \text{ c.c.}\right)^m \rvert \psi_0 \rangle .
\end{equation}
Letting the operator act on the ket gives a superposition of states with definite fermion occupation configuration. However, only even powers of $m$ give states with non-trivial overlap with $\langle \psi_0 \vert $. Therefore, since even powers of $\sigma^z$ are trivial, the Loschmidt amplitude is the same as in the spin chain with $n^{\text{th}}$-nearest neighbour interaction.\par
When including all the interaction terms $1 \le n \le K$, the outcome is unchanged. This is guaranteed by the fact that the potential mixing of contributions [for instance, $\left( c^{\dagger}_{j+1} c_{j} \right)^{nm}$ with $\left( c^{\dagger}_{j+n} c_{j} \right)^m$], cannot produce cancellations, because they come with different coefficients. The claim can also be checked by direct computation, e.g. in the $K=2$ case.\par 
We highlight that the equality of the Loschmidt amplitudes for the two \emph{inequivalent} models stems from two properties:
\begin{itemize}
    \item The initial and final states are identical;
    \item The initial state is an eigenstate of the fermion parity operator.
\end{itemize}

\subsection{Perturbative corrections to the BH entropy}
\label{app:RP1point5}
In \S\ref{app:weaktHooft} we have enriched the BH entropy allowing for a small but finite 't Hooft coupling $0 < \lambda \ll 1$. The matrix model \eqref{eq:MMlambda} was studied in \cite{Alvarez-Gaume:2005dvb}, where it was shown that $b_{n>1}=0$ already retains all the desired qualitative features. The authors of \cite{Alvarez-Gaume:2005dvb} rewrite [$b \equiv b_1 (T, \lambda)$]
\begin{equation}
\begin{aligned}
		e^{\hat{S}_{\lambda}} = \frac{N^3}{4 \sqrt{\pi b}}  & \int_{- \infty} ^{\infty} \frac{\dd \mu}{\mu} ~ e^{- \frac{N^2}{4 b} (\mu-a)^2} ~ \int_{0} ^{\infty} \sigma \dd \sigma \\
		& \times \oint \dd U ~ \exp \left[  - N^2 \left( \frac{\sigma^2}{{4 \mu} } - \mf (\sigma) \right) \right] ,
\end{aligned}
\end{equation}
which makes it clear that the map to the spin chain goes through. In this case, though, a \emph{nested} average over the coupling $J$ is required. First, observe that 
\begin{equation}
	\hat{S}_{\lambda} \sim \ln \left[  \int_{- \infty} ^{\infty}  \frac{\dd \mu}{\mu} ~ e^{- \frac{N^2}{4 b} (\mu-a)^2} ~ \left. e^{\hat{S}} \right\rvert_{a \mapsto \mu} \right] ,
\end{equation}
where $ e^{\hat{S}} \rvert_{a \mapsto \mu}$ stands for \eqref{eq:MM2} with $a$ replaced by $\mu$ on the right-hand side, and the symbol $\sim$ again means that we drop terms that do not affect our discussion. Then, we run the argument of the main text and use \eqref{eq:main} to obtain 
\begin{equation}
\label{eq:SlambdaLmu}
	\hat{S}_{\lambda} \sim  \ln \left[  \int_{- \infty} ^{\infty}  \frac{\dd \mu}{\mu} ~ e^{- \frac{N^2}{4 b} (\mu-a)^2} ~ \left\langle \sqrt{ \hat{\mL}_N} \right\rangle_{2\mu} \right] .
\end{equation}
The outcome of \cite{Alvarez-Gaume:2005dvb} is that \eqref{eq:SlambdaLmu} has a phase structure that agrees with \eqref{eq:MM2} near $b \to 0$ (which corresponds to $\lambda \to 0$ in SYM), but in addition shows the properties predicted in \S\ref{app:weaktHooft}.\par
Our result is to argue that this outcome can be tested on the Heisenberg XX chain, by measuring the thermal Loschmidt echo and then take a nested average of the coupling. Namely, first take a Gaussian average over the spin-spin coupling $J$ centered at $J=0$ with standard deviation $2 \mu$. The parameter $\mu$ is itself sampled from a Gaussian distribution, centered around $\mu = a$ with standard deviation $2b /N^2$. Interestingly, the progress in passing from $\hat{S}$ to $\hat{S}_{\lambda}$, which is a non-trivial step from the SYM side, reduces to `averaging the averaged echo' on the spin chain. That is, in practice it boils down to repeat the measurement several times, sampling both $J$ and the standard deviation with which $J$ is sampled.

\subsection{Complex temperature}
\label{app:RP3}
It is worthwhile to stress that the matrix model \eqref{eq:MMfull} was derived in \cite{Sundborg:1999ue,Polyakov:2001af,Aharony:2003sx} under certain simplifying assumptions. It was pointed out in \cite{Copetti:2020dil} that, adopting most of the same simplifications but allowing for complex couplings $a_n \in \C$ in the matrix model \eqref{eq:MMfull}, leads to physically more realistic predictions about the BH behaviour. This is because it would allow for arbitrary fugacities in the SCI, which in turn give arbitrary fugacities for the BH charges, thus retaining more information in the computation.\par
Conceptually, it is straightforward to introduce complex couplings in our model, even though any hypothetical experimental realization would become more challenging. We focus again on the reduced model \eqref{eq:MM2} to compute $\hat{S}$. Writing $a=\lvert a \rvert e^{ \ii \varphi}$, we use \eqref{eq:main} and \eqref{eq:avgG}, and find that the model of \cite{Copetti:2020dil} satisfies 
\begin{equation}
	 e^{\hat{S}} \sim \left\langle \hat{\mL}_{N} \left( e^{-\ii \varphi/2} J \right) \right\rangle_{2\lvert a\rvert } .
\end{equation}
Namely, we still compute the average of the Loschmidt echo, with random coupling sampled with a Gaussian distribution, except that now the system is at complex temperature $e^{ \ii\varphi/2} \tilde{T}$, i.e. finite temperature and time-evolved. The complex temperature Loschmidt amplitude reduces to the GWW matrix model with complex coupling, in which the integration contour is deformed along a suitable thimble \cite{Alvarez:2016rmo,Santilli:2021eon}.\par
In a laboratory setup, one first fixes the temperature of the system at a value $\tilde{T}_{\R}$ and then lets the system evolve until time $t$, which depends on the phase $\varphi$ of $a$ in such a way that 
\begin{equation}
	\arg\left( \ii t + \tilde{T}_{\R} ^{-1} \right) =- \varphi/2 . 
\end{equation}
The procedure of averaging over $J$ remains unchanged in this scenario.\par

\subsection{Determinant identities from spin chains}
Throughout this appendix we have studied various ways of enriching the XX spin chain, with the aim of refining the correspondence with the Hawking--Page transition. We now comment on a by-product of our construction, which has interesting implications for the spin chain itself, regardless of the applications to BH physics.\par
Consider a spin chain with interaction at next-to-nearest neighbour, but without any nearest-neighbour term. That is, we are in the setup of \S\ref{app:RP1} with $K=2$ and
\begin{equation}
    \tilde{J}_1 =0, \qquad \tilde{J}_2 >0 .
\end{equation}
In this case, the spin chain decouples into two non-interacting XX spin chains, both with $\tilde{J}=\tilde{J}_2$. One copy of the XX chain involves only the spins at even lattice sites, and the other copy involves only the spins at odd lattice sites. Therefore, as a corollary of our analysis in \S\ref{app:RP1} we derive the identity 
\begin{equation}
    \mG_{2N} ^{2\mathrm{nn}} (J) = \left( \mG_N  (J) \right)^2 ,
\end{equation}
where, on the left-hand site, $\mG_{2N} ^{2\mathrm{nn}} $ is the Loschmidt amplitude for the spin chain with only next-to-nearest neighbour ferromagnetic interaction. This holds already at finite $L$, but in the following we consider $L \to \infty$ to simplify the expressions.\par
Utilizing \eqref{eq:GgenToeplitz}, the equality of the amplitudes translates into a mathematical identity:
\begin{equation}
\label{eq:detidI}
    \det_{1\le j,k \le 2N} \left[ I_{j-k} ^{(1,2)} (0,J) \right] = \left( \det_{1\le j,k \le N} \left[ I_{j-k} (J) \right] \right)^2 .
\end{equation}
On the right-hand side, $I_{\nu} (J)$ is the modified Bessel function of first kind, while on the left-hand side $I_{\nu} ^{(1,2)} (J_1,J_2)$ is the generalized modified Bessel function, with integral representation for $\nu \in \Z$
\begin{equation}
    I_{\nu} ^{(1,2)} (J_1,J_2) = \oint \frac{\dd u}{2 \pi \ii u} u^{\nu} ~ e^{\sum_{n=1}^{2} \frac{J_n}{2n} \tr  \left( u^n + u^{-n} \right) } .
\end{equation}\par
Therefore, as an immediate by-product of \cite{Perez-Garcia:2014aba} together with \S\ref{app:RP1} we derive \eqref{eq:detidI} [and its generalization \eqref{eq:KnnequalsGK}]. We stress the following aspects:
\begin{itemize}
    \item The argument is this appendix entails a change in philosophy, where we use an elementary spin chain argument and the map of \cite{Perez-Garcia:2014aba} to write an identity for Toeplitz determinants, which is seemingly less obvious from the point of view of random matrix theory.
    \item This identity involves the ubiquitous GWW model.
\end{itemize}
To strengthen our argument, we now rederive \eqref{eq:detidI} [and its generalization \eqref{eq:KnnequalsGK}] in an independent way, thus providing a solid consistency check of the construction in \S\ref{app:RP1}.\par
To begin with, we note that, in the large $N$ limit, \eqref{eq:detidI} is a direct consequence of Szeg\H{o}'s strong limit theorem \cite{Szegoth}.\par
To study the determinants at finite $N$, we use that the generalized Bessel functions satisfy \cite{Korsch:2006}
\begin{equation}
\label{eq:propBessel}
    I_{\nu} ^{(1,2)} (0,J) = \begin{cases} 0 , & \quad \forall \nu \in 1+2\Z , \\ I_{\nu/2} (J) , & \quad \forall \nu \in 2 \Z . \end{cases}
\end{equation}
The case $N=1$ of \eqref{eq:detidI} reads
\begin{equation}
    I_{0} ^{(1,2)} (0,J)^2 - I_{-1} ^{(1,2)} (0,J) I_{1} ^{(1,2)} (0,J) =  I_{0} (J)^2 ,
\end{equation}
and it is proved as an immediate consequence of \eqref{eq:propBessel}. It is also easy to check explicitly that \eqref{eq:detidI} holds for low value of $N$, thanks to the relations \eqref{eq:propBessel}.\par 
To prove \eqref{eq:detidI} in full generality --- without resorting to the spin chain--- one expands the determinant on the left-hand side,
\begin{equation}
    \det_{1\le j,k \le 2N} \left[ I_{j-k} ^{(1,2)} (0,J) \right] = \sum_{\varpi \in S_{2N}} (-1)^{\varpi} \prod_{j=1}^{2N} I_{j-\varpi (j)} ^{(1,2)} (0,J) 
\end{equation}
and separates the product over $j$ into two products, over $j \in 2 \Z \cap [1,N]$ and $j \in (1+2 \Z) \cap [1,N]$. From \eqref{eq:propBessel}, $I_{j-\varpi (j)} ^{(1,2)} (0,J) \ne 0$ only if both $j$ and $\varpi (j)$ are even or both are odd. Therefore the non-vanishing contributions to the Toeplitz determinant are those from permutations that factorize into $\varpi = \varpi_{\mathrm{e}} \circ \varpi_{\mathrm{o}}$, where $\varpi_{\mathrm{e}}$ (resp. $\varpi_{\mathrm{o}}$) permutes only the even (resp. odd) indices. This yields 
\begin{equation}
\begin{aligned}
    & \sum_{\varpi \in S_{2N}} (-1)^{\varpi} \prod_{j=1}^{2N} I_{j-\varpi (j)} ^{(1,2)} (0,J) \\ 
    =& \left[ \sum_{\varpi_{\mathrm{e}} \in S_{N}} (-1)^{\varpi_{\mathrm{e}}} \prod_{\substack{j=1 \\ j \text{ even}}}^{2N} I_{j-\varpi_{\mathrm{e}} (j)} ^{(1,2)} (0,J) \right] \\ \times & \left[ \sum_{\varpi_{\mathrm{o}} \in S_{N}} (-1)^{\varpi_{\mathrm{o}}} \prod_{\substack{j=1 \\ j \text{ odd}}}^{2N} I_{j-\varpi_{\mathrm{o}} (j)} ^{(1,2)} (0,J) \right] , 
\end{aligned}
\end{equation}
where we remark that this factorization is a consequence of the non-trivial identity \eqref{eq:propBessel}. Recognizing the two terms on the right-hand side as Toeplitz determinants, we arrive at \eqref{eq:detidI}.\par
\medskip
The above argument is easily generalized to chain with $K^{\text{th}}$-nearest neighbour interaction, and no other interaction. Namely, one considers a spin chain with hopping Hamiltonian\footnote{To reduce clutter, we slightly change the notation with respect to \S\ref{app:RP1},  $(\tilde{J}_K /K, \tilde{T}) \mapsto (J,1)$.} 
\begin{equation}
    H_{K\mathrm{nn}} = - \frac{J}{2} \sum_{j=0} ^{L-1}  \left(  \sigma_j ^{-} \sigma_{j+K} ^{+} + \sigma_j ^{+} \sigma_{j+K} ^{-} \right) .
\end{equation}
The same argument as for $K=2$ implies the factorization into $K$ non-interacting XX chains, hence 
\begin{equation}
\label{eq:KnnequalsGK}
    \mG_{KN} ^{K\mathrm{nn}} (J) = \left( \mG_N  (J) \right)^K . 
\end{equation}\par
This translates again into an identity of Toeplitz determinants, involving higher generalized Bessel functions. The generalization of Property \eqref{eq:propBessel} is \cite{Korsch:2006}
\begin{equation}
    I_{\nu} ^{(1,K)} (0,J) = \begin{cases} 0 , & \quad \forall \nu \notin K\Z , \\ I_{\nu/K} (J) , & \quad \forall \nu \in K \Z . \end{cases}
\end{equation}
Using this property, it is easy to rederive \eqref{eq:KnnequalsGK} independently of the spin chain, exactly as in the $K=2$ case.

\section{Comments on planar integrability}
\label{app:integrability}

There exists a connection between spin chains and $\mN=4$ SYM, known since the work of Minahan--Zarembo \cite{Minahan:2002ve}, which goes under the name of planar integrability. This property is the statement that, in the planar limit, correlation functions of certain local operators in $\mN=4$ SYM in flat spacetime equal those of (different) operators in a spin chain \cite{Minahan:2002ve,Beisert:2003yb}. This relation appears in the planar limit of large $N$ with fixed 't Hooft coupling $\lambda$.\par
While the setup is vaguely similar to what we consider here, we ought to emphasize the built-in differences. 
\begin{itemize}
\item Planar integrability, as the name suggests, is a feature that emerges only in the planar limit. On the contrary, given any truncation of the tower of irrelevant operators appearing in \eqref{eq:MMfull} (see \S\ref{app:RP1} for details), we can define a spin chain Hamiltonian such that the averaged Loschmidt echo reproduces the thermal partition function of $\mN=4$ SYM. This holds before taking any large $N$ limit. Moreover, the relation \eqref{eq:main} is exact, even though the large $N$ limit serves to render $\hat{S}$ a good approximation of the SCI.
\item In turn, the bridge between $\mN=4$ SYM and the XX spin chain we find is limited to a map between specific observables, namely the partition function on thermal $\mathbb{S}^3 \times \mathbb{S}^1$ on the SYM side and the averaged Loschmidt echo on the chain side. We \emph{do not} claim a full identification of the two systems, in any regime. This correspondence can be extended to include selected non-local operators in $\mN=4$ SYM, mapped to defects in the XX chain (see \S\ref{app:RP2}), but falls short of yielding a complete dictionary between operators on both sides of the correspondence.
\item Furthermore, for the correspondence to work, we have to introduce disorder in the spin chain by randomizing the coupling and taking the Gaussian average. This situation is radically different from the planar integrability.
\end{itemize}

\section{Outlook: Beyond the Hawking--Page transition}
\label{app:outlook}
We conclude this appendix on a more speculative note. So far, our discussion has focused on a 1d quantum simulation to reproduce the entropy of a holographic BH in AdS$_5$. We now suggest and lay down two open problems, which aim at looking for deeper signatures of BH physics on the spin chain.

\subsection{Operator algebras from the spin chain}
The upshot of our analysis is that the Heisenberg XX spin chain can be used to simulate the Hawking--Page transition. This correspondence has been refined and extended in several directions in \S\ref{app:RP}.\par
It is interesting to ask whether other characteristic properties of the Hawking--Page transition are automatically encoded in the spin chain, if we look beyond the Loschmidt echo. For instance, phase transitions as the one studied presently are accompanied by a drastic change in the large-$N$ algebra of operators \cite{Gesteau:2024dhj}. Due to its relevance for questions concerning the BH horizon and its interior \cite{Leutheusser:2021qhd,Leutheusser:2021frk}, it would be extremely interesting to enlarge our correspondence to include this aspect.\par
To address questions of this kind, the fermionic picture comes in handy. Indeed, it has been recently pointed out that these algebra transitions are detected by fermionic lattices \cite{Basteiro:2024cuh}. If this result could be extended to the fermionic systems obtained in \S\ref{app:fermions}, one would end up with a very concrete quantum mechanical gadget that realizes the criterion of \cite{Gesteau:2024dhj} for the behaviour of the algebra of operators.\par
\medskip
A related problem is to understand the relation between the averaging procedure on the Loschmidt echo and chaos. The presence of a BH should be accompanied by maximal chaos, and it would be interesting to see how this effect appears in the spin chain. Leveraging the connection between quantum chaos and algebras of operators \cite{Gesteau:2023rrx}, establishing that the averaged Loschmidt echo possesses maximal chaos would shed light on how spacetime `emerges' from the spin chain through the Gaussian average over couplings of the measurements.

\subsection{Identifying states}
So far, our conclusions have been about the large-$N$ behaviour of the models, be it the gravitational system, its holographic dual $\mN=4$ SYM, or the spin chain. Nevertheless, the mapping of the effective partition function onto the spin chain is exact and thus holds for finite $N$. One problem, however, is that the contributions discarded in going from \eqref{eq:MMfull} to \eqref{eq:MM1} give finite contributions at finite $N$, so their corrections should be taken into account. This can be done by constructing a more involved spin chain, with interactions beyond nearest neighbours as explained in \S\ref{app:RP1}, which approximates the effect of accounting for higher operators in \eqref{eq:MMfull}. The situation will remain tractable for $N$ finite but large, but will become untamed for small $N$.\par
In principle, it is possible to encode the exact SCI of $\mN=4$ on a spin chain, because it admits a presentation as a unitary matrix model. Therefore, the argument in \S\ref{app:RP1} shows that with sufficiently complicated interactions one should be able to simulate the SCI and extract the microstate count at large $N$ from the spin chain device. The SCI is protected by supersymmetry, so this procedure does not require weak 't Hooft coupling and theoretically allows non-perturbative counting. Unfortunately, the required Hamiltonian seems too complicated to be effectively implemented in practice. Moreover, the microstate counting in this proposal would suffer from the finite-size effects of the lattice, so the empirical result will inevitably differ from the setup it is supposed to simulate.\par
\medskip
The work \cite{Chang:2022mjp} investigated the finite $N$ spectrum of protected operators in $\mN=4$ SYM, building upon \cite{Chang:2013fba}. One significant discovery was the identification of potential black hole microstates even at $N=2$. It is possible to verify the findings of \cite{Chang:2022mjp} using a quantum simulator. At present, it is unclear how to refine our setup to achieve this, as it would likely require considering more complex interactions among qubits to reproduce the significant $1/N$ corrections. Additionally, it would involve probing a more precise quantity than the Loschmidt echo.\par
This paper lays the groundwork for exploring important quantum simulation questions in the future. The exploration will shed light on the nature of black holes.

\bibliography{BHchain}

\begin{thebibliography}{129}%
\makeatletter
\providecommand \@ifxundefined [1]{%
 \@ifx{#1\undefined}
}%
\providecommand \@ifnum [1]{%
 \ifnum #1\expandafter \@firstoftwo
 \else \expandafter \@secondoftwo
 \fi
}%
\providecommand \@ifx [1]{%
 \ifx #1\expandafter \@firstoftwo
 \else \expandafter \@secondoftwo
 \fi
}%
\providecommand \natexlab [1]{#1}%
\providecommand \enquote  [1]{``#1''}%
\providecommand \bibnamefont  [1]{#1}%
\providecommand \bibfnamefont [1]{#1}%
\providecommand \citenamefont [1]{#1}%
\providecommand \href@noop [0]{\@secondoftwo}%
\providecommand \href [0]{\begingroup \@sanitize@url \@href}%
\providecommand \@href[1]{\@@startlink{#1}\@@href}%
\providecommand \@@href[1]{\endgroup#1\@@endlink}%
\providecommand \@sanitize@url [0]{\catcode `\\12\catcode `\$12\catcode `\&12\catcode `\#12\catcode `\^12\catcode `\_12\catcode `\%12\relax}%
\providecommand \@@startlink[1]{}%
\providecommand \@@endlink[0]{}%
\providecommand \url  [0]{\begingroup\@sanitize@url \@url }%
\providecommand \@url [1]{\endgroup\@href {#1}{\urlprefix }}%
\providecommand \urlprefix  [0]{URL }%
\providecommand \Eprint [0]{\href }%
\providecommand \doibase [0]{http://dx.doi.org/}%
\providecommand \selectlanguage [0]{\@gobble}%
\providecommand \bibinfo  [0]{\@secondoftwo}%
\providecommand \bibfield  [0]{\@secondoftwo}%
\providecommand \translation [1]{[#1]}%
\providecommand \BibitemOpen [0]{}%
\providecommand \bibitemStop [0]{}%
\providecommand \bibitemNoStop [0]{.\EOS\space}%
\providecommand \EOS [0]{\spacefactor3000\relax}%
\providecommand \BibitemShut  [1]{\csname bibitem#1\endcsname}%
\let\auto@bib@innerbib\@empty
\bibitem [{\citenamefont {Wheeler}(1989)}]{Wheeler:1989ftm}%
  \BibitemOpen
  \bibfield  {author} {\bibinfo {author} {\bibfnamefont {John~Archibald}\ \bibnamefont {Wheeler}},\ }\bibfield  {title} {\enquote {\bibinfo {title} {{Information, physics, quantum: The search for links}},}\ }in\ \href@noop {} {\emph {\bibinfo {booktitle} {{3rd International Symposium on Foundations of Quantum Mechanics in Light}}}}\ (\bibinfo {year} {1989})\BibitemShut {NoStop}%
\bibitem [{\citenamefont {Maldacena}\ and\ \citenamefont {Susskind}(2013)}]{Maldacena:2013xja}%
  \BibitemOpen
  \bibfield  {author} {\bibinfo {author} {\bibfnamefont {Juan}\ \bibnamefont {Maldacena}}\ and\ \bibinfo {author} {\bibfnamefont {Leonard}\ \bibnamefont {Susskind}},\ }\bibfield  {title} {\enquote {\bibinfo {title} {{Cool horizons for entangled black holes}},}\ }\href {\doibase 10.1002/prop.201300020} {\bibfield  {journal} {\bibinfo  {journal} {Fortsch. Phys.}\ }\textbf {\bibinfo {volume} {61}},\ \bibinfo {pages} {781--811} (\bibinfo {year} {2013})},\ \Eprint {http://arxiv.org/abs/1306.0533} {arXiv:1306.0533 [hep-th]} \BibitemShut {NoStop}%
\bibitem [{\citenamefont {Almheiri}\ \emph {et~al.}(2013)\citenamefont {Almheiri}, \citenamefont {Marolf}, \citenamefont {Polchinski},\ and\ \citenamefont {Sully}}]{Almheiri:2012rt}%
  \BibitemOpen
  \bibfield  {author} {\bibinfo {author} {\bibfnamefont {Ahmed}\ \bibnamefont {Almheiri}}, \bibinfo {author} {\bibfnamefont {Donald}\ \bibnamefont {Marolf}}, \bibinfo {author} {\bibfnamefont {Joseph}\ \bibnamefont {Polchinski}}, \ and\ \bibinfo {author} {\bibfnamefont {James}\ \bibnamefont {Sully}},\ }\bibfield  {title} {\enquote {\bibinfo {title} {{Black Holes: Complementarity or Firewalls?}}}\ }\href {\doibase 10.1007/JHEP02(2013)062} {\bibfield  {journal} {\bibinfo  {journal} {JHEP}\ }\textbf {\bibinfo {volume} {02}},\ \bibinfo {pages} {062} (\bibinfo {year} {2013})},\ \Eprint {http://arxiv.org/abs/1207.3123} {arXiv:1207.3123 [hep-th]} \BibitemShut {NoStop}%
\bibitem [{\citenamefont {Verlinde}\ and\ \citenamefont {Verlinde}(2013)}]{Verlinde:2012cy}%
  \BibitemOpen
  \bibfield  {author} {\bibinfo {author} {\bibfnamefont {Erik}\ \bibnamefont {Verlinde}}\ and\ \bibinfo {author} {\bibfnamefont {Herman}\ \bibnamefont {Verlinde}},\ }\bibfield  {title} {\enquote {\bibinfo {title} {{Black Hole Entanglement and Quantum Error Correction}},}\ }\href {\doibase 10.1007/JHEP10(2013)107} {\bibfield  {journal} {\bibinfo  {journal} {JHEP}\ }\textbf {\bibinfo {volume} {10}},\ \bibinfo {pages} {107} (\bibinfo {year} {2013})},\ \Eprint {http://arxiv.org/abs/1211.6913} {arXiv:1211.6913 [hep-th]} \BibitemShut {NoStop}%
\bibitem [{\citenamefont {Almheiri}\ \emph {et~al.}(2015)\citenamefont {Almheiri}, \citenamefont {Dong},\ and\ \citenamefont {Harlow}}]{Almheiri:2014lwa}%
  \BibitemOpen
  \bibfield  {author} {\bibinfo {author} {\bibfnamefont {Ahmed}\ \bibnamefont {Almheiri}}, \bibinfo {author} {\bibfnamefont {Xi}~\bibnamefont {Dong}}, \ and\ \bibinfo {author} {\bibfnamefont {Daniel}\ \bibnamefont {Harlow}},\ }\bibfield  {title} {\enquote {\bibinfo {title} {{Bulk Locality and Quantum Error Correction in AdS/CFT}},}\ }\href {\doibase 10.1007/JHEP04(2015)163} {\bibfield  {journal} {\bibinfo  {journal} {JHEP}\ }\textbf {\bibinfo {volume} {04}},\ \bibinfo {pages} {163} (\bibinfo {year} {2015})},\ \Eprint {http://arxiv.org/abs/1411.7041} {arXiv:1411.7041 [hep-th]} \BibitemShut {NoStop}%
\bibitem [{\citenamefont {Penington}(2020)}]{Penington:2019npb}%
  \BibitemOpen
  \bibfield  {author} {\bibinfo {author} {\bibfnamefont {Geoffrey}\ \bibnamefont {Penington}},\ }\bibfield  {title} {\enquote {\bibinfo {title} {{Entanglement Wedge Reconstruction and the Information Paradox}},}\ }\href {\doibase 10.1007/JHEP09(2020)002} {\bibfield  {journal} {\bibinfo  {journal} {JHEP}\ }\textbf {\bibinfo {volume} {09}},\ \bibinfo {pages} {002} (\bibinfo {year} {2020})},\ \Eprint {http://arxiv.org/abs/1905.08255} {arXiv:1905.08255 [hep-th]} \BibitemShut {NoStop}%
\bibitem [{\citenamefont {Almheiri}\ \emph {et~al.}(2019)\citenamefont {Almheiri}, \citenamefont {Engelhardt}, \citenamefont {Marolf},\ and\ \citenamefont {Maxfield}}]{Almheiri:2019psf}%
  \BibitemOpen
  \bibfield  {author} {\bibinfo {author} {\bibfnamefont {Ahmed}\ \bibnamefont {Almheiri}}, \bibinfo {author} {\bibfnamefont {Netta}\ \bibnamefont {Engelhardt}}, \bibinfo {author} {\bibfnamefont {Donald}\ \bibnamefont {Marolf}}, \ and\ \bibinfo {author} {\bibfnamefont {Henry}\ \bibnamefont {Maxfield}},\ }\bibfield  {title} {\enquote {\bibinfo {title} {{The entropy of bulk quantum fields and the entanglement wedge of an evaporating black hole}},}\ }\href {\doibase 10.1007/JHEP12(2019)063} {\bibfield  {journal} {\bibinfo  {journal} {JHEP}\ }\textbf {\bibinfo {volume} {12}},\ \bibinfo {pages} {063} (\bibinfo {year} {2019})},\ \Eprint {http://arxiv.org/abs/1905.08762} {arXiv:1905.08762 [hep-th]} \BibitemShut {NoStop}%
\bibitem [{\citenamefont {Jafferis}\ \emph {et~al.}(2022)\citenamefont {Jafferis}, \citenamefont {Zlokapa}, \citenamefont {Lykken}, \citenamefont {Kolchmeyer}, \citenamefont {Davis}, \citenamefont {Lauk}, \citenamefont {Neven},\ and\ \citenamefont {Spiropulu}}]{Jafferis:2022crx}%
  \BibitemOpen
  \bibfield  {author} {\bibinfo {author} {\bibfnamefont {Daniel}\ \bibnamefont {Jafferis}}, \bibinfo {author} {\bibfnamefont {Alexander}\ \bibnamefont {Zlokapa}}, \bibinfo {author} {\bibfnamefont {Joseph~D.}\ \bibnamefont {Lykken}}, \bibinfo {author} {\bibfnamefont {David~K.}\ \bibnamefont {Kolchmeyer}}, \bibinfo {author} {\bibfnamefont {Samantha~I.}\ \bibnamefont {Davis}}, \bibinfo {author} {\bibfnamefont {Nikolai}\ \bibnamefont {Lauk}}, \bibinfo {author} {\bibfnamefont {Hartmut}\ \bibnamefont {Neven}}, \ and\ \bibinfo {author} {\bibfnamefont {Maria}\ \bibnamefont {Spiropulu}},\ }\bibfield  {title} {\enquote {\bibinfo {title} {{Traversable wormhole dynamics on a quantum processor}},}\ }\href {\doibase 10.1038/s41586-022-05424-3} {\bibfield  {journal} {\bibinfo  {journal} {Nature}\ }\textbf {\bibinfo {volume} {612}},\ \bibinfo {pages} {51--55} (\bibinfo {year} {2022})}\BibitemShut {NoStop}%
\bibitem [{\citenamefont {Hawking}\ and\ \citenamefont {Page}(1983)}]{Hawking:1982dh}%
  \BibitemOpen
  \bibfield  {author} {\bibinfo {author} {\bibfnamefont {Stephen~W.}\ \bibnamefont {Hawking}}\ and\ \bibinfo {author} {\bibfnamefont {Don~N.}\ \bibnamefont {Page}},\ }\bibfield  {title} {\enquote {\bibinfo {title} {{Thermodynamics of Black Holes in anti-De Sitter Space}},}\ }\href {\doibase 10.1007/BF01208266} {\bibfield  {journal} {\bibinfo  {journal} {Commun. Math. Phys.}\ }\textbf {\bibinfo {volume} {87}},\ \bibinfo {pages} {577} (\bibinfo {year} {1983})}\BibitemShut {NoStop}%
\bibitem [{\citenamefont {Bekenstein}(1973)}]{Bekenstein:1973ur}%
  \BibitemOpen
  \bibfield  {author} {\bibinfo {author} {\bibfnamefont {Jacob~D.}\ \bibnamefont {Bekenstein}},\ }\bibfield  {title} {\enquote {\bibinfo {title} {{Black holes and entropy}},}\ }\href {\doibase 10.1103/PhysRevD.7.2333} {\bibfield  {journal} {\bibinfo  {journal} {Phys. Rev. D}\ }\textbf {\bibinfo {volume} {7}},\ \bibinfo {pages} {2333--2346} (\bibinfo {year} {1973})}\BibitemShut {NoStop}%
\bibitem [{\citenamefont {Hawking}(1975)}]{Hawking:1975vcx}%
  \BibitemOpen
  \bibfield  {author} {\bibinfo {author} {\bibfnamefont {Stephen~W.}\ \bibnamefont {Hawking}},\ }\bibfield  {title} {\enquote {\bibinfo {title} {{Particle Creation by Black Holes}},}\ }\href {\doibase 10.1007/BF02345020} {\bibfield  {journal} {\bibinfo  {journal} {Commun. Math. Phys.}\ }\textbf {\bibinfo {volume} {43}},\ \bibinfo {pages} {199--220} (\bibinfo {year} {1975})},\ \bibinfo {note} {[Erratum: Commun.Math.Phys. 46, 206 (1976)]}\BibitemShut {NoStop}%
\bibitem [{\citenamefont {Maldacena}(1998)}]{Maldacena:1997re}%
  \BibitemOpen
  \bibfield  {author} {\bibinfo {author} {\bibfnamefont {Juan~Martin}\ \bibnamefont {Maldacena}},\ }\bibfield  {title} {\enquote {\bibinfo {title} {{The Large N limit of superconformal field theories and supergravity}},}\ }\href {\doibase 10.1023/A:1026654312961} {\bibfield  {journal} {\bibinfo  {journal} {Adv. Theor. Math. Phys.}\ }\textbf {\bibinfo {volume} {2}},\ \bibinfo {pages} {231--252} (\bibinfo {year} {1998})},\ \Eprint {http://arxiv.org/abs/hep-th/9711200} {arXiv:hep-th/9711200} \BibitemShut {NoStop}%
\bibitem [{\citenamefont {Gubser}\ \emph {et~al.}(1998)\citenamefont {Gubser}, \citenamefont {Klebanov},\ and\ \citenamefont {Polyakov}}]{Gubser:1998bc}%
  \BibitemOpen
  \bibfield  {author} {\bibinfo {author} {\bibfnamefont {S.~S.}\ \bibnamefont {Gubser}}, \bibinfo {author} {\bibfnamefont {Igor~R.}\ \bibnamefont {Klebanov}}, \ and\ \bibinfo {author} {\bibfnamefont {Alexander~M.}\ \bibnamefont {Polyakov}},\ }\bibfield  {title} {\enquote {\bibinfo {title} {{Gauge theory correlators from noncritical string theory}},}\ }\href {\doibase 10.1016/S0370-2693(98)00377-3} {\bibfield  {journal} {\bibinfo  {journal} {Phys. Lett. B}\ }\textbf {\bibinfo {volume} {428}},\ \bibinfo {pages} {105--114} (\bibinfo {year} {1998})},\ \Eprint {http://arxiv.org/abs/hep-th/9802109} {arXiv:hep-th/9802109} \BibitemShut {NoStop}%
\bibitem [{\citenamefont {Witten}(1998{\natexlab{a}})}]{Witten:1998qj}%
  \BibitemOpen
  \bibfield  {author} {\bibinfo {author} {\bibfnamefont {Edward}\ \bibnamefont {Witten}},\ }\bibfield  {title} {\enquote {\bibinfo {title} {{Anti-de Sitter space and holography}},}\ }\href {\doibase 10.4310/ATMP.1998.v2.n2.a2} {\bibfield  {journal} {\bibinfo  {journal} {Adv. Theor. Math. Phys.}\ }\textbf {\bibinfo {volume} {2}},\ \bibinfo {pages} {253--291} (\bibinfo {year} {1998}{\natexlab{a}})},\ \Eprint {http://arxiv.org/abs/hep-th/9802150} {arXiv:hep-th/9802150} \BibitemShut {NoStop}%
\bibitem [{\citenamefont {Sundborg}(2000)}]{Sundborg:1999ue}%
  \BibitemOpen
  \bibfield  {author} {\bibinfo {author} {\bibfnamefont {Bo}~\bibnamefont {Sundborg}},\ }\bibfield  {title} {\enquote {\bibinfo {title} {{The Hagedorn transition, deconfinement and N=4 SYM theory}},}\ }\href {\doibase 10.1016/S0550-3213(00)00044-4} {\bibfield  {journal} {\bibinfo  {journal} {Nucl. Phys. B}\ }\textbf {\bibinfo {volume} {573}},\ \bibinfo {pages} {349--363} (\bibinfo {year} {2000})},\ \Eprint {http://arxiv.org/abs/hep-th/9908001} {arXiv:hep-th/9908001} \BibitemShut {NoStop}%
\bibitem [{\citenamefont {Polyakov}(2002)}]{Polyakov:2001af}%
  \BibitemOpen
  \bibfield  {author} {\bibinfo {author} {\bibfnamefont {Alexander~M.}\ \bibnamefont {Polyakov}},\ }\bibfield  {title} {\enquote {\bibinfo {title} {{Gauge fields and space-time}},}\ }\href {\doibase 10.1142/S0217751X02013071} {\bibfield  {journal} {\bibinfo  {journal} {Int. J. Mod. Phys. A}\ }\textbf {\bibinfo {volume} {17S1}},\ \bibinfo {pages} {119--136} (\bibinfo {year} {2002})},\ \Eprint {http://arxiv.org/abs/hep-th/0110196} {arXiv:hep-th/0110196} \BibitemShut {NoStop}%
\bibitem [{\citenamefont {Balasubramanian}\ \emph {et~al.}(2002)\citenamefont {Balasubramanian}, \citenamefont {Huang}, \citenamefont {Levi},\ and\ \citenamefont {Naqvi}}]{Balasubramanian:2002sa}%
  \BibitemOpen
  \bibfield  {author} {\bibinfo {author} {\bibfnamefont {Vijay}\ \bibnamefont {Balasubramanian}}, \bibinfo {author} {\bibfnamefont {Min-xin}\ \bibnamefont {Huang}}, \bibinfo {author} {\bibfnamefont {Thomas~S.}\ \bibnamefont {Levi}}, \ and\ \bibinfo {author} {\bibfnamefont {Asad}\ \bibnamefont {Naqvi}},\ }\bibfield  {title} {\enquote {\bibinfo {title} {{Open strings from N=4 superYang-Mills}},}\ }\href {\doibase 10.1088/1126-6708/2002/08/037} {\bibfield  {journal} {\bibinfo  {journal} {JHEP}\ }\textbf {\bibinfo {volume} {08}},\ \bibinfo {pages} {037} (\bibinfo {year} {2002})},\ \Eprint {http://arxiv.org/abs/hep-th/0204196} {arXiv:hep-th/0204196} \BibitemShut {NoStop}%
\bibitem [{\citenamefont {Aharony}\ \emph {et~al.}(2004)\citenamefont {Aharony}, \citenamefont {Marsano}, \citenamefont {Minwalla}, \citenamefont {Papadodimas},\ and\ \citenamefont {Van~Raamsdonk}}]{Aharony:2003sx}%
  \BibitemOpen
  \bibfield  {author} {\bibinfo {author} {\bibfnamefont {Ofer}\ \bibnamefont {Aharony}}, \bibinfo {author} {\bibfnamefont {Joseph}\ \bibnamefont {Marsano}}, \bibinfo {author} {\bibfnamefont {Shiraz}\ \bibnamefont {Minwalla}}, \bibinfo {author} {\bibfnamefont {Kyriakos}\ \bibnamefont {Papadodimas}}, \ and\ \bibinfo {author} {\bibfnamefont {Mark}\ \bibnamefont {Van~Raamsdonk}},\ }\bibfield  {title} {\enquote {\bibinfo {title} {{The Hagedorn - deconfinement phase transition in weakly coupled large N gauge theories}},}\ }\href {\doibase 10.4310/ATMP.2004.v8.n4.a1} {\bibfield  {journal} {\bibinfo  {journal} {Adv. Theor. Math. Phys.}\ }\textbf {\bibinfo {volume} {8}},\ \bibinfo {pages} {603--696} (\bibinfo {year} {2004})},\ \Eprint {http://arxiv.org/abs/hep-th/0310285} {arXiv:hep-th/0310285} \BibitemShut {NoStop}%
\bibitem [{\citenamefont {Liu}(2004)}]{Liu:2004vy}%
  \BibitemOpen
  \bibfield  {author} {\bibinfo {author} {\bibfnamefont {Hong}\ \bibnamefont {Liu}},\ }\bibfield  {title} {\enquote {\bibinfo {title} {{Fine structure of Hagedorn transitions}},}\ }\href@noop {} {\  (\bibinfo {year} {2004})},\ \Eprint {http://arxiv.org/abs/hep-th/0408001} {arXiv:hep-th/0408001} \BibitemShut {NoStop}%
\bibitem [{\citenamefont {Berenstein}(2004)}]{Berenstein:2004kk}%
  \BibitemOpen
  \bibfield  {author} {\bibinfo {author} {\bibfnamefont {David}\ \bibnamefont {Berenstein}},\ }\bibfield  {title} {\enquote {\bibinfo {title} {{A Toy model for the AdS / CFT correspondence}},}\ }\href {\doibase 10.1088/1126-6708/2004/07/018} {\bibfield  {journal} {\bibinfo  {journal} {JHEP}\ }\textbf {\bibinfo {volume} {07}},\ \bibinfo {pages} {018} (\bibinfo {year} {2004})},\ \Eprint {http://arxiv.org/abs/hep-th/0403110} {arXiv:hep-th/0403110} \BibitemShut {NoStop}%
\bibitem [{\citenamefont {Alvarez-Gaume}\ \emph {et~al.}(2005)\citenamefont {Alvarez-Gaume}, \citenamefont {Gomez}, \citenamefont {Liu},\ and\ \citenamefont {Wadia}}]{Alvarez-Gaume:2005dvb}%
  \BibitemOpen
  \bibfield  {author} {\bibinfo {author} {\bibfnamefont {Luis}\ \bibnamefont {Alvarez-Gaume}}, \bibinfo {author} {\bibfnamefont {Cesar}\ \bibnamefont {Gomez}}, \bibinfo {author} {\bibfnamefont {Hong}\ \bibnamefont {Liu}}, \ and\ \bibinfo {author} {\bibfnamefont {Spenta}\ \bibnamefont {Wadia}},\ }\bibfield  {title} {\enquote {\bibinfo {title} {{Finite temperature effective action, AdS(5) black holes, and 1/N expansion}},}\ }\href {\doibase 10.1103/PhysRevD.71.124023} {\bibfield  {journal} {\bibinfo  {journal} {Phys. Rev. D}\ }\textbf {\bibinfo {volume} {71}},\ \bibinfo {pages} {124023} (\bibinfo {year} {2005})},\ \Eprint {http://arxiv.org/abs/hep-th/0502227} {arXiv:hep-th/0502227} \BibitemShut {NoStop}%
\bibitem [{\citenamefont {Festuccia}\ and\ \citenamefont {Liu}(2006)}]{Festuccia:2005pi}%
  \BibitemOpen
  \bibfield  {author} {\bibinfo {author} {\bibfnamefont {Guido}\ \bibnamefont {Festuccia}}\ and\ \bibinfo {author} {\bibfnamefont {Hong}\ \bibnamefont {Liu}},\ }\bibfield  {title} {\enquote {\bibinfo {title} {{Excursions beyond the horizon: Black hole singularities in Yang-Mills theories. I.}}}\ }\href {\doibase 10.1088/1126-6708/2006/04/044} {\bibfield  {journal} {\bibinfo  {journal} {JHEP}\ }\textbf {\bibinfo {volume} {04}},\ \bibinfo {pages} {044} (\bibinfo {year} {2006})},\ \Eprint {http://arxiv.org/abs/hep-th/0506202} {arXiv:hep-th/0506202} \BibitemShut {NoStop}%
\bibitem [{\citenamefont {Basu}\ and\ \citenamefont {Wadia}(2006)}]{Basu:2005pj}%
  \BibitemOpen
  \bibfield  {author} {\bibinfo {author} {\bibfnamefont {Pallab}\ \bibnamefont {Basu}}\ and\ \bibinfo {author} {\bibfnamefont {Spenta~R.}\ \bibnamefont {Wadia}},\ }\bibfield  {title} {\enquote {\bibinfo {title} {{R-charged AdS(5) black holes and large N unitary matrix models}},}\ }\href {\doibase 10.1103/PhysRevD.73.045022} {\bibfield  {journal} {\bibinfo  {journal} {Phys. Rev. D}\ }\textbf {\bibinfo {volume} {73}},\ \bibinfo {pages} {045022} (\bibinfo {year} {2006})},\ \Eprint {http://arxiv.org/abs/hep-th/0506203} {arXiv:hep-th/0506203} \BibitemShut {NoStop}%
\bibitem [{\citenamefont {Alvarez-Gaume}\ \emph {et~al.}(2006)\citenamefont {Alvarez-Gaume}, \citenamefont {Basu}, \citenamefont {Marino},\ and\ \citenamefont {Wadia}}]{Alvarez-Gaume:2006fwd}%
  \BibitemOpen
  \bibfield  {author} {\bibinfo {author} {\bibfnamefont {Luis}\ \bibnamefont {Alvarez-Gaume}}, \bibinfo {author} {\bibfnamefont {Pallab}\ \bibnamefont {Basu}}, \bibinfo {author} {\bibfnamefont {Marcos}\ \bibnamefont {Marino}}, \ and\ \bibinfo {author} {\bibfnamefont {Spenta~R.}\ \bibnamefont {Wadia}},\ }\bibfield  {title} {\enquote {\bibinfo {title} {{Blackhole/String Transition for the Small Schwarzschild Blackhole of AdS(5)x S**5 and Critical Unitary Matrix Models}},}\ }\href {\doibase 10.1140/epjc/s10052-006-0049-x} {\bibfield  {journal} {\bibinfo  {journal} {Eur. Phys. J. C}\ }\textbf {\bibinfo {volume} {48}},\ \bibinfo {pages} {647--665} (\bibinfo {year} {2006})},\ \Eprint {http://arxiv.org/abs/hep-th/0605041} {arXiv:hep-th/0605041} \BibitemShut {NoStop}%
\bibitem [{\citenamefont {Biswas}\ \emph {et~al.}(2007)\citenamefont {Biswas}, \citenamefont {Gaiotto}, \citenamefont {Lahiri},\ and\ \citenamefont {Minwalla}}]{Biswas:2006tj}%
  \BibitemOpen
  \bibfield  {author} {\bibinfo {author} {\bibfnamefont {Indranil}\ \bibnamefont {Biswas}}, \bibinfo {author} {\bibfnamefont {Davide}\ \bibnamefont {Gaiotto}}, \bibinfo {author} {\bibfnamefont {Subhaneil}\ \bibnamefont {Lahiri}}, \ and\ \bibinfo {author} {\bibfnamefont {Shiraz}\ \bibnamefont {Minwalla}},\ }\bibfield  {title} {\enquote {\bibinfo {title} {{Supersymmetric states of N=4 Yang-Mills from giant gravitons}},}\ }\href {\doibase 10.1088/1126-6708/2007/12/006} {\bibfield  {journal} {\bibinfo  {journal} {JHEP}\ }\textbf {\bibinfo {volume} {12}},\ \bibinfo {pages} {006} (\bibinfo {year} {2007})},\ \Eprint {http://arxiv.org/abs/hep-th/0606087} {arXiv:hep-th/0606087} \BibitemShut {NoStop}%
\bibitem [{\citenamefont {Chang}\ and\ \citenamefont {Yin}(2013)}]{Chang:2013fba}%
  \BibitemOpen
  \bibfield  {author} {\bibinfo {author} {\bibfnamefont {Chi-Ming}\ \bibnamefont {Chang}}\ and\ \bibinfo {author} {\bibfnamefont {Xi}~\bibnamefont {Yin}},\ }\bibfield  {title} {\enquote {\bibinfo {title} {{1/16 BPS states in $\mathcal N=$ 4 super-Yang-Mills theory}},}\ }\href {\doibase 10.1103/PhysRevD.88.106005} {\bibfield  {journal} {\bibinfo  {journal} {Phys. Rev. D}\ }\textbf {\bibinfo {volume} {88}},\ \bibinfo {pages} {106005} (\bibinfo {year} {2013})},\ \Eprint {http://arxiv.org/abs/1305.6314} {arXiv:1305.6314} \BibitemShut {NoStop}%
\bibitem [{\citenamefont {Hosseini}\ \emph {et~al.}(2017)\citenamefont {Hosseini}, \citenamefont {Hristov},\ and\ \citenamefont {Zaffaroni}}]{Hosseini:2017mds}%
  \BibitemOpen
  \bibfield  {author} {\bibinfo {author} {\bibfnamefont {Seyed~Morteza}\ \bibnamefont {Hosseini}}, \bibinfo {author} {\bibfnamefont {Kiril}\ \bibnamefont {Hristov}}, \ and\ \bibinfo {author} {\bibfnamefont {Alberto}\ \bibnamefont {Zaffaroni}},\ }\bibfield  {title} {\enquote {\bibinfo {title} {{An extremization principle for the entropy of rotating BPS black holes in AdS$_{5}$}},}\ }\href {\doibase 10.1007/JHEP07(2017)106} {\bibfield  {journal} {\bibinfo  {journal} {JHEP}\ }\textbf {\bibinfo {volume} {07}},\ \bibinfo {pages} {106} (\bibinfo {year} {2017})},\ \Eprint {http://arxiv.org/abs/1705.05383} {arXiv:1705.05383} \BibitemShut {NoStop}%
\bibitem [{\citenamefont {Cabo-Bizet}\ \emph {et~al.}(2019{\natexlab{a}})\citenamefont {Cabo-Bizet}, \citenamefont {Cassani}, \citenamefont {Martelli},\ and\ \citenamefont {Murthy}}]{Cabo-Bizet:2018ehj}%
  \BibitemOpen
  \bibfield  {author} {\bibinfo {author} {\bibfnamefont {Alejandro}\ \bibnamefont {Cabo-Bizet}}, \bibinfo {author} {\bibfnamefont {Davide}\ \bibnamefont {Cassani}}, \bibinfo {author} {\bibfnamefont {Dario}\ \bibnamefont {Martelli}}, \ and\ \bibinfo {author} {\bibfnamefont {Sameer}\ \bibnamefont {Murthy}},\ }\bibfield  {title} {\enquote {\bibinfo {title} {{Microscopic origin of the Bekenstein-Hawking entropy of supersymmetric AdS$_{5}$ black holes}},}\ }\href {\doibase 10.1007/JHEP10(2019)062} {\bibfield  {journal} {\bibinfo  {journal} {JHEP}\ }\textbf {\bibinfo {volume} {10}},\ \bibinfo {pages} {062} (\bibinfo {year} {2019}{\natexlab{a}})},\ \Eprint {http://arxiv.org/abs/1810.11442} {arXiv:1810.11442} \BibitemShut {NoStop}%
\bibitem [{\citenamefont {Choi}\ \emph {et~al.}(2018{\natexlab{a}})\citenamefont {Choi}, \citenamefont {Kim}, \citenamefont {Kim},\ and\ \citenamefont {Nahmgoong}}]{Choi:2018hmj}%
  \BibitemOpen
  \bibfield  {author} {\bibinfo {author} {\bibfnamefont {Sunjin}\ \bibnamefont {Choi}}, \bibinfo {author} {\bibfnamefont {Joonho}\ \bibnamefont {Kim}}, \bibinfo {author} {\bibfnamefont {Seok}\ \bibnamefont {Kim}}, \ and\ \bibinfo {author} {\bibfnamefont {June}\ \bibnamefont {Nahmgoong}},\ }\bibfield  {title} {\enquote {\bibinfo {title} {{Large AdS black holes from QFT}},}\ }\href@noop {} {\  (\bibinfo {year} {2018}{\natexlab{a}})},\ \Eprint {http://arxiv.org/abs/1810.12067} {arXiv:1810.12067} \BibitemShut {NoStop}%
\bibitem [{\citenamefont {Choi}\ \emph {et~al.}(2018{\natexlab{b}})\citenamefont {Choi}, \citenamefont {Kim}, \citenamefont {Kim},\ and\ \citenamefont {Nahmgoong}}]{Choi:2018vbz}%
  \BibitemOpen
  \bibfield  {author} {\bibinfo {author} {\bibfnamefont {Sunjin}\ \bibnamefont {Choi}}, \bibinfo {author} {\bibfnamefont {Joonho}\ \bibnamefont {Kim}}, \bibinfo {author} {\bibfnamefont {Seok}\ \bibnamefont {Kim}}, \ and\ \bibinfo {author} {\bibfnamefont {June}\ \bibnamefont {Nahmgoong}},\ }\bibfield  {title} {\enquote {\bibinfo {title} {{Comments on deconfinement in AdS/CFT}},}\ }\href@noop {} {\  (\bibinfo {year} {2018}{\natexlab{b}})},\ \Eprint {http://arxiv.org/abs/1811.08646} {arXiv:1811.08646} \BibitemShut {NoStop}%
\bibitem [{\citenamefont {Benini}\ and\ \citenamefont {Milan}(2020{\natexlab{a}})}]{Benini:2018mlo}%
  \BibitemOpen
  \bibfield  {author} {\bibinfo {author} {\bibfnamefont {Francesco}\ \bibnamefont {Benini}}\ and\ \bibinfo {author} {\bibfnamefont {Paolo}\ \bibnamefont {Milan}},\ }\bibfield  {title} {\enquote {\bibinfo {title} {{A Bethe Ansatz type formula for the superconformal index}},}\ }\href {\doibase 10.1007/s00220-019-03679-y} {\bibfield  {journal} {\bibinfo  {journal} {Commun. Math. Phys.}\ }\textbf {\bibinfo {volume} {376}},\ \bibinfo {pages} {1413--1440} (\bibinfo {year} {2020}{\natexlab{a}})},\ \Eprint {http://arxiv.org/abs/1811.04107} {arXiv:1811.04107} \BibitemShut {NoStop}%
\bibitem [{\citenamefont {Benini}\ and\ \citenamefont {Milan}(2020{\natexlab{b}})}]{Benini:2018ywd}%
  \BibitemOpen
  \bibfield  {author} {\bibinfo {author} {\bibfnamefont {Francesco}\ \bibnamefont {Benini}}\ and\ \bibinfo {author} {\bibfnamefont {Paolo}\ \bibnamefont {Milan}},\ }\bibfield  {title} {\enquote {\bibinfo {title} {{Black Holes in 4D $\mathcal{N}$=4 Super-Yang-Mills Field Theory}},}\ }\href {\doibase 10.1103/PhysRevX.10.021037} {\bibfield  {journal} {\bibinfo  {journal} {Phys. Rev. X}\ }\textbf {\bibinfo {volume} {10}},\ \bibinfo {pages} {021037} (\bibinfo {year} {2020}{\natexlab{b}})},\ \Eprint {http://arxiv.org/abs/1812.09613} {arXiv:1812.09613} \BibitemShut {NoStop}%
\bibitem [{\citenamefont {Honda}(2019)}]{Honda:2019cio}%
  \BibitemOpen
  \bibfield  {author} {\bibinfo {author} {\bibfnamefont {Masazumi}\ \bibnamefont {Honda}},\ }\bibfield  {title} {\enquote {\bibinfo {title} {{Quantum Black Hole Entropy from 4d Supersymmetric Cardy formula}},}\ }\href {\doibase 10.1103/PhysRevD.100.026008} {\bibfield  {journal} {\bibinfo  {journal} {Phys. Rev. D}\ }\textbf {\bibinfo {volume} {100}},\ \bibinfo {pages} {026008} (\bibinfo {year} {2019})},\ \Eprint {http://arxiv.org/abs/1901.08091} {arXiv:1901.08091} \BibitemShut {NoStop}%
\bibitem [{\citenamefont {Arabi~Ardehali}(2019)}]{ArabiArdehali:2019tdm}%
  \BibitemOpen
  \bibfield  {author} {\bibinfo {author} {\bibfnamefont {Arash}\ \bibnamefont {Arabi~Ardehali}},\ }\bibfield  {title} {\enquote {\bibinfo {title} {{Cardy-like asymptotics of the 4d $ \mathcal{N}=4 $ index and AdS$_{5}$ blackholes}},}\ }\href {\doibase 10.1007/JHEP06(2019)134} {\bibfield  {journal} {\bibinfo  {journal} {JHEP}\ }\textbf {\bibinfo {volume} {06}},\ \bibinfo {pages} {134} (\bibinfo {year} {2019})},\ \Eprint {http://arxiv.org/abs/1902.06619} {arXiv:1902.06619} \BibitemShut {NoStop}%
\bibitem [{\citenamefont {Cabo-Bizet}\ \emph {et~al.}(2019{\natexlab{b}})\citenamefont {Cabo-Bizet}, \citenamefont {Cassani}, \citenamefont {Martelli},\ and\ \citenamefont {Murthy}}]{Cabo-Bizet:2019osg}%
  \BibitemOpen
  \bibfield  {author} {\bibinfo {author} {\bibfnamefont {Alejandro}\ \bibnamefont {Cabo-Bizet}}, \bibinfo {author} {\bibfnamefont {Davide}\ \bibnamefont {Cassani}}, \bibinfo {author} {\bibfnamefont {Dario}\ \bibnamefont {Martelli}}, \ and\ \bibinfo {author} {\bibfnamefont {Sameer}\ \bibnamefont {Murthy}},\ }\bibfield  {title} {\enquote {\bibinfo {title} {{The asymptotic growth of states of the 4d $ \mathcal{N}=1 $ superconformal index}},}\ }\href {\doibase 10.1007/JHEP08(2019)120} {\bibfield  {journal} {\bibinfo  {journal} {JHEP}\ }\textbf {\bibinfo {volume} {08}},\ \bibinfo {pages} {120} (\bibinfo {year} {2019}{\natexlab{b}})},\ \Eprint {http://arxiv.org/abs/1904.05865} {arXiv:1904.05865} \BibitemShut {NoStop}%
\bibitem [{\citenamefont {Gonz\'alez~Lezcano}\ and\ \citenamefont {Pando~Zayas}(2020)}]{GonzalezLezcano:2019nca}%
  \BibitemOpen
  \bibfield  {author} {\bibinfo {author} {\bibfnamefont {Alfredo}\ \bibnamefont {Gonz\'alez~Lezcano}}\ and\ \bibinfo {author} {\bibfnamefont {Leopoldo~A.}\ \bibnamefont {Pando~Zayas}},\ }\bibfield  {title} {\enquote {\bibinfo {title} {{Microstate counting via Bethe Ans\"atze in the 4d $ \mathcal{N} $ = 1 superconformal index}},}\ }\href {\doibase 10.1007/JHEP03(2020)088} {\bibfield  {journal} {\bibinfo  {journal} {JHEP}\ }\textbf {\bibinfo {volume} {03}},\ \bibinfo {pages} {088} (\bibinfo {year} {2020})},\ \Eprint {http://arxiv.org/abs/1907.12841} {arXiv:1907.12841} \BibitemShut {NoStop}%
\bibitem [{\citenamefont {Larsen}\ \emph {et~al.}(2020)\citenamefont {Larsen}, \citenamefont {Nian},\ and\ \citenamefont {Zeng}}]{Larsen:2019oll}%
  \BibitemOpen
  \bibfield  {author} {\bibinfo {author} {\bibfnamefont {Finn}\ \bibnamefont {Larsen}}, \bibinfo {author} {\bibfnamefont {Jun}\ \bibnamefont {Nian}}, \ and\ \bibinfo {author} {\bibfnamefont {Yangwenxiao}\ \bibnamefont {Zeng}},\ }\bibfield  {title} {\enquote {\bibinfo {title} {{AdS$_{5}$ black hole entropy near the BPS limit}},}\ }\href {\doibase 10.1007/JHEP06(2020)001} {\bibfield  {journal} {\bibinfo  {journal} {JHEP}\ }\textbf {\bibinfo {volume} {06}},\ \bibinfo {pages} {001} (\bibinfo {year} {2020})},\ \Eprint {http://arxiv.org/abs/1907.02505} {arXiv:1907.02505 [hep-th]} \BibitemShut {NoStop}%
\bibitem [{\citenamefont {Cabo-Bizet}\ and\ \citenamefont {Murthy}(2020)}]{Cabo-Bizet:2019eaf}%
  \BibitemOpen
  \bibfield  {author} {\bibinfo {author} {\bibfnamefont {Alejandro}\ \bibnamefont {Cabo-Bizet}}\ and\ \bibinfo {author} {\bibfnamefont {Sameer}\ \bibnamefont {Murthy}},\ }\bibfield  {title} {\enquote {\bibinfo {title} {{Supersymmetric phases of 4d $ \mathcal{N} $ = 4 SYM at large $N$}},}\ }\href {\doibase 10.1007/JHEP09(2020)184} {\bibfield  {journal} {\bibinfo  {journal} {JHEP}\ }\textbf {\bibinfo {volume} {09}},\ \bibinfo {pages} {184} (\bibinfo {year} {2020})},\ \Eprint {http://arxiv.org/abs/1909.09597} {arXiv:1909.09597} \BibitemShut {NoStop}%
\bibitem [{\citenamefont {Arabi~Ardehali}\ \emph {et~al.}(2020)\citenamefont {Arabi~Ardehali}, \citenamefont {Hong},\ and\ \citenamefont {Liu}}]{ArabiArdehali:2019orz}%
  \BibitemOpen
  \bibfield  {author} {\bibinfo {author} {\bibfnamefont {Arash}\ \bibnamefont {Arabi~Ardehali}}, \bibinfo {author} {\bibfnamefont {Junho}\ \bibnamefont {Hong}}, \ and\ \bibinfo {author} {\bibfnamefont {James~T.}\ \bibnamefont {Liu}},\ }\bibfield  {title} {\enquote {\bibinfo {title} {{Asymptotic growth of the 4d $ \mathcal{N} $ = 4 index and partially deconfined phases}},}\ }\href {\doibase 10.1007/JHEP07(2020)073} {\bibfield  {journal} {\bibinfo  {journal} {JHEP}\ }\textbf {\bibinfo {volume} {07}},\ \bibinfo {pages} {073} (\bibinfo {year} {2020})},\ \Eprint {http://arxiv.org/abs/1912.04169} {arXiv:1912.04169} \BibitemShut {NoStop}%
\bibitem [{\citenamefont {Benini}\ \emph {et~al.}(2020)\citenamefont {Benini}, \citenamefont {Colombo}, \citenamefont {Soltani}, \citenamefont {Zaffaroni},\ and\ \citenamefont {Zhang}}]{Benini:2020gjh}%
  \BibitemOpen
  \bibfield  {author} {\bibinfo {author} {\bibfnamefont {Francesco}\ \bibnamefont {Benini}}, \bibinfo {author} {\bibfnamefont {Edoardo}\ \bibnamefont {Colombo}}, \bibinfo {author} {\bibfnamefont {Saman}\ \bibnamefont {Soltani}}, \bibinfo {author} {\bibfnamefont {Alberto}\ \bibnamefont {Zaffaroni}}, \ and\ \bibinfo {author} {\bibfnamefont {Ziruo}\ \bibnamefont {Zhang}},\ }\bibfield  {title} {\enquote {\bibinfo {title} {{Superconformal indices at large $N$ and the entropy of AdS$_5$ $\times$ SE$_5$ black holes}},}\ }\href {\doibase 10.1088/1361-6382/abb39b} {\bibfield  {journal} {\bibinfo  {journal} {Class. Quant. Grav.}\ }\textbf {\bibinfo {volume} {37}},\ \bibinfo {pages} {215021} (\bibinfo {year} {2020})},\ \Eprint {http://arxiv.org/abs/2005.12308} {arXiv:2005.12308} \BibitemShut {NoStop}%
\bibitem [{\citenamefont {David}\ \emph {et~al.}(2020)\citenamefont {David}, \citenamefont {Nian},\ and\ \citenamefont {Pando~Zayas}}]{David:2020ems}%
  \BibitemOpen
  \bibfield  {author} {\bibinfo {author} {\bibfnamefont {Marina}\ \bibnamefont {David}}, \bibinfo {author} {\bibfnamefont {Jun}\ \bibnamefont {Nian}}, \ and\ \bibinfo {author} {\bibfnamefont {Leopoldo~A.}\ \bibnamefont {Pando~Zayas}},\ }\bibfield  {title} {\enquote {\bibinfo {title} {{Gravitational Cardy Limit and AdS Black Hole Entropy}},}\ }\href {\doibase 10.1007/JHEP11(2020)041} {\bibfield  {journal} {\bibinfo  {journal} {JHEP}\ }\textbf {\bibinfo {volume} {11}},\ \bibinfo {pages} {041} (\bibinfo {year} {2020})},\ \Eprint {http://arxiv.org/abs/2005.10251} {arXiv:2005.10251} \BibitemShut {NoStop}%
\bibitem [{\citenamefont {Cabo-Bizet}\ \emph {et~al.}(2020)\citenamefont {Cabo-Bizet}, \citenamefont {Cassani}, \citenamefont {Martelli},\ and\ \citenamefont {Murthy}}]{Cabo-Bizet:2020nkr}%
  \BibitemOpen
  \bibfield  {author} {\bibinfo {author} {\bibfnamefont {Alejandro}\ \bibnamefont {Cabo-Bizet}}, \bibinfo {author} {\bibfnamefont {Davide}\ \bibnamefont {Cassani}}, \bibinfo {author} {\bibfnamefont {Dario}\ \bibnamefont {Martelli}}, \ and\ \bibinfo {author} {\bibfnamefont {Sameer}\ \bibnamefont {Murthy}},\ }\bibfield  {title} {\enquote {\bibinfo {title} {{The large-$N$ limit of the 4d $ \mathcal{N} $ = 1 superconformal index}},}\ }\href {\doibase 10.1007/JHEP11(2020)150} {\bibfield  {journal} {\bibinfo  {journal} {JHEP}\ }\textbf {\bibinfo {volume} {11}},\ \bibinfo {pages} {150} (\bibinfo {year} {2020})},\ \Eprint {http://arxiv.org/abs/2005.10654} {arXiv:2005.10654} \BibitemShut {NoStop}%
\bibitem [{\citenamefont {Agarwal}\ \emph {et~al.}(2021)\citenamefont {Agarwal}, \citenamefont {Choi}, \citenamefont {Kim}, \citenamefont {Kim},\ and\ \citenamefont {Nahmgoong}}]{Agarwal:2020zwm}%
  \BibitemOpen
  \bibfield  {author} {\bibinfo {author} {\bibfnamefont {Prarit}\ \bibnamefont {Agarwal}}, \bibinfo {author} {\bibfnamefont {Sunjin}\ \bibnamefont {Choi}}, \bibinfo {author} {\bibfnamefont {Joonho}\ \bibnamefont {Kim}}, \bibinfo {author} {\bibfnamefont {Seok}\ \bibnamefont {Kim}}, \ and\ \bibinfo {author} {\bibfnamefont {June}\ \bibnamefont {Nahmgoong}},\ }\bibfield  {title} {\enquote {\bibinfo {title} {{AdS black holes and finite N indices}},}\ }\href {\doibase 10.1103/PhysRevD.103.126006} {\bibfield  {journal} {\bibinfo  {journal} {Phys. Rev. D}\ }\textbf {\bibinfo {volume} {103}},\ \bibinfo {pages} {126006} (\bibinfo {year} {2021})},\ \Eprint {http://arxiv.org/abs/2005.11240} {arXiv:2005.11240} \BibitemShut {NoStop}%
\bibitem [{\citenamefont {Gonz\'alez~Lezcano}\ \emph {et~al.}(2021)\citenamefont {Gonz\'alez~Lezcano}, \citenamefont {Hong}, \citenamefont {Liu},\ and\ \citenamefont {Pando~Zayas}}]{GonzalezLezcano:2020yeb}%
  \BibitemOpen
  \bibfield  {author} {\bibinfo {author} {\bibfnamefont {Alfredo}\ \bibnamefont {Gonz\'alez~Lezcano}}, \bibinfo {author} {\bibfnamefont {Junho}\ \bibnamefont {Hong}}, \bibinfo {author} {\bibfnamefont {James~T.}\ \bibnamefont {Liu}}, \ and\ \bibinfo {author} {\bibfnamefont {Leopoldo~A.}\ \bibnamefont {Pando~Zayas}},\ }\bibfield  {title} {\enquote {\bibinfo {title} {{Sub-leading Structures in Superconformal Indices: Subdominant Saddles and Logarithmic Contributions}},}\ }\href {\doibase 10.1007/JHEP01(2021)001} {\bibfield  {journal} {\bibinfo  {journal} {JHEP}\ }\textbf {\bibinfo {volume} {01}},\ \bibinfo {pages} {001} (\bibinfo {year} {2021})},\ \Eprint {http://arxiv.org/abs/2007.12604} {arXiv:2007.12604} \BibitemShut {NoStop}%
\bibitem [{\citenamefont {Copetti}\ \emph {et~al.}(2022)\citenamefont {Copetti}, \citenamefont {Grassi}, \citenamefont {Komargodski},\ and\ \citenamefont {Tizzano}}]{Copetti:2020dil}%
  \BibitemOpen
  \bibfield  {author} {\bibinfo {author} {\bibfnamefont {Christian}\ \bibnamefont {Copetti}}, \bibinfo {author} {\bibfnamefont {Alba}\ \bibnamefont {Grassi}}, \bibinfo {author} {\bibfnamefont {Zohar}\ \bibnamefont {Komargodski}}, \ and\ \bibinfo {author} {\bibfnamefont {Luigi}\ \bibnamefont {Tizzano}},\ }\bibfield  {title} {\enquote {\bibinfo {title} {{Delayed deconfinement and the Hawking-Page transition}},}\ }\href {\doibase 10.1007/JHEP04(2022)132} {\bibfield  {journal} {\bibinfo  {journal} {JHEP}\ }\textbf {\bibinfo {volume} {04}},\ \bibinfo {pages} {132} (\bibinfo {year} {2022})},\ \Eprint {http://arxiv.org/abs/2008.04950} {arXiv:2008.04950 [hep-th]} \BibitemShut {NoStop}%
\bibitem [{\citenamefont {Goldstein}\ \emph {et~al.}(2021)\citenamefont {Goldstein}, \citenamefont {Jejjala}, \citenamefont {Lei}, \citenamefont {van Leuven},\ and\ \citenamefont {Li}}]{Goldstein:2020yvj}%
  \BibitemOpen
  \bibfield  {author} {\bibinfo {author} {\bibfnamefont {Kevin}\ \bibnamefont {Goldstein}}, \bibinfo {author} {\bibfnamefont {Vishnu}\ \bibnamefont {Jejjala}}, \bibinfo {author} {\bibfnamefont {Yang}\ \bibnamefont {Lei}}, \bibinfo {author} {\bibfnamefont {Sam}\ \bibnamefont {van Leuven}}, \ and\ \bibinfo {author} {\bibfnamefont {Wei}\ \bibnamefont {Li}},\ }\bibfield  {title} {\enquote {\bibinfo {title} {{Residues, modularity, and the Cardy limit of the 4d $ \mathcal{N} $ = 4 superconformal index}},}\ }\href {\doibase 10.1007/JHEP04(2021)216} {\bibfield  {journal} {\bibinfo  {journal} {JHEP}\ }\textbf {\bibinfo {volume} {04}},\ \bibinfo {pages} {216} (\bibinfo {year} {2021})},\ \Eprint {http://arxiv.org/abs/2011.06605} {arXiv:2011.06605} \BibitemShut {NoStop}%
\bibitem [{\citenamefont {Amariti}\ \emph {et~al.}(2021{\natexlab{a}})\citenamefont {Amariti}, \citenamefont {Fazzi},\ and\ \citenamefont {Segati}}]{Amariti:2020jyx}%
  \BibitemOpen
  \bibfield  {author} {\bibinfo {author} {\bibfnamefont {Antonio}\ \bibnamefont {Amariti}}, \bibinfo {author} {\bibfnamefont {Marco}\ \bibnamefont {Fazzi}}, \ and\ \bibinfo {author} {\bibfnamefont {Alessia}\ \bibnamefont {Segati}},\ }\bibfield  {title} {\enquote {\bibinfo {title} {{The SCI of $ \mathcal{N} $ = 4 USp(2N$_{c}$) and SO(N$_{c}$) SYM as a matrix integral}},}\ }\href {\doibase 10.1007/JHEP06(2021)132} {\bibfield  {journal} {\bibinfo  {journal} {JHEP}\ }\textbf {\bibinfo {volume} {06}},\ \bibinfo {pages} {132} (\bibinfo {year} {2021}{\natexlab{a}})},\ \Eprint {http://arxiv.org/abs/2012.15208} {arXiv:2012.15208 [hep-th]} \BibitemShut {NoStop}%
\bibitem [{\citenamefont {Amariti}\ \emph {et~al.}(2021{\natexlab{b}})\citenamefont {Amariti}, \citenamefont {Fazzi},\ and\ \citenamefont {Segati}}]{Amariti:2021ubd}%
  \BibitemOpen
  \bibfield  {author} {\bibinfo {author} {\bibfnamefont {Antonio}\ \bibnamefont {Amariti}}, \bibinfo {author} {\bibfnamefont {Marco}\ \bibnamefont {Fazzi}}, \ and\ \bibinfo {author} {\bibfnamefont {Alessia}\ \bibnamefont {Segati}},\ }\bibfield  {title} {\enquote {\bibinfo {title} {{Expanding on the Cardy-like limit of the SCI of 4d $ \mathcal{N} $ = 1 ABCD SCFTs}},}\ }\href {\doibase 10.1007/JHEP07(2021)141} {\bibfield  {journal} {\bibinfo  {journal} {JHEP}\ }\textbf {\bibinfo {volume} {07}},\ \bibinfo {pages} {141} (\bibinfo {year} {2021}{\natexlab{b}})},\ \Eprint {http://arxiv.org/abs/2103.15853} {arXiv:2103.15853 [hep-th]} \BibitemShut {NoStop}%
\bibitem [{\citenamefont {Cassani}\ and\ \citenamefont {Komargodski}(2021)}]{Cassani:2021fyv}%
  \BibitemOpen
  \bibfield  {author} {\bibinfo {author} {\bibfnamefont {Davide}\ \bibnamefont {Cassani}}\ and\ \bibinfo {author} {\bibfnamefont {Zohar}\ \bibnamefont {Komargodski}},\ }\bibfield  {title} {\enquote {\bibinfo {title} {{EFT and the SUSY Index on the 2nd Sheet}},}\ }\href {\doibase 10.21468/SciPostPhys.11.1.004} {\bibfield  {journal} {\bibinfo  {journal} {SciPost Phys.}\ }\textbf {\bibinfo {volume} {11}},\ \bibinfo {pages} {004} (\bibinfo {year} {2021})},\ \Eprint {http://arxiv.org/abs/2104.01464} {arXiv:2104.01464} \BibitemShut {NoStop}%
\bibitem [{\citenamefont {Aharony}\ \emph {et~al.}(2021)\citenamefont {Aharony}, \citenamefont {Benini}, \citenamefont {Mamroud},\ and\ \citenamefont {Milan}}]{Aharony:2021zkr}%
  \BibitemOpen
  \bibfield  {author} {\bibinfo {author} {\bibfnamefont {Ofer}\ \bibnamefont {Aharony}}, \bibinfo {author} {\bibfnamefont {Francesco}\ \bibnamefont {Benini}}, \bibinfo {author} {\bibfnamefont {Ohad}\ \bibnamefont {Mamroud}}, \ and\ \bibinfo {author} {\bibfnamefont {Paolo}\ \bibnamefont {Milan}},\ }\bibfield  {title} {\enquote {\bibinfo {title} {{A gravity interpretation for the Bethe Ansatz expansion of the $\mathcal{N}=4$ SYM index}},}\ }\href {\doibase 10.1103/PhysRevD.104.086026} {\bibfield  {journal} {\bibinfo  {journal} {Phys. Rev. D}\ }\textbf {\bibinfo {volume} {104}},\ \bibinfo {pages} {086026} (\bibinfo {year} {2021})},\ \Eprint {http://arxiv.org/abs/2104.13932} {arXiv:2104.13932} \BibitemShut {NoStop}%
\bibitem [{\citenamefont {Uhlemann}(2021)}]{Uhlemann:2021nhu}%
  \BibitemOpen
  \bibfield  {author} {\bibinfo {author} {\bibfnamefont {Christoph~F.}\ \bibnamefont {Uhlemann}},\ }\bibfield  {title} {\enquote {\bibinfo {title} {{Islands and Page curves in 4d from Type IIB}},}\ }\href {\doibase 10.1007/JHEP08(2021)104} {\bibfield  {journal} {\bibinfo  {journal} {JHEP}\ }\textbf {\bibinfo {volume} {08}},\ \bibinfo {pages} {104} (\bibinfo {year} {2021})},\ \Eprint {http://arxiv.org/abs/2105.00008} {arXiv:2105.00008 [hep-th]} \BibitemShut {NoStop}%
\bibitem [{\citenamefont {Hong}(2021)}]{Hong:2021bzg}%
  \BibitemOpen
  \bibfield  {author} {\bibinfo {author} {\bibfnamefont {Junho}\ \bibnamefont {Hong}},\ }\bibfield  {title} {\enquote {\bibinfo {title} {{The topologically twisted index of $ \mathcal{N} $ = 4 SU(N) Super-Yang-Mills theory and a black hole Farey tail}},}\ }\href {\doibase 10.1007/JHEP10(2021)145} {\bibfield  {journal} {\bibinfo  {journal} {JHEP}\ }\textbf {\bibinfo {volume} {10}},\ \bibinfo {pages} {145} (\bibinfo {year} {2021})},\ \Eprint {http://arxiv.org/abs/2108.02355} {arXiv:2108.02355} \BibitemShut {NoStop}%
\bibitem [{\citenamefont {Ezroura}\ \emph {et~al.}(2022)\citenamefont {Ezroura}, \citenamefont {Larsen}, \citenamefont {Liu},\ and\ \citenamefont {Zeng}}]{Ezroura:2021vrt}%
  \BibitemOpen
  \bibfield  {author} {\bibinfo {author} {\bibfnamefont {Nizar}\ \bibnamefont {Ezroura}}, \bibinfo {author} {\bibfnamefont {Finn}\ \bibnamefont {Larsen}}, \bibinfo {author} {\bibfnamefont {Zhihan}\ \bibnamefont {Liu}}, \ and\ \bibinfo {author} {\bibfnamefont {Yangwenxiao}\ \bibnamefont {Zeng}},\ }\bibfield  {title} {\enquote {\bibinfo {title} {{The phase diagram of BPS black holes in AdS$_{5}$}},}\ }\href {\doibase 10.1007/JHEP09(2022)033} {\bibfield  {journal} {\bibinfo  {journal} {JHEP}\ }\textbf {\bibinfo {volume} {09}},\ \bibinfo {pages} {033} (\bibinfo {year} {2022})},\ \Eprint {http://arxiv.org/abs/2108.11542} {arXiv:2108.11542 [hep-th]} \BibitemShut {NoStop}%
\bibitem [{\citenamefont {Imamura}(2021)}]{Imamura:2021ytr}%
  \BibitemOpen
  \bibfield  {author} {\bibinfo {author} {\bibfnamefont {Yosuke}\ \bibnamefont {Imamura}},\ }\bibfield  {title} {\enquote {\bibinfo {title} {{Finite-N superconformal index via the AdS/CFT correspondence}},}\ }\href {\doibase 10.1093/ptep/ptab141} {\bibfield  {journal} {\bibinfo  {journal} {PTEP}\ }\textbf {\bibinfo {volume} {2021}},\ \bibinfo {pages} {123B05} (\bibinfo {year} {2021})},\ \Eprint {http://arxiv.org/abs/2108.12090} {arXiv:2108.12090 [hep-th]} \BibitemShut {NoStop}%
\bibitem [{\citenamefont {Gaiotto}\ and\ \citenamefont {Lee}(2021)}]{Gaiotto:2021xce}%
  \BibitemOpen
  \bibfield  {author} {\bibinfo {author} {\bibfnamefont {Davide}\ \bibnamefont {Gaiotto}}\ and\ \bibinfo {author} {\bibfnamefont {Ji~Hoon}\ \bibnamefont {Lee}},\ }\bibfield  {title} {\enquote {\bibinfo {title} {{The Giant Graviton Expansion}},}\ }\href@noop {} {\  (\bibinfo {year} {2021})},\ \Eprint {http://arxiv.org/abs/2109.02545} {arXiv:2109.02545 [hep-th]} \BibitemShut {NoStop}%
\bibitem [{\citenamefont {Cabo-Bizet}(2022)}]{Cabo-Bizet:2021jar}%
  \BibitemOpen
  \bibfield  {author} {\bibinfo {author} {\bibfnamefont {Alejandro}\ \bibnamefont {Cabo-Bizet}},\ }\bibfield  {title} {\enquote {\bibinfo {title} {{Quantum phases of 4d SU(N) $ \mathcal{N} $ = 4 SYM}},}\ }\href {\doibase 10.1007/JHEP10(2022)052} {\bibfield  {journal} {\bibinfo  {journal} {JHEP}\ }\textbf {\bibinfo {volume} {10}},\ \bibinfo {pages} {052} (\bibinfo {year} {2022})},\ \Eprint {http://arxiv.org/abs/2111.14942} {arXiv:2111.14942 [hep-th]} \BibitemShut {NoStop}%
\bibitem [{\citenamefont {Murthy}(2023)}]{Murthy:2022ien}%
  \BibitemOpen
  \bibfield  {author} {\bibinfo {author} {\bibfnamefont {Sameer}\ \bibnamefont {Murthy}},\ }\bibfield  {title} {\enquote {\bibinfo {title} {{Unitary matrix models, free fermions, and the giant graviton expansion}},}\ }\href {\doibase 10.4310/PAMQ.2023.v19.n1.a12} {\bibfield  {journal} {\bibinfo  {journal} {Pure Appl. Math. Quart.}\ }\textbf {\bibinfo {volume} {19}},\ \bibinfo {pages} {299--340} (\bibinfo {year} {2023})},\ \Eprint {http://arxiv.org/abs/2202.06897} {arXiv:2202.06897 [hep-th]} \BibitemShut {NoStop}%
\bibitem [{\citenamefont {Boruch}\ \emph {et~al.}(2022)\citenamefont {Boruch}, \citenamefont {Heydeman}, \citenamefont {Iliesiu},\ and\ \citenamefont {Turiaci}}]{Boruch:2022tno}%
  \BibitemOpen
  \bibfield  {author} {\bibinfo {author} {\bibfnamefont {Jan}\ \bibnamefont {Boruch}}, \bibinfo {author} {\bibfnamefont {Matthew~T.}\ \bibnamefont {Heydeman}}, \bibinfo {author} {\bibfnamefont {Luca~V.}\ \bibnamefont {Iliesiu}}, \ and\ \bibinfo {author} {\bibfnamefont {Gustavo~J.}\ \bibnamefont {Turiaci}},\ }\bibfield  {title} {\enquote {\bibinfo {title} {{BPS and near-BPS black holes in $AdS_5$ and their spectrum in $\mathcal{N}=4$ SYM}},}\ }\href@noop {} {\  (\bibinfo {year} {2022})},\ \Eprint {http://arxiv.org/abs/2203.01331} {arXiv:2203.01331} \BibitemShut {NoStop}%
\bibitem [{\citenamefont {Honda}\ and\ \citenamefont {Yoda}(2022)}]{Honda:2022hvy}%
  \BibitemOpen
  \bibfield  {author} {\bibinfo {author} {\bibfnamefont {Masazumi}\ \bibnamefont {Honda}}\ and\ \bibinfo {author} {\bibfnamefont {Takuya}\ \bibnamefont {Yoda}},\ }\bibfield  {title} {\enquote {\bibinfo {title} {{String theory, $\mathcal{N}=4$ SYM and Riemann hypothesis}},}\ }\href@noop {} {\  (\bibinfo {year} {2022})},\ \Eprint {http://arxiv.org/abs/2203.17091} {arXiv:2203.17091} \BibitemShut {NoStop}%
\bibitem [{\citenamefont {Lee}(2022)}]{Lee:2022vig}%
  \BibitemOpen
  \bibfield  {author} {\bibinfo {author} {\bibfnamefont {Ji~Hoon}\ \bibnamefont {Lee}},\ }\bibfield  {title} {\enquote {\bibinfo {title} {{Exact stringy microstates from gauge theories}},}\ }\href {\doibase 10.1007/JHEP11(2022)137} {\bibfield  {journal} {\bibinfo  {journal} {JHEP}\ }\textbf {\bibinfo {volume} {11}},\ \bibinfo {pages} {137} (\bibinfo {year} {2022})},\ \Eprint {http://arxiv.org/abs/2204.09286} {arXiv:2204.09286 [hep-th]} \BibitemShut {NoStop}%
\bibitem [{\citenamefont {Huang}(2022)}]{Huang:2022bry}%
  \BibitemOpen
  \bibfield  {author} {\bibinfo {author} {\bibfnamefont {Min-xin}\ \bibnamefont {Huang}},\ }\bibfield  {title} {\enquote {\bibinfo {title} {{Modular anomaly equation for Schur index of $ \mathcal{N} $ = 4 super-Yang-Mills}},}\ }\href {\doibase 10.1007/JHEP08(2022)049} {\bibfield  {journal} {\bibinfo  {journal} {JHEP}\ }\textbf {\bibinfo {volume} {08}},\ \bibinfo {pages} {049} (\bibinfo {year} {2022})},\ \Eprint {http://arxiv.org/abs/2205.00818} {arXiv:2205.00818 [hep-th]} \BibitemShut {NoStop}%
\bibitem [{\citenamefont {Imamura}(2022)}]{Imamura:2022aua}%
  \BibitemOpen
  \bibfield  {author} {\bibinfo {author} {\bibfnamefont {Yosuke}\ \bibnamefont {Imamura}},\ }\bibfield  {title} {\enquote {\bibinfo {title} {{Analytic continuation for giant gravitons}},}\ }\href {\doibase 10.1093/ptep/ptac127} {\bibfield  {journal} {\bibinfo  {journal} {PTEP}\ }\textbf {\bibinfo {volume} {2022}},\ \bibinfo {pages} {103B02} (\bibinfo {year} {2022})},\ \Eprint {http://arxiv.org/abs/2205.14615} {arXiv:2205.14615 [hep-th]} \BibitemShut {NoStop}%
\bibitem [{\citenamefont {Holguin}\ and\ \citenamefont {Wang}(2022)}]{Holguin:2022drf}%
  \BibitemOpen
  \bibfield  {author} {\bibinfo {author} {\bibfnamefont {Adolfo}\ \bibnamefont {Holguin}}\ and\ \bibinfo {author} {\bibfnamefont {Shannon}\ \bibnamefont {Wang}},\ }\bibfield  {title} {\enquote {\bibinfo {title} {{Giant gravitons, Harish-Chandra integrals, and BPS states in symplectic and orthogonal $ \mathcal{N} $ = 4 SYM}},}\ }\href {\doibase 10.1007/JHEP10(2022)078} {\bibfield  {journal} {\bibinfo  {journal} {JHEP}\ }\textbf {\bibinfo {volume} {10}},\ \bibinfo {pages} {078} (\bibinfo {year} {2022})},\ \Eprint {http://arxiv.org/abs/2206.00020} {arXiv:2206.00020 [hep-th]} \BibitemShut {NoStop}%
\bibitem [{\citenamefont {Eleftheriou}(2023)}]{Eleftheriou:2022kkv}%
  \BibitemOpen
  \bibfield  {author} {\bibinfo {author} {\bibfnamefont {Giorgos}\ \bibnamefont {Eleftheriou}},\ }\bibfield  {title} {\enquote {\bibinfo {title} {{Root of unity asymptotics for Schur indices of 4d Lagrangian theories}},}\ }\href {\doibase 10.1007/JHEP01(2023)081} {\bibfield  {journal} {\bibinfo  {journal} {JHEP}\ }\textbf {\bibinfo {volume} {01}},\ \bibinfo {pages} {081} (\bibinfo {year} {2023})},\ \Eprint {http://arxiv.org/abs/2207.14271} {arXiv:2207.14271 [hep-th]} \BibitemShut {NoStop}%
\bibitem [{\citenamefont {Choi}\ \emph {et~al.}(2023)\citenamefont {Choi}, \citenamefont {Kim}, \citenamefont {Lee},\ and\ \citenamefont {Lee}}]{Choi:2022ovw}%
  \BibitemOpen
  \bibfield  {author} {\bibinfo {author} {\bibfnamefont {Sunjin}\ \bibnamefont {Choi}}, \bibinfo {author} {\bibfnamefont {Seok}\ \bibnamefont {Kim}}, \bibinfo {author} {\bibfnamefont {Eunwoo}\ \bibnamefont {Lee}}, \ and\ \bibinfo {author} {\bibfnamefont {Jehyun}\ \bibnamefont {Lee}},\ }\bibfield  {title} {\enquote {\bibinfo {title} {{From giant gravitons to black holes}},}\ }\href {\doibase 10.1007/JHEP11(2023)086} {\bibfield  {journal} {\bibinfo  {journal} {JHEP}\ }\textbf {\bibinfo {volume} {11}},\ \bibinfo {pages} {086} (\bibinfo {year} {2023})},\ \Eprint {http://arxiv.org/abs/2207.05172} {arXiv:2207.05172 [hep-th]} \BibitemShut {NoStop}%
\bibitem [{\citenamefont {Chang}\ and\ \citenamefont {Lin}(2023)}]{Chang:2022mjp}%
  \BibitemOpen
  \bibfield  {author} {\bibinfo {author} {\bibfnamefont {Chi-Ming}\ \bibnamefont {Chang}}\ and\ \bibinfo {author} {\bibfnamefont {Ying-Hsuan}\ \bibnamefont {Lin}},\ }\bibfield  {title} {\enquote {\bibinfo {title} {{Words to describe a black hole}},}\ }\href {\doibase 10.1007/JHEP02(2023)109} {\bibfield  {journal} {\bibinfo  {journal} {JHEP}\ }\textbf {\bibinfo {volume} {02}},\ \bibinfo {pages} {109} (\bibinfo {year} {2023})},\ \Eprint {http://arxiv.org/abs/2209.06728} {arXiv:2209.06728 [hep-th]} \BibitemShut {NoStop}%
\bibitem [{\citenamefont {Chen}\ \emph {et~al.}(2023{\natexlab{a}})\citenamefont {Chen}, \citenamefont {Liu}, \citenamefont {Tian}, \citenamefont {Wu},\ and\ \citenamefont {Zhang}}]{Chen:2022tfy}%
  \BibitemOpen
  \bibfield  {author} {\bibinfo {author} {\bibfnamefont {Qian}\ \bibnamefont {Chen}}, \bibinfo {author} {\bibfnamefont {Yuxuan}\ \bibnamefont {Liu}}, \bibinfo {author} {\bibfnamefont {Yu}~\bibnamefont {Tian}}, \bibinfo {author} {\bibfnamefont {Xiaoning}\ \bibnamefont {Wu}}, \ and\ \bibinfo {author} {\bibfnamefont {Hongbao}\ \bibnamefont {Zhang}},\ }\bibfield  {title} {\enquote {\bibinfo {title} {{Quench dynamics in holographic first-order phase transition}},}\ }\href {\doibase 10.1103/PhysRevD.108.106017} {\bibfield  {journal} {\bibinfo  {journal} {Phys. Rev. D}\ }\textbf {\bibinfo {volume} {108}},\ \bibinfo {pages} {106017} (\bibinfo {year} {2023}{\natexlab{a}})},\ \Eprint {http://arxiv.org/abs/2211.11291} {arXiv:2211.11291 [hep-th]} \BibitemShut {NoStop}%
\bibitem [{\citenamefont {Liu}\ and\ \citenamefont {Rajappa}(2023)}]{Liu:2022olj}%
  \BibitemOpen
  \bibfield  {author} {\bibinfo {author} {\bibfnamefont {James~T.}\ \bibnamefont {Liu}}\ and\ \bibinfo {author} {\bibfnamefont {Neville~Joshua}\ \bibnamefont {Rajappa}},\ }\bibfield  {title} {\enquote {\bibinfo {title} {{Finite N indices and the giant graviton expansion}},}\ }\href {\doibase 10.1007/JHEP04(2023)078} {\bibfield  {journal} {\bibinfo  {journal} {JHEP}\ }\textbf {\bibinfo {volume} {04}},\ \bibinfo {pages} {078} (\bibinfo {year} {2023})},\ \Eprint {http://arxiv.org/abs/2212.05408} {arXiv:2212.05408 [hep-th]} \BibitemShut {NoStop}%
\bibitem [{\citenamefont {Lin}(2023)}]{Lin:2022gbu}%
  \BibitemOpen
  \bibfield  {author} {\bibinfo {author} {\bibfnamefont {Hai}\ \bibnamefont {Lin}},\ }\bibfield  {title} {\enquote {\bibinfo {title} {{Coherent state operators, giant gravitons, and gauge-gravity correspondence}},}\ }\href {\doibase 10.1016/j.aop.2023.169248} {\bibfield  {journal} {\bibinfo  {journal} {Annals Phys.}\ }\textbf {\bibinfo {volume} {451}},\ \bibinfo {pages} {169248} (\bibinfo {year} {2023})},\ \Eprint {http://arxiv.org/abs/2212.14002} {arXiv:2212.14002 [hep-th]} \BibitemShut {NoStop}%
\bibitem [{\citenamefont {Eniceicu}(2023)}]{Eniceicu:2023uvd}%
  \BibitemOpen
  \bibfield  {author} {\bibinfo {author} {\bibfnamefont {Dan~S.}\ \bibnamefont {Eniceicu}},\ }\bibfield  {title} {\enquote {\bibinfo {title} {{Comments on the Giant-Graviton Expansion of the Superconformal Index}},}\ }\href@noop {} {\  (\bibinfo {year} {2023})},\ \Eprint {http://arxiv.org/abs/2302.04887} {arXiv:2302.04887 [hep-th]} \BibitemShut {NoStop}%
\bibitem [{\citenamefont {Beccaria}\ and\ \citenamefont {Cabo-Bizet}(2023)}]{Beccaria:2023zjw}%
  \BibitemOpen
  \bibfield  {author} {\bibinfo {author} {\bibfnamefont {Matteo}\ \bibnamefont {Beccaria}}\ and\ \bibinfo {author} {\bibfnamefont {Alejandro}\ \bibnamefont {Cabo-Bizet}},\ }\bibfield  {title} {\enquote {\bibinfo {title} {{On the brane expansion of the Schur index}},}\ }\href {\doibase 10.1007/JHEP08(2023)073} {\bibfield  {journal} {\bibinfo  {journal} {JHEP}\ }\textbf {\bibinfo {volume} {08}},\ \bibinfo {pages} {073} (\bibinfo {year} {2023})},\ \Eprint {http://arxiv.org/abs/2305.17730} {arXiv:2305.17730 [hep-th]} \BibitemShut {NoStop}%
\bibitem [{\citenamefont {Chang}\ \emph {et~al.}(2023{\natexlab{a}})\citenamefont {Chang}, \citenamefont {Feng}, \citenamefont {Lin},\ and\ \citenamefont {Tao}}]{Chang:2023zqk}%
  \BibitemOpen
  \bibfield  {author} {\bibinfo {author} {\bibfnamefont {Chi-Ming}\ \bibnamefont {Chang}}, \bibinfo {author} {\bibfnamefont {Li}~\bibnamefont {Feng}}, \bibinfo {author} {\bibfnamefont {Ying-Hsuan}\ \bibnamefont {Lin}}, \ and\ \bibinfo {author} {\bibfnamefont {Yi-Xiao}\ \bibnamefont {Tao}},\ }\bibfield  {title} {\enquote {\bibinfo {title} {{Decoding stringy near-supersymmetric black holes}},}\ }\href@noop {} {\  (\bibinfo {year} {2023}{\natexlab{a}})},\ \Eprint {http://arxiv.org/abs/2306.04673} {arXiv:2306.04673 [hep-th]} \BibitemShut {NoStop}%
\bibitem [{\citenamefont {Chen}\ \emph {et~al.}(2023{\natexlab{b}})\citenamefont {Chen}, \citenamefont {Heydeman}, \citenamefont {Wang},\ and\ \citenamefont {Zhang}}]{Chen:2023lzq}%
  \BibitemOpen
  \bibfield  {author} {\bibinfo {author} {\bibfnamefont {Yiming}\ \bibnamefont {Chen}}, \bibinfo {author} {\bibfnamefont {Matthew}\ \bibnamefont {Heydeman}}, \bibinfo {author} {\bibfnamefont {Yifan}\ \bibnamefont {Wang}}, \ and\ \bibinfo {author} {\bibfnamefont {Mengyang}\ \bibnamefont {Zhang}},\ }\bibfield  {title} {\enquote {\bibinfo {title} {{Probing supersymmetric black holes with surface defects}},}\ }\href {\doibase 10.1007/JHEP10(2023)136} {\bibfield  {journal} {\bibinfo  {journal} {JHEP}\ }\textbf {\bibinfo {volume} {10}},\ \bibinfo {pages} {136} (\bibinfo {year} {2023}{\natexlab{b}})},\ \Eprint {http://arxiv.org/abs/2306.05463} {arXiv:2306.05463 [hep-th]} \BibitemShut {NoStop}%
\bibitem [{\citenamefont {Ekhammar}\ \emph {et~al.}(2023)\citenamefont {Ekhammar}, \citenamefont {Minahan},\ and\ \citenamefont {Thull}}]{Ekhammar:2023glu}%
  \BibitemOpen
  \bibfield  {author} {\bibinfo {author} {\bibfnamefont {Simon}\ \bibnamefont {Ekhammar}}, \bibinfo {author} {\bibfnamefont {Joseph~A.}\ \bibnamefont {Minahan}}, \ and\ \bibinfo {author} {\bibfnamefont {Charles}\ \bibnamefont {Thull}},\ }\bibfield  {title} {\enquote {\bibinfo {title} {{The asymptotic form of the Hagedorn temperature in planar super Yang-Mills}},}\ }\href {\doibase 10.1088/1751-8121/acf9d0} {\bibfield  {journal} {\bibinfo  {journal} {J. Phys. A}\ }\textbf {\bibinfo {volume} {56}},\ \bibinfo {pages} {435401} (\bibinfo {year} {2023})},\ \Eprint {http://arxiv.org/abs/2306.09883} {arXiv:2306.09883 [hep-th]} \BibitemShut {NoStop}%
\bibitem [{\citenamefont {Bigazzi}\ \emph {et~al.}(2023)\citenamefont {Bigazzi}, \citenamefont {Canneti},\ and\ \citenamefont {Cotrone}}]{Bigazzi:2023hxt}%
  \BibitemOpen
  \bibfield  {author} {\bibinfo {author} {\bibfnamefont {Francesco}\ \bibnamefont {Bigazzi}}, \bibinfo {author} {\bibfnamefont {Tommaso}\ \bibnamefont {Canneti}}, \ and\ \bibinfo {author} {\bibfnamefont {Aldo~L.}\ \bibnamefont {Cotrone}},\ }\bibfield  {title} {\enquote {\bibinfo {title} {{Higher order corrections to the Hagedorn temperature at strong coupling}},}\ }\href {\doibase 10.1007/JHEP10(2023)056} {\bibfield  {journal} {\bibinfo  {journal} {JHEP}\ }\textbf {\bibinfo {volume} {10}},\ \bibinfo {pages} {056} (\bibinfo {year} {2023})},\ \Eprint {http://arxiv.org/abs/2306.17126} {arXiv:2306.17126 [hep-th]} \BibitemShut {NoStop}%
\bibitem [{\citenamefont {Beccaria}\ and\ \citenamefont {Cabo-Bizet}(2024)}]{Beccaria:2023hip}%
  \BibitemOpen
  \bibfield  {author} {\bibinfo {author} {\bibfnamefont {Matteo}\ \bibnamefont {Beccaria}}\ and\ \bibinfo {author} {\bibfnamefont {Alejandro}\ \bibnamefont {Cabo-Bizet}},\ }\bibfield  {title} {\enquote {\bibinfo {title} {{Large black hole entropy from the giant brane expansion}},}\ }\href {\doibase 10.1007/JHEP04(2024)146} {\bibfield  {journal} {\bibinfo  {journal} {JHEP}\ }\textbf {\bibinfo {volume} {04}},\ \bibinfo {pages} {146} (\bibinfo {year} {2024})},\ \Eprint {http://arxiv.org/abs/2308.05191} {arXiv:2308.05191 [hep-th]} \BibitemShut {NoStop}%
\bibitem [{\citenamefont {Arabi~Ardehali}\ \emph {et~al.}(2023)\citenamefont {Arabi~Ardehali}, \citenamefont {Martone},\ and\ \citenamefont {Rossell\'o}}]{ArabiArdehali:2023bpq}%
  \BibitemOpen
  \bibfield  {author} {\bibinfo {author} {\bibfnamefont {Arash}\ \bibnamefont {Arabi~Ardehali}}, \bibinfo {author} {\bibfnamefont {Mario}\ \bibnamefont {Martone}}, \ and\ \bibinfo {author} {\bibfnamefont {Mart\'\i{}}\ \bibnamefont {Rossell\'o}},\ }\bibfield  {title} {\enquote {\bibinfo {title} {{High-temperature expansion of the Schur index and modularity}},}\ }\href@noop {} {\  (\bibinfo {year} {2023})},\ \Eprint {http://arxiv.org/abs/2308.09738} {arXiv:2308.09738 [hep-th]} \BibitemShut {NoStop}%
\bibitem [{\citenamefont {Chang}\ \emph {et~al.}(2023{\natexlab{b}})\citenamefont {Chang}, \citenamefont {Lin},\ and\ \citenamefont {Wu}}]{Chang:2023ywj}%
  \BibitemOpen
  \bibfield  {author} {\bibinfo {author} {\bibfnamefont {Chi-Ming}\ \bibnamefont {Chang}}, \bibinfo {author} {\bibfnamefont {Ying-Hsuan}\ \bibnamefont {Lin}}, \ and\ \bibinfo {author} {\bibfnamefont {Jingxiang}\ \bibnamefont {Wu}},\ }\bibfield  {title} {\enquote {\bibinfo {title} {{On $\frac18$-BPS black holes and the chiral algebra of $\mathcal{N}=4$ SYM}},}\ }\href@noop {} {\  (\bibinfo {year} {2023}{\natexlab{b}})},\ \Eprint {http://arxiv.org/abs/2310.20086} {arXiv:2310.20086 [hep-th]} \BibitemShut {NoStop}%
\bibitem [{\citenamefont {Lee}(2024)}]{Lee:2023iil}%
  \BibitemOpen
  \bibfield  {author} {\bibinfo {author} {\bibfnamefont {Ji~Hoon}\ \bibnamefont {Lee}},\ }\bibfield  {title} {\enquote {\bibinfo {title} {{Trace relations and open string vacua}},}\ }\href {\doibase 10.1007/JHEP02(2024)224} {\bibfield  {journal} {\bibinfo  {journal} {JHEP}\ }\textbf {\bibinfo {volume} {02}},\ \bibinfo {pages} {224} (\bibinfo {year} {2024})},\ \Eprint {http://arxiv.org/abs/2312.00242} {arXiv:2312.00242 [hep-th]} \BibitemShut {NoStop}%
\bibitem [{\citenamefont {Eleftheriou}\ \emph {et~al.}(2023)\citenamefont {Eleftheriou}, \citenamefont {Murthy},\ and\ \citenamefont {Rossell\'o}}]{Eleftheriou:2023jxr}%
  \BibitemOpen
  \bibfield  {author} {\bibinfo {author} {\bibfnamefont {Giorgos}\ \bibnamefont {Eleftheriou}}, \bibinfo {author} {\bibfnamefont {Sameer}\ \bibnamefont {Murthy}}, \ and\ \bibinfo {author} {\bibfnamefont {Mart\'\i{}}\ \bibnamefont {Rossell\'o}},\ }\bibfield  {title} {\enquote {\bibinfo {title} {{The giant graviton expansion in $\mathbf{\text{AdS}_5 \times S^5}$}},}\ }\href@noop {} {\  (\bibinfo {year} {2023})},\ \Eprint {http://arxiv.org/abs/2312.14921} {arXiv:2312.14921 [hep-th]} \BibitemShut {NoStop}%
\bibitem [{\citenamefont {Polchinski}\ and\ \citenamefont {Strassler}(2002)}]{Polchinski:2001tt}%
  \BibitemOpen
  \bibfield  {author} {\bibinfo {author} {\bibfnamefont {Joseph}\ \bibnamefont {Polchinski}}\ and\ \bibinfo {author} {\bibfnamefont {Matthew~J.}\ \bibnamefont {Strassler}},\ }\bibfield  {title} {\enquote {\bibinfo {title} {{Hard scattering and gauge / string duality}},}\ }\href {\doibase 10.1103/PhysRevLett.88.031601} {\bibfield  {journal} {\bibinfo  {journal} {Phys. Rev. Lett.}\ }\textbf {\bibinfo {volume} {88}},\ \bibinfo {pages} {031601} (\bibinfo {year} {2002})},\ \Eprint {http://arxiv.org/abs/hep-th/0109174} {arXiv:hep-th/0109174} \BibitemShut {NoStop}%
\bibitem [{\citenamefont {Bern}\ \emph {et~al.}(2004)\citenamefont {Bern}, \citenamefont {Dixon},\ and\ \citenamefont {Kosower}}]{Bern:2004kq}%
  \BibitemOpen
  \bibfield  {author} {\bibinfo {author} {\bibfnamefont {Z.}~\bibnamefont {Bern}}, \bibinfo {author} {\bibfnamefont {Lance~J.}\ \bibnamefont {Dixon}}, \ and\ \bibinfo {author} {\bibfnamefont {D.~A.}\ \bibnamefont {Kosower}},\ }\bibfield  {title} {\enquote {\bibinfo {title} {{N=4 super-Yang-Mills theory, QCD and collider physics}},}\ }\href {\doibase 10.1016/j.crhy.2004.09.007} {\bibfield  {journal} {\bibinfo  {journal} {Comptes Rendus Physique}\ }\textbf {\bibinfo {volume} {5}},\ \bibinfo {pages} {955--964} (\bibinfo {year} {2004})},\ \Eprint {http://arxiv.org/abs/hep-th/0410021} {arXiv:hep-th/0410021} \BibitemShut {NoStop}%
\bibitem [{\citenamefont {Caron-Huot}\ and\ \citenamefont {Moore}(2008)}]{Caron-Huot:2008dyw}%
  \BibitemOpen
  \bibfield  {author} {\bibinfo {author} {\bibfnamefont {Simon}\ \bibnamefont {Caron-Huot}}\ and\ \bibinfo {author} {\bibfnamefont {Guy~D.}\ \bibnamefont {Moore}},\ }\bibfield  {title} {\enquote {\bibinfo {title} {{Heavy quark diffusion in QCD and N=4 SYM at next-to-leading order}},}\ }\href {\doibase 10.1088/1126-6708/2008/02/081} {\bibfield  {journal} {\bibinfo  {journal} {JHEP}\ }\textbf {\bibinfo {volume} {02}},\ \bibinfo {pages} {081} (\bibinfo {year} {2008})},\ \Eprint {http://arxiv.org/abs/0801.2173} {arXiv:0801.2173 [hep-ph]} \BibitemShut {NoStop}%
\bibitem [{\citenamefont {Witten}(1998{\natexlab{b}})}]{Witten:1998zw}%
  \BibitemOpen
  \bibfield  {author} {\bibinfo {author} {\bibfnamefont {Edward}\ \bibnamefont {Witten}},\ }\bibfield  {title} {\enquote {\bibinfo {title} {{Anti-de Sitter space, thermal phase transition, and confinement in gauge theories}},}\ }\href {\doibase 10.4310/ATMP.1998.v2.n3.a3} {\bibfield  {journal} {\bibinfo  {journal} {Adv. Theor. Math. Phys.}\ }\textbf {\bibinfo {volume} {2}},\ \bibinfo {pages} {505--532} (\bibinfo {year} {1998}{\natexlab{b}})},\ \Eprint {http://arxiv.org/abs/hep-th/9803131} {arXiv:hep-th/9803131} \BibitemShut {NoStop}%
\bibitem [{\citenamefont {Gross}\ and\ \citenamefont {Witten}(1980)}]{Gross:1980he}%
  \BibitemOpen
  \bibfield  {author} {\bibinfo {author} {\bibfnamefont {David~J.}\ \bibnamefont {Gross}}\ and\ \bibinfo {author} {\bibfnamefont {Edward}\ \bibnamefont {Witten}},\ }\bibfield  {title} {\enquote {\bibinfo {title} {{Possible Third Order Phase Transition in the Large N Lattice Gauge Theory}},}\ }\href {\doibase 10.1103/PhysRevD.21.446} {\bibfield  {journal} {\bibinfo  {journal} {Phys. Rev. D}\ }\textbf {\bibinfo {volume} {21}},\ \bibinfo {pages} {446--453} (\bibinfo {year} {1980})}\BibitemShut {NoStop}%
\bibitem [{\citenamefont {Wadia}(1980)}]{Wadia:1980cp}%
  \BibitemOpen
  \bibfield  {author} {\bibinfo {author} {\bibfnamefont {Spenta~R.}\ \bibnamefont {Wadia}},\ }\bibfield  {title} {\enquote {\bibinfo {title} {{$N$ = Infinity Phase Transition in a Class of Exactly Soluble Model Lattice Gauge Theories}},}\ }\href {\doibase 10.1016/0370-2693(80)90353-6} {\bibfield  {journal} {\bibinfo  {journal} {Phys. Lett. B}\ }\textbf {\bibinfo {volume} {93}},\ \bibinfo {pages} {403--410} (\bibinfo {year} {1980})}\BibitemShut {NoStop}%
\bibitem [{\citenamefont {Wadia}(2012)}]{Wadia:2012fr}%
  \BibitemOpen
  \bibfield  {author} {\bibinfo {author} {\bibfnamefont {Spenta~R.}\ \bibnamefont {Wadia}},\ }\href@noop {} {\enquote {\bibinfo {title} {{A Study of U(N) Lattice Gauge Theory in 2-dimensions}},}\ } (\bibinfo {year} {2012}),\ \Eprint {http://arxiv.org/abs/1212.2906} {arXiv:1212.2906 [hep-th]} \BibitemShut {NoStop}%
\bibitem [{\citenamefont {Gorin}\ \emph {et~al.}(2006)\citenamefont {Gorin}, \citenamefont {Prosen}, \citenamefont {Seligman},\ and\ \citenamefont {{\v{Z}}nidari{\v{c}}}}]{Gorin:2006}%
  \BibitemOpen
  \bibfield  {author} {\bibinfo {author} {\bibfnamefont {Thomas}\ \bibnamefont {Gorin}}, \bibinfo {author} {\bibfnamefont {Toma{\v{z}}}\ \bibnamefont {Prosen}}, \bibinfo {author} {\bibfnamefont {Thomas~H.}\ \bibnamefont {Seligman}}, \ and\ \bibinfo {author} {\bibfnamefont {Marko}\ \bibnamefont {{\v{Z}}nidari{\v{c}}}},\ }\bibfield  {title} {\enquote {\bibinfo {title} {{Dynamics of Loschmidt echoes and fidelity decay}},}\ }\href {\doibase 10.1016/j.physrep.2006.09.003} {\bibfield  {journal} {\bibinfo  {journal} {Phys. Rep.}\ }\textbf {\bibinfo {volume} {435}},\ \bibinfo {pages} {33--156} (\bibinfo {year} {2006})},\ \Eprint {http://arxiv.org/abs/quant-ph/0607050} {arXiv:quant-ph/0607050 [quant-ph]} \BibitemShut {NoStop}%
\bibitem [{\citenamefont {del Campo}\ \emph {et~al.}(2018)\citenamefont {del Campo}, \citenamefont {Molina-Vilaplana}, \citenamefont {Santos},\ and\ \citenamefont {Sonner}}]{delCampo:2017ftn}%
  \BibitemOpen
  \bibfield  {author} {\bibinfo {author} {\bibfnamefont {A.}~\bibnamefont {del Campo}}, \bibinfo {author} {\bibfnamefont {J.}~\bibnamefont {Molina-Vilaplana}}, \bibinfo {author} {\bibfnamefont {L.~F.}\ \bibnamefont {Santos}}, \ and\ \bibinfo {author} {\bibfnamefont {J.}~\bibnamefont {Sonner}},\ }\bibfield  {title} {\enquote {\bibinfo {title} {{Decay of a Thermofield-Double State in Chaotic Quantum Systems}},}\ }\href {\doibase 10.1140/epjst/e2018-00083-5} {\bibfield  {journal} {\bibinfo  {journal} {Eur. Phys. J. ST}\ }\textbf {\bibinfo {volume} {227}},\ \bibinfo {pages} {247--258} (\bibinfo {year} {2018})},\ \Eprint {http://arxiv.org/abs/1709.10105} {arXiv:1709.10105 [cond-mat.str-el]} \BibitemShut {NoStop}%
\bibitem [{\citenamefont {Chenu}\ \emph {et~al.}(2018)\citenamefont {Chenu}, \citenamefont {Egusquiza}, \citenamefont {Molina-Vilaplana},\ and\ \citenamefont {Del~Campo}}]{Chenu:2017qdv}%
  \BibitemOpen
  \bibfield  {author} {\bibinfo {author} {\bibfnamefont {Aur\'elia}\ \bibnamefont {Chenu}}, \bibinfo {author} {\bibfnamefont {Inigo~L.}\ \bibnamefont {Egusquiza}}, \bibinfo {author} {\bibfnamefont {Javier}\ \bibnamefont {Molina-Vilaplana}}, \ and\ \bibinfo {author} {\bibfnamefont {Adolfo}\ \bibnamefont {Del~Campo}},\ }\bibfield  {title} {\enquote {\bibinfo {title} {{Quantum work statistics, Loschmidt echo and information scrambling}},}\ }\href {\doibase 10.1038/s41598-018-30982-w} {\bibfield  {journal} {\bibinfo  {journal} {Sci. Rep.}\ }\textbf {\bibinfo {volume} {8}},\ \bibinfo {pages} {12634} (\bibinfo {year} {2018})},\ \Eprint {http://arxiv.org/abs/1711.01277} {arXiv:1711.01277 [quant-ph]} \BibitemShut {NoStop}%
\bibitem [{\citenamefont {Chenu}\ \emph {et~al.}(2019)\citenamefont {Chenu}, \citenamefont {Molina-Vilaplana},\ and\ \citenamefont {Del~Campo}}]{Chenu:2018spm}%
  \BibitemOpen
  \bibfield  {author} {\bibinfo {author} {\bibfnamefont {Aur\'elia}\ \bibnamefont {Chenu}}, \bibinfo {author} {\bibfnamefont {Javier}\ \bibnamefont {Molina-Vilaplana}}, \ and\ \bibinfo {author} {\bibfnamefont {Adolfo}\ \bibnamefont {Del~Campo}},\ }\bibfield  {title} {\enquote {\bibinfo {title} {{Work Statistics, Loschmidt Echo and Information Scrambling in Chaotic Quantum Systems}},}\ }\href {\doibase 10.22331/q-2019-03-04-127} {\bibfield  {journal} {\bibinfo  {journal} {Quantum}\ }\textbf {\bibinfo {volume} {3}},\ \bibinfo {pages} {127} (\bibinfo {year} {2019})},\ \Eprint {http://arxiv.org/abs/1804.09188} {arXiv:1804.09188 [quant-ph]} \BibitemShut {NoStop}%
\bibitem [{\citenamefont {Almheiri}\ and\ \citenamefont {Lin}(2022)}]{Almheiri:2021jwq}%
  \BibitemOpen
  \bibfield  {author} {\bibinfo {author} {\bibfnamefont {Ahmed}\ \bibnamefont {Almheiri}}\ and\ \bibinfo {author} {\bibfnamefont {Henry~W.}\ \bibnamefont {Lin}},\ }\bibfield  {title} {\enquote {\bibinfo {title} {{The entanglement wedge of unknown couplings}},}\ }\href {\doibase 10.1007/JHEP08(2022)062} {\bibfield  {journal} {\bibinfo  {journal} {JHEP}\ }\textbf {\bibinfo {volume} {08}},\ \bibinfo {pages} {062} (\bibinfo {year} {2022})},\ \Eprint {http://arxiv.org/abs/2111.06298} {arXiv:2111.06298 [hep-th]} \BibitemShut {NoStop}%
\bibitem [{\citenamefont {P\'erez-Garc\'\i{}a}\ \emph {et~al.}(2022)\citenamefont {P\'erez-Garc\'\i{}a}, \citenamefont {Santilli},\ and\ \citenamefont {Tierz}}]{Perez-Garcia:2022geq}%
  \BibitemOpen
  \bibfield  {author} {\bibinfo {author} {\bibfnamefont {David}\ \bibnamefont {P\'erez-Garc\'\i{}a}}, \bibinfo {author} {\bibfnamefont {Leonardo}\ \bibnamefont {Santilli}}, \ and\ \bibinfo {author} {\bibfnamefont {Miguel}\ \bibnamefont {Tierz}},\ }\bibfield  {title} {\enquote {\bibinfo {title} {{Dynamical quantum phase transitions from random matrix theory}},}\ }\href@noop {} {\  (\bibinfo {year} {2022})},\ \Eprint {http://arxiv.org/abs/2208.01659} {arXiv:2208.01659 [quant-ph]} \BibitemShut {NoStop}%
\bibitem [{\citenamefont {Lieb}\ \emph {et~al.}(1961)\citenamefont {Lieb}, \citenamefont {Schultz},\ and\ \citenamefont {Mattis}}]{Lieb:1961fr}%
  \BibitemOpen
  \bibfield  {author} {\bibinfo {author} {\bibfnamefont {Elliott~H.}\ \bibnamefont {Lieb}}, \bibinfo {author} {\bibfnamefont {Theodore}\ \bibnamefont {Schultz}}, \ and\ \bibinfo {author} {\bibfnamefont {Daniel}\ \bibnamefont {Mattis}},\ }\bibfield  {title} {\enquote {\bibinfo {title} {{Two soluble models of an antiferromagnetic chain}},}\ }\href {\doibase 10.1016/0003-4916(61)90115-4} {\bibfield  {journal} {\bibinfo  {journal} {Annals Phys.}\ }\textbf {\bibinfo {volume} {16}},\ \bibinfo {pages} {407--466} (\bibinfo {year} {1961})}\BibitemShut {NoStop}%
\bibitem [{\citenamefont {Bogoliubov}\ \emph {et~al.}(2011)\citenamefont {Bogoliubov}, \citenamefont {Pronko},\ and\ \citenamefont {Timonen}}]{BPT}%
  \BibitemOpen
  \bibfield  {author} {\bibinfo {author} {\bibfnamefont {N.~M.}\ \bibnamefont {Bogoliubov}}, \bibinfo {author} {\bibfnamefont {A.~G.}\ \bibnamefont {Pronko}}, \ and\ \bibinfo {author} {\bibfnamefont {J.}~\bibnamefont {Timonen}},\ }\bibfield  {title} {\enquote {\bibinfo {title} {Scaling of many-particle correlations in a dissipative sandpile},}\ }\href@noop {} {\  (\bibinfo {year} {2011})},\ \Eprint {http://arxiv.org/abs/1102.5639} {arXiv:1102.5639} \BibitemShut {NoStop}%
\bibitem [{\citenamefont {P\'erez-Garc\'ia}\ and\ \citenamefont {Tierz}(2014)}]{Perez-Garcia:2013lba}%
  \BibitemOpen
  \bibfield  {author} {\bibinfo {author} {\bibfnamefont {David}\ \bibnamefont {P\'erez-Garc\'ia}}\ and\ \bibinfo {author} {\bibfnamefont {Miguel}\ \bibnamefont {Tierz}},\ }\bibfield  {title} {\enquote {\bibinfo {title} {{Mapping between the Heisenberg XX Spin Chain and Low-Energy QCD}},}\ }\href {\doibase 10.1103/PhysRevX.4.021050} {\bibfield  {journal} {\bibinfo  {journal} {Phys. Rev. X}\ }\textbf {\bibinfo {volume} {4}},\ \bibinfo {pages} {021050} (\bibinfo {year} {2014})},\ \Eprint {http://arxiv.org/abs/1305.3877} {arXiv:1305.3877} \BibitemShut {NoStop}%
\bibitem [{\citenamefont {St{\'{e}}phan}(2014)}]{Stephan:2013}%
  \BibitemOpen
  \bibfield  {author} {\bibinfo {author} {\bibfnamefont {Jean-Marie}\ \bibnamefont {St{\'{e}}phan}},\ }\bibfield  {title} {\enquote {\bibinfo {title} {{Emptiness formation probability, Toeplitz determinants, and conformal field theory}},}\ }\href {\doibase 10.1088/1742-5468/2014/05/p05010} {\bibfield  {journal} {\bibinfo  {journal} {J. Stat. Mech.}\ }\textbf {\bibinfo {volume} {2014}},\ \bibinfo {pages} {P05010} (\bibinfo {year} {2014})},\ \Eprint {http://arxiv.org/abs/1303.5499} {arXiv:1303.5499 [cond-mat.stat-mech]} \BibitemShut {NoStop}%
\bibitem [{\citenamefont {Pozsgay}(2013)}]{Pozsgay:2013}%
  \BibitemOpen
  \bibfield  {author} {\bibinfo {author} {\bibfnamefont {Bal{\'{a}}zs}\ \bibnamefont {Pozsgay}},\ }\bibfield  {title} {\enquote {\bibinfo {title} {{The dynamical free energy and the Loschmidt echo for a class of quantum quenches in the Heisenberg spin chain}},}\ }\href {\doibase 10.1088/1742-5468/2013/10/p10028} {\bibfield  {journal} {\bibinfo  {journal} {J. Stat. Mech.}\ }\textbf {\bibinfo {volume} {2013}},\ \bibinfo {pages} {P10028} (\bibinfo {year} {2013})},\ \Eprint {http://arxiv.org/abs/1308.3087} {arXiv:1308.3087 [cond-mat.stat-mech]} \BibitemShut {NoStop}%
\bibitem [{\citenamefont {P\'erez-Garc\'ia}\ and\ \citenamefont {Tierz}(2016)}]{Perez-Garcia:2014aba}%
  \BibitemOpen
  \bibfield  {author} {\bibinfo {author} {\bibfnamefont {David}\ \bibnamefont {P\'erez-Garc\'ia}}\ and\ \bibinfo {author} {\bibfnamefont {Miguel}\ \bibnamefont {Tierz}},\ }\bibfield  {title} {\enquote {\bibinfo {title} {{Chern-Simons theory encoded on a spin chain}},}\ }\href {\doibase 10.1088/1742-5468/2016/01/013103} {\bibfield  {journal} {\bibinfo  {journal} {J. Stat. Mech.}\ }\textbf {\bibinfo {volume} {1601}},\ \bibinfo {pages} {013103} (\bibinfo {year} {2016})},\ \Eprint {http://arxiv.org/abs/1403.6780} {arXiv:1403.6780} \BibitemShut {NoStop}%
\bibitem [{\citenamefont {Viti}\ \emph {et~al.}(2016)\citenamefont {Viti}, \citenamefont {St{\'{e}}phan}, \citenamefont {Dubail},\ and\ \citenamefont {Haque}}]{Viti:2016}%
  \BibitemOpen
  \bibfield  {author} {\bibinfo {author} {\bibfnamefont {Jacopo}\ \bibnamefont {Viti}}, \bibinfo {author} {\bibfnamefont {Jean-Marie}\ \bibnamefont {St{\'{e}}phan}}, \bibinfo {author} {\bibfnamefont {J{\'{e}}r{\^{o}}me}\ \bibnamefont {Dubail}}, \ and\ \bibinfo {author} {\bibfnamefont {Masudul}\ \bibnamefont {Haque}},\ }\bibfield  {title} {\enquote {\bibinfo {title} {{Inhomogeneous quenches in a free fermionic chain: Exact results}},}\ }\href {\doibase 10.1209/0295-5075/115/40011} {\bibfield  {journal} {\bibinfo  {journal} {{EPL}}\ }\textbf {\bibinfo {volume} {115}},\ \bibinfo {pages} {40011} (\bibinfo {year} {2016})},\ \Eprint {http://arxiv.org/abs/1507.08132} {arXiv:1507.08132} \BibitemShut {NoStop}%
\bibitem [{\citenamefont {Krapivsky}\ \emph {et~al.}(2018)\citenamefont {Krapivsky}, \citenamefont {Luck},\ and\ \citenamefont {Mallick}}]{Krapivsky:2017sua}%
  \BibitemOpen
  \bibfield  {author} {\bibinfo {author} {\bibfnamefont {P.~L.}\ \bibnamefont {Krapivsky}}, \bibinfo {author} {\bibfnamefont {J.~M.}\ \bibnamefont {Luck}}, \ and\ \bibinfo {author} {\bibfnamefont {K.}~\bibnamefont {Mallick}},\ }\bibfield  {title} {\enquote {\bibinfo {title} {{Quantum return probability of a system of $N$ non-interacting lattice fermions}},}\ }\href {\doibase 10.1088/1742-5468/aaa79a} {\bibfield  {journal} {\bibinfo  {journal} {J. Stat. Mech.}\ }\textbf {\bibinfo {volume} {1802}},\ \bibinfo {pages} {023104} (\bibinfo {year} {2018})},\ \Eprint {http://arxiv.org/abs/1710.08178} {arXiv:1710.08178 [cond-mat.mes-hall]} \BibitemShut {NoStop}%
\bibitem [{\citenamefont {St{\'{e}}phan}(2017)}]{Stephan:2017}%
  \BibitemOpen
  \bibfield  {author} {\bibinfo {author} {\bibfnamefont {Jean-Marie}\ \bibnamefont {St{\'{e}}phan}},\ }\bibfield  {title} {\enquote {\bibinfo {title} {{Return probability after a quench from a domain wall initial state in the spin-1/2 {XXZ} chain}},}\ }\href {\doibase 10.1088/1742-5468/aa8c19} {\bibfield  {journal} {\bibinfo  {journal} {J. Stat. Mech.}\ }\textbf {\bibinfo {volume} {2017}},\ \bibinfo {pages} {103108} (\bibinfo {year} {2017})},\ \Eprint {http://arxiv.org/abs/1707.06625} {arXiv:1707.06625} \BibitemShut {NoStop}%
\bibitem [{\citenamefont {Santilli}\ and\ \citenamefont {Tierz}(2020)}]{Santilli:2019wvq}%
  \BibitemOpen
  \bibfield  {author} {\bibinfo {author} {\bibfnamefont {Leonardo}\ \bibnamefont {Santilli}}\ and\ \bibinfo {author} {\bibfnamefont {Miguel}\ \bibnamefont {Tierz}},\ }\bibfield  {title} {\enquote {\bibinfo {title} {{Phase transition in complex-time Loschmidt echo of short and long range spin chain}},}\ }\href {\doibase 10.1088/1742-5468/ab837b} {\bibfield  {journal} {\bibinfo  {journal} {J. Stat. Mech.}\ }\textbf {\bibinfo {volume} {2006}},\ \bibinfo {pages} {063102} (\bibinfo {year} {2020})},\ \Eprint {http://arxiv.org/abs/1902.06649} {arXiv:1902.06649} \BibitemShut {NoStop}%
\bibitem [{\citenamefont {St\'ephan}(2022)}]{Stephan:2021yvk}%
  \BibitemOpen
  \bibfield  {author} {\bibinfo {author} {\bibfnamefont {Jean-Marie}\ \bibnamefont {St\'ephan}},\ }\bibfield  {title} {\enquote {\bibinfo {title} {{Exact time evolution formulae in the XXZ spin chain with domain wall initial state}},}\ }\href {\doibase 10.1088/1751-8121/ac5fe8} {\bibfield  {journal} {\bibinfo  {journal} {J. Phys. A}\ }\textbf {\bibinfo {volume} {55}},\ \bibinfo {pages} {204003} (\bibinfo {year} {2022})},\ \Eprint {http://arxiv.org/abs/2112.12092} {arXiv:2112.12092 [cond-mat.stat-mech]} \BibitemShut {NoStop}%
\bibitem [{\citenamefont {Parez}(2022)}]{Parez:2022sgc}%
  \BibitemOpen
  \bibfield  {author} {\bibinfo {author} {\bibfnamefont {Gilles}\ \bibnamefont {Parez}},\ }\bibfield  {title} {\enquote {\bibinfo {title} {{Symmetry-resolved R\'enyi fidelities and quantum phase transitions}},}\ }\href {\doibase 10.1103/PhysRevB.106.235101} {\bibfield  {journal} {\bibinfo  {journal} {Phys. Rev. B}\ }\textbf {\bibinfo {volume} {106}},\ \bibinfo {pages} {235101} (\bibinfo {year} {2022})},\ \Eprint {http://arxiv.org/abs/2208.09457} {arXiv:2208.09457 [cond-mat.stat-mech]} \BibitemShut {NoStop}%
\bibitem [{\citenamefont {Vleeshouwers}\ and\ \citenamefont {Gritsev}(2023)}]{Vleeshouwers:2022eou}%
  \BibitemOpen
  \bibfield  {author} {\bibinfo {author} {\bibfnamefont {Ward~L.}\ \bibnamefont {Vleeshouwers}}\ and\ \bibinfo {author} {\bibfnamefont {Vladimir}\ \bibnamefont {Gritsev}},\ }\bibfield  {title} {\enquote {\bibinfo {title} {{Unitary matrix integrals, symmetric polynomials, and long-range random walks}},}\ }\href {\doibase 10.1088/1751-8121/acc21f} {\bibfield  {journal} {\bibinfo  {journal} {J. Phys. A}\ }\textbf {\bibinfo {volume} {56}},\ \bibinfo {pages} {185002} (\bibinfo {year} {2023})},\ \Eprint {http://arxiv.org/abs/2209.06538} {arXiv:2209.06538 [cond-mat.stat-mech]} \BibitemShut {NoStop}%
\bibitem [{\citenamefont {Jordan}\ and\ \citenamefont {Wigner}(1928)}]{Jordan:1928wi}%
  \BibitemOpen
  \bibfield  {author} {\bibinfo {author} {\bibfnamefont {Pascual}\ \bibnamefont {Jordan}}\ and\ \bibinfo {author} {\bibfnamefont {Eugene~P.}\ \bibnamefont {Wigner}},\ }\bibfield  {title} {\enquote {\bibinfo {title} {{About the Pauli exclusion principle}},}\ }\href {\doibase 10.1007/BF01331938} {\bibfield  {journal} {\bibinfo  {journal} {Z. Phys.}\ }\textbf {\bibinfo {volume} {47}},\ \bibinfo {pages} {631--651} (\bibinfo {year} {1928})}\BibitemShut {NoStop}%
\bibitem [{\citenamefont {Kinney}\ \emph {et~al.}(2007)\citenamefont {Kinney}, \citenamefont {Maldacena}, \citenamefont {Minwalla},\ and\ \citenamefont {Raju}}]{Kinney:2005ej}%
  \BibitemOpen
  \bibfield  {author} {\bibinfo {author} {\bibfnamefont {Justin}\ \bibnamefont {Kinney}}, \bibinfo {author} {\bibfnamefont {Juan~Martin}\ \bibnamefont {Maldacena}}, \bibinfo {author} {\bibfnamefont {Shiraz}\ \bibnamefont {Minwalla}}, \ and\ \bibinfo {author} {\bibfnamefont {Suvrat}\ \bibnamefont {Raju}},\ }\bibfield  {title} {\enquote {\bibinfo {title} {{An Index for 4 dimensional super conformal theories}},}\ }\href {\doibase 10.1007/s00220-007-0258-7} {\bibfield  {journal} {\bibinfo  {journal} {Commun. Math. Phys.}\ }\textbf {\bibinfo {volume} {275}},\ \bibinfo {pages} {209--254} (\bibinfo {year} {2007})},\ \Eprint {http://arxiv.org/abs/hep-th/0510251} {arXiv:hep-th/0510251} \BibitemShut {NoStop}%
\bibitem [{\citenamefont {Romelsberger}(2006)}]{Romelsberger:2005eg}%
  \BibitemOpen
  \bibfield  {author} {\bibinfo {author} {\bibfnamefont {Christian}\ \bibnamefont {Romelsberger}},\ }\bibfield  {title} {\enquote {\bibinfo {title} {{Counting chiral primaries in N = 1, d=4 superconformal field theories}},}\ }\href {\doibase 10.1016/j.nuclphysb.2006.03.037} {\bibfield  {journal} {\bibinfo  {journal} {Nucl. Phys. B}\ }\textbf {\bibinfo {volume} {747}},\ \bibinfo {pages} {329--353} (\bibinfo {year} {2006})},\ \Eprint {http://arxiv.org/abs/hep-th/0510060} {arXiv:hep-th/0510060} \BibitemShut {NoStop}%
\bibitem [{\citenamefont {Romelsberger}(2007)}]{Romelsberger:2007ec}%
  \BibitemOpen
  \bibfield  {author} {\bibinfo {author} {\bibfnamefont {Christian}\ \bibnamefont {Romelsberger}},\ }\bibfield  {title} {\enquote {\bibinfo {title} {{Calculating the Superconformal Index and Seiberg Duality}},}\ }\href@noop {} {\  (\bibinfo {year} {2007})},\ \Eprint {http://arxiv.org/abs/0707.3702} {arXiv:0707.3702} \BibitemShut {NoStop}%
\bibitem [{\citenamefont {Rastelli}\ and\ \citenamefont {Razamat}(2017)}]{Rastelli:2016tbz}%
  \BibitemOpen
  \bibfield  {author} {\bibinfo {author} {\bibfnamefont {Leonardo}\ \bibnamefont {Rastelli}}\ and\ \bibinfo {author} {\bibfnamefont {Shlomo~S.}\ \bibnamefont {Razamat}},\ }\bibfield  {title} {\enquote {\bibinfo {title} {{The supersymmetric index in four dimensions}},}\ }\href {\doibase 10.1088/1751-8121/aa76a6} {\bibfield  {journal} {\bibinfo  {journal} {J. Phys. A}\ }\textbf {\bibinfo {volume} {50}},\ \bibinfo {pages} {443013} (\bibinfo {year} {2017})},\ \Eprint {http://arxiv.org/abs/1608.02965} {arXiv:1608.02965} \BibitemShut {NoStop}%
\bibitem [{\citenamefont {Witten}(1982)}]{Witten:1982im}%
  \BibitemOpen
  \bibfield  {author} {\bibinfo {author} {\bibfnamefont {Edward}\ \bibnamefont {Witten}},\ }\bibfield  {title} {\enquote {\bibinfo {title} {{Supersymmetry and Morse theory}},}\ }\href@noop {} {\bibfield  {journal} {\bibinfo  {journal} {J. Diff. Geom.}\ }\textbf {\bibinfo {volume} {17}},\ \bibinfo {pages} {661--692} (\bibinfo {year} {1982})}\BibitemShut {NoStop}%
\bibitem [{\citenamefont {'t~Hooft}(1978)}]{tHooft:1977nqb}%
  \BibitemOpen
  \bibfield  {author} {\bibinfo {author} {\bibfnamefont {Gerard}\ \bibnamefont {'t~Hooft}},\ }\bibfield  {title} {\enquote {\bibinfo {title} {{On the Phase Transition Towards Permanent Quark Confinement}},}\ }\href {\doibase 10.1016/0550-3213(78)90153-0} {\bibfield  {journal} {\bibinfo  {journal} {Nucl. Phys. B}\ }\textbf {\bibinfo {volume} {138}},\ \bibinfo {pages} {1--25} (\bibinfo {year} {1978})}\BibitemShut {NoStop}%
\bibitem [{\citenamefont {Polyakov}(1978)}]{Polyakov:1978vu}%
  \BibitemOpen
  \bibfield  {author} {\bibinfo {author} {\bibfnamefont {Alexander~M.}\ \bibnamefont {Polyakov}},\ }\bibfield  {title} {\enquote {\bibinfo {title} {{Thermal Properties of Gauge Fields and Quark Liberation}},}\ }\href {\doibase 10.1016/0370-2693(78)90737-2} {\bibfield  {journal} {\bibinfo  {journal} {Phys. Lett. B}\ }\textbf {\bibinfo {volume} {72}},\ \bibinfo {pages} {477--480} (\bibinfo {year} {1978})}\BibitemShut {NoStop}%
\bibitem [{\citenamefont {Hartnoll}\ and\ \citenamefont {Kumar}(2006)}]{Hartnoll:2006}%
  \BibitemOpen
  \bibfield  {author} {\bibinfo {author} {\bibfnamefont {Sean~A.}\ \bibnamefont {Hartnoll}}\ and\ \bibinfo {author} {\bibfnamefont {S.~Prem}\ \bibnamefont {Kumar}},\ }\bibfield  {title} {\enquote {\bibinfo {title} {{Higher rank Wilson loops from a matrix model}},}\ }\href {\doibase 10.1088/1126-6708/2006/08/026} {\bibfield  {journal} {\bibinfo  {journal} {JHEP}\ }\textbf {\bibinfo {volume} {08}},\ \bibinfo {pages} {026} (\bibinfo {year} {2006})},\ \Eprint {http://arxiv.org/abs/hep-th/0605027} {arXiv:hep-th/0605027} \BibitemShut {NoStop}%
\bibitem [{\citenamefont {Santilli}(2021)}]{Santilli:2021rcp}%
  \BibitemOpen
  \bibfield  {author} {\bibinfo {author} {\bibfnamefont {Leonardo}\ \bibnamefont {Santilli}},\ }\bibfield  {title} {\enquote {\bibinfo {title} {{Phases of five-dimensional supersymmetric gauge theories}},}\ }\href {\doibase 10.1007/JHEP07(2021)088} {\bibfield  {journal} {\bibinfo  {journal} {JHEP}\ }\textbf {\bibinfo {volume} {07}},\ \bibinfo {pages} {088} (\bibinfo {year} {2021})},\ \Eprint {http://arxiv.org/abs/2103.14049} {arXiv:2103.14049} \BibitemShut {NoStop}%
\bibitem [{\citenamefont {Suzuki}(1971)}]{Suzuki1971}%
  \BibitemOpen
  \bibfield  {author} {\bibinfo {author} {\bibfnamefont {Masuo}\ \bibnamefont {Suzuki}},\ }\bibfield  {title} {\enquote {\bibinfo {title} {{Relationship among Exactly Soluble Models of Critical Phenomena. I*): 2D Ising Model, Dimer Problem and the Generalized XY-Model}},}\ }\href {\doibase 10.1143/PTP.46.1337} {\bibfield  {journal} {\bibinfo  {journal} {Progress of Theoretical Physics}\ }\textbf {\bibinfo {volume} {46}},\ \bibinfo {pages} {1337--1359} (\bibinfo {year} {1971})}\BibitemShut {NoStop}%
\bibitem [{\citenamefont {Derzhko}(2008)}]{derzhko2008jordan}%
  \BibitemOpen
  \bibfield  {author} {\bibinfo {author} {\bibfnamefont {Oleg}\ \bibnamefont {Derzhko}},\ }\bibfield  {title} {\enquote {\bibinfo {title} {{Jordan-Wigner fermionization and the theory of low-dimensional quantum spin models. Dynamic properties}},}\ }in\ \href {\doibase 10.1142/9789812709455_0002} {\emph {\bibinfo {booktitle} {{Condensed Matter Physics in the Prime of the 21st Century: Phenomena, Materials, Ideas, Methods}}}}\ (\bibinfo  {publisher} {World Scientific},\ \bibinfo {year} {2008})\ pp.\ \bibinfo {pages} {35--87},\ \Eprint {http://arxiv.org/abs/0812.4750} {arXiv:0812.4750 [cond-mat.str-el]} \BibitemShut {NoStop}%
\bibitem [{\citenamefont {\'Alvarez}\ \emph {et~al.}(2016)\citenamefont {\'Alvarez}, \citenamefont {Mart\'\i{}nez~Alonso},\ and\ \citenamefont {Medina}}]{Alvarez:2016rmo}%
  \BibitemOpen
  \bibfield  {author} {\bibinfo {author} {\bibfnamefont {Gabriel}\ \bibnamefont {\'Alvarez}}, \bibinfo {author} {\bibfnamefont {Luis}\ \bibnamefont {Mart\'\i{}nez~Alonso}}, \ and\ \bibinfo {author} {\bibfnamefont {Elena}\ \bibnamefont {Medina}},\ }\bibfield  {title} {\enquote {\bibinfo {title} {{Complex saddles in the Gross-Witten-Wadia matrix model}},}\ }\href {\doibase 10.1103/PhysRevD.94.105010} {\bibfield  {journal} {\bibinfo  {journal} {Phys. Rev. D}\ }\textbf {\bibinfo {volume} {94}},\ \bibinfo {pages} {105010} (\bibinfo {year} {2016})},\ \Eprint {http://arxiv.org/abs/1610.09948} {arXiv:1610.09948 [hep-th]} \BibitemShut {NoStop}%
\bibitem [{\citenamefont {Santilli}\ and\ \citenamefont {Tierz}(2022)}]{Santilli:2021eon}%
  \BibitemOpen
  \bibfield  {author} {\bibinfo {author} {\bibfnamefont {Leonardo}\ \bibnamefont {Santilli}}\ and\ \bibinfo {author} {\bibfnamefont {Miguel}\ \bibnamefont {Tierz}},\ }\bibfield  {title} {\enquote {\bibinfo {title} {{Multiple phases and meromorphic deformations of unitary matrix models}},}\ }\href {\doibase 10.1016/j.nuclphysb.2022.115694} {\bibfield  {journal} {\bibinfo  {journal} {Nucl. Phys. B}\ }\textbf {\bibinfo {volume} {976}},\ \bibinfo {pages} {115694} (\bibinfo {year} {2022})},\ \Eprint {http://arxiv.org/abs/2102.11305} {arXiv:2102.11305} \BibitemShut {NoStop}%
\bibitem [{\citenamefont {Szeg\H{o}}(1952)}]{Szegoth}%
  \BibitemOpen
  \bibfield  {author} {\bibinfo {author} {\bibfnamefont {Gabor}\ \bibnamefont {Szeg\H{o}}},\ }\bibfield  {title} {\enquote {\bibinfo {title} {{On certain Hermitian forms associated with the Fourier series of a positive function}},}\ }\href@noop {} {\bibfield  {journal} {\bibinfo  {journal} {Comm. S\'em. Math. Univ. Lund}\ }\textbf {\bibinfo {volume} {Tome Suppl\'ementaire}},\ \bibinfo {pages} {228–238} (\bibinfo {year} {1952})}\BibitemShut {NoStop}%
\bibitem [{\citenamefont {Korsch}\ \emph {et~al.}(2006)\citenamefont {Korsch}, \citenamefont {Klumpp},\ and\ \citenamefont {Witthaut}}]{Korsch:2006}%
  \BibitemOpen
  \bibfield  {author} {\bibinfo {author} {\bibfnamefont {H.~J.}\ \bibnamefont {Korsch}}, \bibinfo {author} {\bibfnamefont {A.}~\bibnamefont {Klumpp}}, \ and\ \bibinfo {author} {\bibfnamefont {D.}~\bibnamefont {Witthaut}},\ }\bibfield  {title} {\enquote {\bibinfo {title} {{On two-dimensional Bessel functions}},}\ }\href {\doibase 10.1088/0305-4470/39/48/008} {\bibfield  {journal} {\bibinfo  {journal} {J. Phys. A}\ }\textbf {\bibinfo {volume} {39}},\ \bibinfo {pages} {14947} (\bibinfo {year} {2006})},\ \Eprint {http://arxiv.org/abs/quant-ph/0608216} {arXiv:quant-ph/0608216 [quant-ph]} \BibitemShut {NoStop}%
\bibitem [{\citenamefont {Minahan}\ and\ \citenamefont {Zarembo}(2003)}]{Minahan:2002ve}%
  \BibitemOpen
  \bibfield  {author} {\bibinfo {author} {\bibfnamefont {J.~A.}\ \bibnamefont {Minahan}}\ and\ \bibinfo {author} {\bibfnamefont {K.}~\bibnamefont {Zarembo}},\ }\bibfield  {title} {\enquote {\bibinfo {title} {{The Bethe ansatz for N=4 superYang-Mills}},}\ }\href {\doibase 10.1088/1126-6708/2003/03/013} {\bibfield  {journal} {\bibinfo  {journal} {JHEP}\ }\textbf {\bibinfo {volume} {03}},\ \bibinfo {pages} {013} (\bibinfo {year} {2003})},\ \Eprint {http://arxiv.org/abs/hep-th/0212208} {arXiv:hep-th/0212208} \BibitemShut {NoStop}%
\bibitem [{\citenamefont {Beisert}\ and\ \citenamefont {Staudacher}(2003)}]{Beisert:2003yb}%
  \BibitemOpen
  \bibfield  {author} {\bibinfo {author} {\bibfnamefont {Niklas}\ \bibnamefont {Beisert}}\ and\ \bibinfo {author} {\bibfnamefont {Matthias}\ \bibnamefont {Staudacher}},\ }\bibfield  {title} {\enquote {\bibinfo {title} {{The N=4 SYM integrable super spin chain}},}\ }\href {\doibase 10.1016/j.nuclphysb.2003.08.015} {\bibfield  {journal} {\bibinfo  {journal} {Nucl. Phys. B}\ }\textbf {\bibinfo {volume} {670}},\ \bibinfo {pages} {439--463} (\bibinfo {year} {2003})},\ \Eprint {http://arxiv.org/abs/hep-th/0307042} {arXiv:hep-th/0307042} \BibitemShut {NoStop}%
\bibitem [{\citenamefont {Gesteau}\ and\ \citenamefont {Santilli}(2024)}]{Gesteau:2024dhj}%
  \BibitemOpen
  \bibfield  {author} {\bibinfo {author} {\bibfnamefont {Elliott}\ \bibnamefont {Gesteau}}\ and\ \bibinfo {author} {\bibfnamefont {Leonardo}\ \bibnamefont {Santilli}},\ }\bibfield  {title} {\enquote {\bibinfo {title} {{Explicit large $N$ von Neumann algebras from matrix models}},}\ }\href@noop {} {\  (\bibinfo {year} {2024})},\ \Eprint {http://arxiv.org/abs/2402.10262} {arXiv:2402.10262 [hep-th]} \BibitemShut {NoStop}%
\bibitem [{\citenamefont {Leutheusser}\ and\ \citenamefont {Liu}(2023{\natexlab{a}})}]{Leutheusser:2021qhd}%
  \BibitemOpen
  \bibfield  {author} {\bibinfo {author} {\bibfnamefont {Samuel}\ \bibnamefont {Leutheusser}}\ and\ \bibinfo {author} {\bibfnamefont {Hong}\ \bibnamefont {Liu}},\ }\bibfield  {title} {\enquote {\bibinfo {title} {{Causal connectability between quantum systems and the black hole interior in holographic duality}},}\ }\href {\doibase 10.1103/PhysRevD.108.086019} {\bibfield  {journal} {\bibinfo  {journal} {Phys. Rev. D}\ }\textbf {\bibinfo {volume} {108}},\ \bibinfo {pages} {086019} (\bibinfo {year} {2023}{\natexlab{a}})},\ \Eprint {http://arxiv.org/abs/2110.05497} {arXiv:2110.05497 [hep-th]} \BibitemShut {NoStop}%
\bibitem [{\citenamefont {Leutheusser}\ and\ \citenamefont {Liu}(2023{\natexlab{b}})}]{Leutheusser:2021frk}%
  \BibitemOpen
  \bibfield  {author} {\bibinfo {author} {\bibfnamefont {Samuel}\ \bibnamefont {Leutheusser}}\ and\ \bibinfo {author} {\bibfnamefont {Hong}\ \bibnamefont {Liu}},\ }\bibfield  {title} {\enquote {\bibinfo {title} {{Emergent times in holographic duality}},}\ }\href {\doibase 10.1103/PhysRevD.108.086020} {\bibfield  {journal} {\bibinfo  {journal} {Phys. Rev. D}\ }\textbf {\bibinfo {volume} {108}},\ \bibinfo {pages} {086020} (\bibinfo {year} {2023}{\natexlab{b}})},\ \Eprint {http://arxiv.org/abs/2112.12156} {arXiv:2112.12156 [hep-th]} \BibitemShut {NoStop}%
\bibitem [{\citenamefont {Basteiro}\ \emph {et~al.}(2024)\citenamefont {Basteiro}, \citenamefont {Di~Giulio}, \citenamefont {Erdmenger},\ and\ \citenamefont {Xian}}]{Basteiro:2024cuh}%
  \BibitemOpen
  \bibfield  {author} {\bibinfo {author} {\bibfnamefont {Pablo}\ \bibnamefont {Basteiro}}, \bibinfo {author} {\bibfnamefont {Giuseppe}\ \bibnamefont {Di~Giulio}}, \bibinfo {author} {\bibfnamefont {Johanna}\ \bibnamefont {Erdmenger}}, \ and\ \bibinfo {author} {\bibfnamefont {Zhuo-Yu}\ \bibnamefont {Xian}},\ }\bibfield  {title} {\enquote {\bibinfo {title} {{Entanglement in Interacting Majorana Chains and Transitions of von Neumann Algebras}},}\ }\href {\doibase 10.1103/PhysRevLett.132.161604} {\bibfield  {journal} {\bibinfo  {journal} {Phys. Rev. Lett.}\ }\textbf {\bibinfo {volume} {132}},\ \bibinfo {pages} {161604} (\bibinfo {year} {2024})},\ \Eprint {http://arxiv.org/abs/2401.04764} {arXiv:2401.04764 [hep-th]} \BibitemShut {NoStop}%
\bibitem [{\citenamefont {Gesteau}(2023)}]{Gesteau:2023rrx}%
  \BibitemOpen
  \bibfield  {author} {\bibinfo {author} {\bibfnamefont {Elliott}\ \bibnamefont {Gesteau}},\ }\bibfield  {title} {\enquote {\bibinfo {title} {{Emergent spacetime and the ergodic hierarchy}},}\ }\href@noop {} {\  (\bibinfo {year} {2023})},\ \Eprint {http://arxiv.org/abs/2310.13733} {arXiv:2310.13733 [hep-th]} \BibitemShut {NoStop}%
\end{thebibliography}%
\end{document}